\documentclass[prb,twocolumn,preprintnumbers,amsmath,amssymb]{revtex4}

\usepackage{graphicx}

\begin{document}
\title{ \bf  Quantum phases, Supersolids and quantum phase transitions of interacting bosons in frustrated lattices }
\author{  Yan Chen$^{1}$ and Jinwu Ye $^{2,3}$  }
\affiliation{
$^{1}$Department of Physics, State Key Laboratory of Surface Physics and Lab of Advanced Materials, Fudan
University, Shanghai, 200433, China\\
$^{2}$ Department of Physics and Astronomy, Mississippi State
University, P. O. Box 5167, Mississippi State, MS, 39762   \\
$^{3}$ Department of Physics, Capital Normal University,
Beijing, 100048 China }
\date{\today}

\begin{abstract}

    By using the dual vortex method (DVM), we develop systematically a simple and effective scheme to use the vortex degree of freedoms on dual lattices to characterize
    the symmetry breaking patterns of the boson insulating states in the direct lattices.
    Then we apply our scheme to study quantum phases and phase transitions in an  extended boson Hubbard model
    slightly away from $ 1/3 $ ( $ 2/3 $ ) filling on frustrated lattices such as triangular and kagome lattice.
    In a triangular lattice at $ 1/3 $, we find a X-CDW, a stripe CDW
    phase which was found perviously by a density operator formalism (DOF). Most importantly, we also find a new CDW-VB phase which has
    both local CDW and local VB orders, in sharp contrast to a bubble CDW phase found previously by the DOF.
    In the Kagome lattice at $ 1/3 $, we find a VBS phase and a 6 fold-CDW phase. Most importantly, we also identify a CDW-VB phase which has
    both local CDW and local VB orders which was found in previous QMC simulations.
    We also study several other phases which are not found by the DVM.
    By analyzing carefully the saddle point structures of the dual gauge fields in the translational symmetry breaking sides
    and  pushing the effective actions slightly
    away from the commensurate filling $ f=1/3 $( $ 2/3 $ ), we classified all the possible types of supersolids and analyze their stability conditions.
    In a triangular lattice, there are X-CDW supersolid, stripe CDW  supersolid, but absence of any valence bond
    supersolid ( VB-SS ). There are also  a new kind of supersolid: CDW-VB supersolid.
    In a Kagome lattice,  there are 6 fold-CDW supersolid, stripe CDW  supersolid, but absence of any valence bond
    supersolid ( VB-SS ). There are also  a new kind of supersolid: CDW-VB supersolid.
    We show that independent of the types of the SS, the quantum phase transitions from solids to
    supersolids driven by a chemical potential are in the
    same universality class as that from a Mott insulator to a superfluid,
    therefore have exact exponents $ z=2, \nu=1/2, \eta=0 $ ( with logarithmic corrections ).
    Excitation spectra of all these insulating phases and supersolid phases are also studied.
    Implications on QMC simulations with both nearest neighbor  and next nearest neighbor interactions in both lattices are given.
    Some possible intrinsic problems of the DOF in identifying the insulating phases are also pointed out.
\end{abstract}

\maketitle

\section{ Introduction }

     The EBHM with various  kinds of interactions,
     at various kinds lattices ( bi-partisan or frustrated ) at various kinds of filling factors
     ( commensurate or in-commensurate ) is described by the following Hamiltonian \cite{gan,pq1,1d,five,sandvik,frusrev}:
\begin{eqnarray}
  H  & = & -t \sum_{ < ij > } ( b^{\dagger}_{i} b_{j} + h.c. ) - \mu \sum_{i} n_{i} +
       \frac{U}{2} \sum_{i} n_{i} ( n_{i} -1 )     \nonumber  \\
      & + &  V_{1} \sum_{ <ij> } n_{i} n_{j}  + V_{2} \sum_{ <<ik>> } n_{i} n_{k} + \cdots
\label{boson}
\end{eqnarray}
    where $ n_{i} = b^{\dagger}_{i} b_{i} $ is the boson density, $
    t $ is the hopping amplitude,
    $ U, V_{1}, V_{2} $ are onsite, nearest neighbor ( $ NN $ ) and next nearest neighbor ( $ NNN $ ) interactions
    between the bosons. The $ \cdots $ may include further neighbor interactions and
    possible ring-exchange interactions.  A supersolid in Eqn.\ref{boson} is defined as to have both
    off-diagonal long range order $ < b_{i} > \neq 0 $ and diagonal
    charge density wave in the boson density $ n_{i} $ which breaks the lattice symmetry.
    In bipartite lattices, the sign of the hopping $ t $  in Eqn.\ref{boson}
    makes no differences. However, in frustrated lattices, the sign of the hopping $ t $  in Eqn.\ref{boson} makes crucial differences \cite{gan}.
    These differences lead to many new and interesting physics in frustrated
    lattices which are quite different than those in bipartite lattices studied in \cite{pq1,univ}.

    The EBHM Eqn.\ref{boson} can be easily realized in cold atom experiments \cite{boson,bloch}. The on-site interaction
    $ U $ can be tuned by the Feshbach resonance. Various kinds of optical lattices  can be realized by suitably choosing
    the geometry of the laser beams forming the optical lattices.
    For example, using three coplanar beams of equal intensity having
    the three vectors making a $ 120 ^{\circ}$ angle with each other \cite{honeyexp},
    the potential wells have their minima in a honeycomb or a triangular lattice.
    Four beams travel along the three fold symmetry axes of a regular
    tetrahedron, the potential wells have their minima in a body-centered-cubic lattice \cite{honeyexp}.
    The authors in  \cite{kalatticeexp} proposed to create a kagome optical lattice using superlattice
    techniques. There are many possible ways to generate longer range interaction $ V_{1}, V_{2},....$ of
    ultra-cold atoms loaded in optical lattices (1)
     Very exciting perspectives have been opened by recent
     experiments on cooling and trapping of $ ^{52}Cr $ atoms \cite{cromium}
     and fermionic polar molecules $ ^{40}K+ ^{87} Rb $\cite{junpolar,polarlayer}.
     The bosonic polar molecules $ ^{39} K+ ^{87} Rb $ is also
     expected to be realized in the near future.
     Being magnetically  or electrically polarized, the $ ^{52}Cr $  atoms or polar molecules
     interact with each other via long-rang anisotropic
     dipole-dipole interactions. Loading the $ ^{52}Cr $ or the polar molecules on a 2d optical lattice
     with the dipole moments perpendicular to the trapping plane can be mapped to
     Eqn.\ref{boson} with long-range  repulsive interactions $ \sim p^{2}/r^{3}
     $ where $ p $ is the dipole moment.  The CDW supersolid phases described  by QMC simulations in \cite{squarehard} and
     by the dual vortex method (DVM)  in \cite{univ} was numerically found to be stable in large parameter regimes  in this system \cite{polarsszoller}.
     Possible techniques to generate long-range interactions in a gas of groundstate alkali atoms by
     weakly admixing excited Rydberg states with laser light was proposed in \cite{mixture}.
     Very recently, the cold bosons or fermions in optical lattice can interact with a very long range interaction due to the exchange of cavity photons
     \cite{orbital,phased}.  The generation of the ring exchange interaction has been discussed in \cite{qs}.

    Recently, the EBHM Eqn.\ref{boson} in a square lattice
    at generic commensurate filling factors $ f=p/q $ ( $ p, q $ are relative prime numbers ) were
    systematically studied in \cite{pq1} by a dual vortex method ( DVM ). The DVM method was applied by one of the authors to study quantum phases and phase transitions
    of the EBHM Eqn.\ref{boson} on bipartite lattices such as honeycome and square lattice  at
    half and near half fillings  and its possible
    experimental applications in cold atoms, adatom absorptions on substrates and possible Cooper-pair supersolids
    in high temperature superconductors. It was pointed out in \cite{univ} that the DVM developed in \cite{pq1} holds in the superfluid ( SF ) and
     the valence bond solid ( VBS )  side where the saddle point of the dual gauge field can be taken as uniform,
     but cares are needed to apply the DVM developed in \cite{pq1} in the CDW side where the saddle point of the dual gauge field can {\em not} be taken as
     uniform anymore.  When studying phases and phase traditions slightly away from  $ 1/2 $ filling, special cares are needed to choose
     a correct saddle point of the dual gauge field  in the CDW side, so a different effective action
     is needed in the CDW side to make the theory self-consistent \cite{univ}.
     In \cite{univ}, by extending the DVM explicitly to the lattice symmetry breaking side
     by choosing the corresponding self-consistent saddle points of the dual gauge field,
     then by pushing the DVM to slightly away from commensurate filling
     factors, the author mapped out the global phase diagram at $ T=0 $ of
     the chemical potential $ \mu $  versus the ratio of the kinetic energy  over the interaction $ t/V_1 $.
     It was found that in the insulating side, different kinds of supersolids
     are generic possible states slightly away from half filling. A novel kind of supersolid called valence bond supersolid  ( VB-SS ) was proposed.
    It was also shown that the transition from a CDW  ( Valence bond ) solid to a CDW ( valence bond ) supersolid
    driven by the chemical potential $ \mu $
    is in the same universality class as the Mott to the SF with the exact exponents $ z=2, \nu=1/2, \eta=0 $ subject to a logarithmic correction.
    It was also suggested that the density order formalism (DOF)  constructed in \cite{pq1} to characterize the symmetry breaking  patterns  in the insulating states
    are not physically transparent.
    It is very hard to generalize the DOF to a honeycomb lattice and other lattices.
    The subsequent QMC simulations on the EBHM of hard and soft core bosons
    with the nearest neighbor ($ NN $) $ V_1 $ interaction on a honeycomb lattice  \cite{honey}
    indeed found a stable X-supersolid phase  in the soft core case slightly above the half filling.

    The EBHM Eqn.\ref{boson} in frustrated lattices have also been studied by various methods such as the spin wave expansion \cite{gan},
    QMC simulations \cite{ss12,ss3,ss4} and the DVM \cite{tri,ka1}.
    In the hard core limit $ U = \infty $, Eqn.\ref{boson} can be mapped to a $ XXZ $ quantum spin model \cite{five,univ}.
    Using a  $ 1/S $ spin wave expansion on the resulting quantum spin model,  the authors in \cite{gan} found that a SS state is more robust in
    a triangular lattice with only $ t $ and $ V_{1} $ terms in
    Eqn.\ref{boson} and  a SS state with $ \sqrt{3} \times \sqrt{3} $
    pattern is stable even at half filling $ f=1/2 $.  Indeed, this discovery was confirmed by several recent QMC simulations
    in the EBHM of hard core bosons with the nearest neighbor ($ NN $) $ V_1 $ interaction \cite{ss12,ss3,ss4}.
    However, due to too strong quantum fluctuations, the authors in \cite{gan} found that the
    spin wave expansion does not work anymore in Kagome lattice. So a different
    analytical method is needed to study possible supersolid phases in a Kagome lattice.

    Indeed, the DVM method was extended to study the EBHM  in a triangular lattice at $ 1/3 $ and $ 1/2 $ fillings in \cite{tri}.
    Starting from the SF side, the authors derived the effective action Eqn.\ref{tri} to describe
    quantum phase transitions from the SF to some insulating states with several kinds of lattice symmetry breaking patterns.
    By generalizing the density operator formalism (DOF) in the square lattice constructed in \cite{pq1} to a triangular lattice,
    they identified an X-solid phase at  $ f=1/3 $ in the Ising limit of the effective action, a stripe solid phase in one easy-plane limit,
    especially a bubble solid phase in the other easy-plane limit.
    They suggested that the transition from the SF to the stripe solid could be second order through the so-called
    de-confined quantum critical point \cite{senthil}.
    Then subsequent QMC simulations  in \cite{trinnn} on the EBHM Eqn.\ref{bosonnnn} of hard core bosons with a next nearest neighbor ($ NN $N) interaction
    indeed found the stripe solid phase as a ground state at $ f=1/3 $ in some parameter regimes.
    The QMC in \cite{trinnn} seems also to find a meta-stable bubble phase which has higher energy than the stripe phase at the {\sl same} parameter regimes.
    However, they found the transition from
    the SF to the stripe solid phase at $ f=1/3 $ is a strong first order transition in contrast to the claim made in \cite{tri}.

    Later, the DVM was applied to study the EBHM on a Kagome lattice in \cite{ka1}. From the Magnetic space group (MSG),
    starting from the SF side, the authors derived the effective action Eqn.\ref{ka} at $ \delta f=0 $ to describe quantum phase transitions
    from the SF to some insulating states with some lattice symmetry breaking patterns.
    Realizing it is very difficult to generalize the DOF in the square lattice constructed in \cite{pq1} to the Kagome lattice,
    they tried to draw the vortex fields on the dual dice lattice to identify possible symmetry breaking patterns in the insulating side.
    Unfortunately, because the vortex fields are gauge dependent, so may not be used to characterize symmetry breaking patterns effectively.
    There are also two QMC simulations on the EBHM of hard core bosons with nearest neighbor ($ NN $) interaction in a Kagome lattice \cite{ka2,ka3}.
    Both groups found that the solid state at  $ f=1/3 $ has both CDW and VB order. The VB order
    corresponds to boson hopping around a hexagon ( see Fig.14b). However, they did not see any stable supersolid phase away from the $ 1/3 $ filling
    in the hard core case.  The QMC in Ref.\cite{ka3} suggested that the the SF to the CDW+VB solid transition is a 2nd order transition through the
    deconfined quantum critical point \cite{senthil}. However, using a double-peaked histogram of the boson kinetic energy,
    the QMC in Ref.\cite{ka2} concluded that a weakly 1st order transition.

    In this paper, we will use the DVM developed in \cite{pq1,univ}
    to study quantum phases, especially various kinds of insulating phases and phase transitions in the two most common
    frustrated lattices such as triangular and kagome lattices at and slightly away $ 1/3 $ and $ 2/3 $ fillings.
    Most importantly, we will develop a systematic way to determine the symmetry breaking patterns of the insulating states
    in the direct lattices in terms of the vortex degree of freedoms only at the dual lattice points.
    These vortex degree of freedoms are  the gauge invariant physical vortex densities,
    the kinetic energies and vortex currents defined in Eqn.\ref{density},\ref{bond} which are the central results achieved in this paper.
    The importance of gauge invariant electron Green functions has been
    discussed previously in the context of Angle-Resolved Photo-Emission
    Spectroscopy (ARPES) of high temperature superconductors \cite{gauge1,gauge2}.
    In the continuum limit discussed in \cite{gauge1,gauge2}, the
    gauge invariant  propagator becomes path dependent. For
    simplicity, a straight line connecting the two points in the  propagator is chosen.
    In all the lattice models discussed in this paper, all the gauge
    invariant vortex quantities are defined on every point and every
    link in the dual lattice, so any ambiguity associated with paths disappears.
    These vortex quantities can determine the boson density in every lattice point and boson
    kinetic energy in every link in the direct lattice without any
    ambiguity.
    We contrast different symmetry breaking patterns
    of the insulating phases in both Ising and Easy plane limits in the two lattices.
    We will also compare our results with some previous Quantum Monte-Carlo
    (QMC) results in the two lattices \cite{trinnn,ka2} and also give important implications to possible future
     QMC simulations on the EBHM  Eqn.\ref{boson} with both hard and soft cores,
    with both the $ NN $ and the $ NNN $ interactions.
    The method developed here should be very general and can be used to characterize uniquely the symmetry
    breaking patterns in any lattices at any  filling factors
    by using the vortex degree of freedoms only at the corresponding dual lattice points.

    In a triangular lattice, we found the X-solid phase in the Ising limit $ v > 0 $,
    the stripe phase in one of the easy plane limit $ v <0, w< 0 $, both of which were
    found previously by the density operator formalism (DOF) in \cite{tri}. However, we identify a novel phase with both CDW and VBS orders
    shown in Fig.7 in another easy plane limit $ v<0, w>0 $.
    This finding differs from the bubble solid phase shown in Fig.25 in the same easy plane limit found by the DOF in \cite{tri}.
    This disagreement calls for the reexamination of the correctness
    of the density operator formalism in terms of the dual vortex degree of freedoms in any general lattices at general filling  factors.
    We then apply our method to a Kagome lattice. We find that the quantum phases and phase transitions in a Kagome lattice in both Ising and easy plane limit are
    dramatically different from those in bipartite lattices \cite{pq1,univ} studied previously \cite{tri} and the triangular
    lattice studied in the Sec.II.
    So far, the density operator formalism in \cite{univ,tri} has not been generalized to the Kagome
    lattice. However, by using our simple and effective method, we identify a triangular valence bond (TVB) solid phase in the Ising limit $ v > 0
    $. For the first time, the Ising limit can be used to describe a VBS phase in a concrete
    model. The SF to the triangle valence  bond order  (TVB)  is a weak  first order one.
    In the easy-plane limit, we identify a 6-fold CDW phase in one of the easy plane limit $ v <0, w< 0 $,
    the CDW-VB phase first discovered in the QMC in \cite{ka2,ka3} in the other easy plane limit $ v<0, w>0 $.
    The firm identification of this CDW-VBS phase in the easy-plane limit  $ v<0, w>0 $ is important, because one is sure that the effective
    action Eqn.\ref{ka} in terms of the two vortex fields $ \phi_0, \phi_1 $ in the easy-plane limit  $ v<0, w>0 $ indeed
    describes the SF to the CDW-VBS transition.
    The same transition was also studied by the QMC \cite{ka2,ka3} in a simple microscopic EBHM Eqn.\ref{bosonnn} in a Kagome lattice.
    This EBHM Eqn.\ref{bosonnn} is very simple and has no ring exchange interaction which is usually needed to stabilize a VBS
    order in a bipartite lattice \cite{sandvik}. The 6-fold CDW to the SF transition, also
    the CDW+VB to the SF transition  at commensurate filling $ f=1/3 $ in Kagome lattice are a
    weakly first order one. This finding from the DVM can shed considerable lights on the QMC results in a kagome lattice \cite{ka2,ka3}.
    The results in \cite{ka2} showed that this transition is a very weak first order one instead of a second order one through
    the DCQCP found in \cite{ka1}. Note that this 6-fold CDW phase has the same symmetry breaking patterns as the CDW-VB phase, however,
    the crucial difference is that the former has a local CDW order, while the latter has a local VBS order.
    There is a first order transition between the 6-fold CDW phase and the CDW-VB
    phase just like a liquid-gas phase transition across which there is no symmetry breakings.
    It is interesting to note that one can achieve different CDW+VBS states in the easy plane limit $ v< 0, w>0 $ in both triangular
    and Kagome lattices from the DVM. We also study the excitation spectra in these insulating phases. These results are summarized in the Table 1.

\vspace{0.25cm}
\begin{widetext}
\begin{tabular}{ | c | c | c  | c | c | }  \hline
  Frustrated lattice   &  Ising, $ v > 0 $  & EP1,$ v < 0, w < 0  $ & EP2,$ v < 0, w >0  $  & Some additional phases   \\  \hline
  Triangular    &  X-CDW, d=3   & Stripe-CDW, d=3  & CDW-VBS, d=18     &   TVB, d=3        \\  \hline
  Kagome   &  TVB,  d=3        &   6-fold CDW, d=3    &  CDW-VBS, d=3        &  Stripe-CDW, d=3    \\   \hline
\end{tabular}
\par
\vspace{0.25cm}
{\footnotesize  {Table 1: Various quantum phases  achieved by the DVM in a Triangular and a Kagome lattices at filling factor $ f=1/3 $ or $ f=2/3 $
 and their corresponding degeneracies.
 EP means the easy-plane limit. TVB means the triangular valence bond. Slightly away from the $ f=1/3 $ or $ f=2/3 $ fillings, there are the corresponding supersolids
 with one important exception: There is no corresponding TVB supersolid in either a Triangular or a Kagome lattice.
 Some additional phases are simple and interesting phases not found by the DVM.  } }
\vspace{0.25cm}
\end{widetext}


    Then we study possible quantum phases and quantum phase transitions slightly away from $ 1/3 (2/3) $ fillings in the two
    frustrated lattices. When pushing the effective action slightly
    away from the commensurate fillings inside various insulating phases, we find that one has to choose the
    corresponding saddle point structure of the dual gauge field  carefully and self-consistently to construct the effective
    actions away from the commensurate fillings on the insulating sides.
    We classified all the possible types of supersolids (SS) and analyze their stability conditions.
    In a triangular lattice,  there are X-CDW SS and stripe CDW SS which are stable in both hard core and soft core cases,
    but absence of any triangular valence bond (TVB) supersolid.  There are also  a new kind of supersolid:
    CDW-VB supersolid which maybe stable only in the soft core case.
    This novel kind of supersolid has the coexistence of three kinds of orders: CDW, VBS and superfluid order.
    The supersolid in the triangular lattice near $ 1/2 $ filling  can be considered
    as doping the adjacent solid at $ 1/3 $ filling by interstitials ( SS-i in Fig.9) or
    as doping the adjacent solid at $ 2/3 $ filling by vacancies ( SS-v in
    Fig.9). The supersolid at exactly $ 1/2 $ is the coexistence of SS-i and SS-v with any possible
    ratio.  There are  only two kinds of supersolids: vacancy type or interstitial type. There is no other kinds of supersolids.
    The $ 1/2 $ filling at triangular lattice may not be a " commensurate " filling  as thought previously.
    Only $ 1/3 (2/3) $ are commensurate fillings. Therefore the
    SS in triangular lattice discussed in \cite{gan,ss12,ss3,ss4} has the same origin as that in the square lattice discussed in \cite{univ}.
    In a Kagome lattice, in the Ising limit, there can only be a direct first-order
    transition from the triangular valence bond (TVB) to the SF,
    there is no immediate TVB supersolid intervening between the TVB
    and the SF, in sharp contrast to the possible VB-SS intervening
    between the VBS and the SF discovered in bipartite lattices
    \cite{univ}.  However, in the Easy-plane limit, different kinds of supersolids are generic
    states slightly away from  $ 1/3 $ ( $ 2/3 $ ) filling.
    In addition to a 6-fold CDW supersolid, the stripe supersolid ( stripe-SS), we also find a new kind of supersolid:
    CDW-VB supersolid. The stripe-SS should be stable even in
    the hard core limit, while the 6-fold CDW SS and CDW-VB-SS maybe stable only in  the soft core limit.
    The absence of the VB-SS  in frustrated lattices is in sharp contrast to that in  bipartite lattices
    discovered in \cite{univ}. The quantum phase transitions from solids to
    the adjacent supersolids ( if they cab be stabilized  ) driven by a chemical potential are in the
    same universality class as that from a Mott insulator to a superfluid,
    therefore have exact exponents $ z=2, \nu=1/2, \eta=0 $ ( with logarithmic corrections ).
    The superfluid density  in the SS scales as $ \rho_{s} \sim |\rho- 1/3 |^{(d+z-2)\nu }=|
    \rho-1/3|= \delta f $ with logarithmic corrections. In the
    anisotropic stripe SS case, the superfluid densities along different directions should scale the same way
    with different coefficients.  We also make the implications of our results achieved by the DVM on the available and future QMC simulations on EBHM
    with both $ NN $ and $ NNN $ interactions in both lattices.

    The rest of the paper are organized as follows. In the section II, we study the EBHM in a triangular lattice. We first outline the method to
    characterize the symmetry breaking patterns in the insulating sides in the direct lattices using the vortex degree of freedoms only at the
    corresponding  dual lattices.
    Then by using this method,  we identify the X-CDW phase in the Ising limit in Sec.II-A, the stripe CDW phase
    in the first easy-plane limit in Sect.II-B-1, the new and the most
    interesting CDW-VB solid phase in the other easy-plane limit in
    Sect.II-B-2 and speculate the VBS phase in Sect.II-D.
    In the Sect. II-C, we contrast explicitly the orders of the CDW+VBS phase against that of the bubble phase in the direct lattice.
    We also construct effective actions to study possible supersolid phases
    and their stabilities slightly away from the $ 1/3 (2/3) $  filling. We compare our results with
    QMC simulations on various EBHM in a triangular lattice in Sec.II-E.
    In the section III, we study the EBHM in a Kagome lattice.
    We identify the VBS phase in the Ising limit in Sec.III-A, the 6 fold CDW phase in the first easy-plane limit in
    Sect.III-B-1, the most interesting CDW-VB  phase in the other easy-plane limit in
    Sect.III-B-2, the stripe CDW phase in Sect.III-C. We also construct effective actions to study possible supersolid phases
    and their stabilities slightly away from the $ 1/3 $ filling. We then compare our results with QMC simulations on various EBHM
    in a Kagome lattice in Sec.III-D. We summarize our results
    and give some future perspectives on comparing the DVM and the QMC simulations in the concluding Sect.IV.
    In the appendix A and B, we apply the same method to identify insulating phases in square lattice and honeycomb lattice respectively,
    also the excitation spectra in all these phases. In appendix C, we compare our method with the density operator formalism used in \cite{pq1,tri}
    and commented on possible intrinsic problems of the formalism.
    In appendix D, we list specific vortex fields values used in Sect. II-B-2 for the CDW-VB phase in a triangular lattice.
    A very short version was reported in \cite{dualchenye}.


\section{ Solid and supersolid near $ 1/3 ( 2/3 ) $ on a triangular
Lattice }

    Triangular lattice is the simplest frustrated lattice ( Fig.1a).
    From spin wave expansion, the authors in \cite{gan} found that a SS state is more robust in
    a triangular lattice with only $ t $ and $ V_{1} $ terms in
    Eqn.\ref{boson} and  a SS state with $ \sqrt{3} \times \sqrt{3} $
    pattern is stable even at half filling $ q=2 $.
    Indeed, this discovery was confirmed by several recent QMC simulations \cite{ss12,ss3,ss4}.

\begin{figure}
\includegraphics[width=7cm]{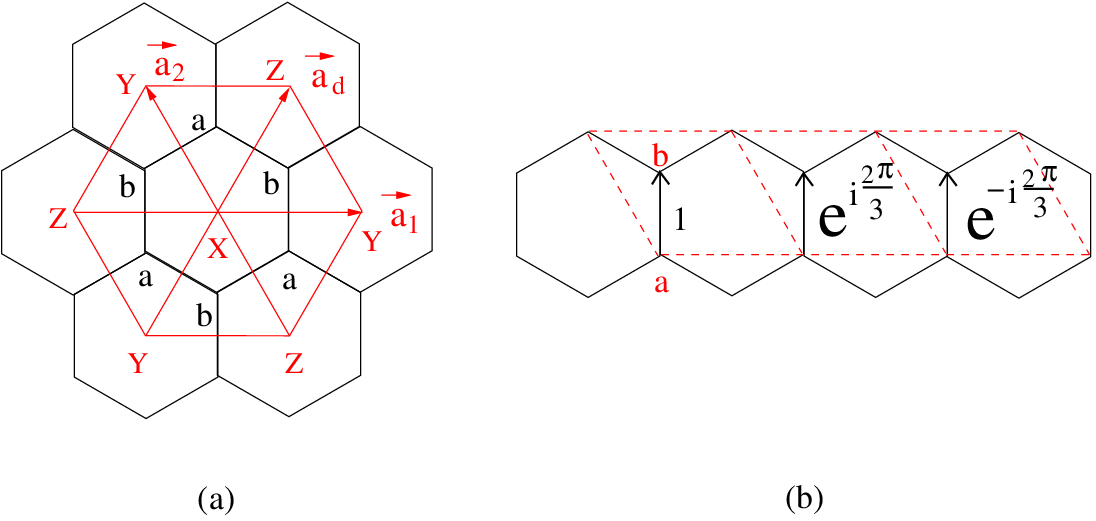}
\caption{ (a) Bosons at filling factor $ f $ are hopping on a
triangular lattice ( red line ) which has three sublattices $ X, Y,
Z $. Its dual lattice is a honeycomb lattice ( black line ) which
has two sublattices $ a $ and $ b $. (b) The bond phase factors in a
dual honeycomb lattice at $ f=1/3 $.  The direction of the gauge
field is important.
 At $ f=2/3 $, the bond phase factors are just complex  conjugate of  those at $ f=1/3 $. }
\label{trigauge}
\end{figure}

     We will investigate the phases, especially various kinds insulating phases and quantum phase transitions at or slightly away from the $ 1/3 (2/3) $  filling.
     With the gauge chosen in Fig.1b, the general effective vortex action in terms of the three eigen-modes $ \xi_{l}, l=0,1,2 $ ( See Eqn.\ref{fields} below )
     invariant  under all the PSG transformations upto sixth order terms was written down in \cite{tri}.
     In the permutative  representation basis $ \phi_{l}, l=0,1,2 $ given by
\begin{eqnarray}
      \xi_{0} & = & \frac{1}{\sqrt{3}} ( \phi_0 + e^{i \frac{2 \pi}{3}} \phi_{1} + \phi_{2}
      )     \nonumber \\
       \xi_{1} & = & \frac{1}{\sqrt{3}} (e^{i \frac{2 \pi}{3}} \phi_0 +  \phi_{1} + \phi_{2}
      )     \nonumber \\
       \xi_{2} & = & \frac{1}{\sqrt{3}} ( \phi_0 +  \phi_{1} + e^{i \frac{2 \pi}{3}} \phi_{2}
      ),
\label{permut}
\end{eqnarray}
      the effective vortex action  $ {\cal L}_{SF} ={\cal L}_{0} + {\cal L}_{1} + {\cal L}_{2} $ upto sixth order can be simplified  to:
\begin{eqnarray}
    {\cal L}_{0} & = &  \sum_{l} | (  \partial_{\mu} - i A_{\mu} ) \phi_{l} |^{2} + r | \phi_{l} |^{2}
    +  ( \epsilon_{\mu \nu \lambda} \partial_{\nu} A_{\lambda} )^{2}/4
       \nonumber  \\
    {\cal L}_{1} &  =  & u ( | \phi_{0} |^{2} + | \phi_{1} |^{2}+ | \phi_{2} |^{2}  )^{2}
       - v [ ( | \phi_{0} |^{2} - |\phi_{1} |^{2} )^{2}
                        \nonumber    \\
          & + & ( | \phi_{1} |^{2} - |\phi_{2} |^{2} )^{2}
          +  ( | \phi_{2} |^{2} - |\phi_{0} |^{2} )^{2} ]
                                   \nonumber  \\
     {\cal L}_{2} & = &  w [ (\phi^{*}_{0} \phi_{1} )^{3}+ (\phi^{*}_{1} \phi_{2}
           )^{3}+ (\phi^{*}_{2} \phi_{0} )^{3} +h.c.]
\label{tri0}
\end{eqnarray}
      where $ A_{\mu} $ is a non-compact  $ U(1) $ gauge field.

     In the superfluid side, moving {\em slightly} away from the $ 1/3 $
     filling $ f=1/3 $ corresponds to adding a small {\em mean} dual magnetic field $ \delta f= f-1/3 $
     in the action Eqn.\ref{tri0}:
\begin{eqnarray}
    {\cal L}_{0} & = &  \sum_{l} | (  \partial_{\mu} - i A_{\mu} ) \phi_{l} |^{2} + r | \phi_{l} |^{2}
    +  ( \epsilon_{\mu \nu \lambda} \partial_{\nu} A_{\lambda} - 2 \pi \delta f \delta_{\mu
    \tau})^{2}/4
       \nonumber  \\
    {\cal L}_{1} &  =  & u ( | \phi_{0} |^{2} + | \phi_{1} |^{2}+ | \phi_{2} |^{2}  )^{2}
       - v [ ( | \phi_{0} |^{2} - |\phi_{1} |^{2} )^{2}
                        \nonumber    \\
          & + & ( | \phi_{1} |^{2} - |\phi_{2} |^{2} )^{2}
          +  ( | \phi_{2} |^{2} - |\phi_{0} |^{2} )^{2} ]
                                   \nonumber  \\
     {\cal L}_{2} & = &  w [ (\phi^{*}_{0} \phi_{1} )^{3}+ (\phi^{*}_{1} \phi_{2}
           )^{3}+ (\phi^{*}_{2} \phi_{0} )^{3} +h.c.]
\label{tri}
\end{eqnarray}

     Because the duality transformation is a non-local
     transformation, the relations between the phenomenological
     parameters in Eqn.\ref{tri} and the microscopic parameters in
     Eqn.\ref{boson} are highly non-local and not known.
     Fortunately,  we are still able to classify some
     phases and phase transitions from Eqn.\ref{tri} without knowing
     these relations.  When $ r > 0 $, the $  < \phi_{l} > =0 $ for every $ l=1,2,3 $, the system is in the SF
     state. When $ r > 0 $, the  $ < \phi_{l} > \neq 0 $ for at least one $ l $, the system is in the insulating state.

    In order to develop a systematic way to use the vortex degree of
    freedoms in the honeycomb lattice to describe the symmetry
    breaking patters of the bosons in the triangular lattice, one
    has to derive the relation \cite{dof} (namely Eqn.\ref{fields} below) between the total vortex fields
    in the honeycomb lattice and the order parameters $ \phi_{l} $
    in Eqn.\ref{tri}. Then one need to first study \cite{five} the energy spectrum of
    the vortices hopping in the honeycomb lattice in the presence of  $ f=p/q $ flux quantum per hexagon shown in Fig.1a.
    For the gauge chosen in Fig.1b, the vortex hopping Hamiltonian is:
\begin{eqnarray}
     H_{v} & = & - t_{v} \sum_{\vec{x}}[ |\vec{x}+ \vec{\delta}
    \rangle e^{i 2 \pi f a_{1} } \langle \vec{x} |      +  |\vec{x}+ \vec{\delta} \rangle \langle
    \vec{x} + \vec{a}_{2} |         \nonumber  \\
    & + & |\vec{x}+ \vec{\delta} \rangle \langle \vec{x} +
    \vec{a}_{1}+ \vec{a}_{2} | +h.c. ]
\end{eqnarray}
    where  $ \vec{x}= a_1 \vec{a}_1 + a_2 \vec{a}_2 $ and $ \vec{x} + \vec{\delta },  \vec{\delta }= 1/3 \vec{a}_1 + 2/3 \vec{a}_2 $
    belong to the sublattice $a$ and $b$ respectively in Fig.1b.
    Following the Sec.3 of Ref.\cite{five}, one can derive the Harper's equation at the filling factor $ f=1/q $ corresponding to $ H_v $:
\begin{eqnarray}
    c^{a}_{m-1}( k_x,k_y )  & +  & e^{i k_y}( 1+ e^{i ( k_x + 2 \pi f m ) } ) c^{a}_{m}( k_x,k_y)     \nonumber  \\
                            &  = &  \epsilon( k_x,k_y)   c^{b}_{m}( k_x,k_y),   \nonumber  \\
     c^{b}_{m+1}( k_x,k_y ) & +  & e^{-i k_y}( 1+ e^{-i ( k_x + 2 \pi f m ) } ) c^{b}_{m}( k_x,k_y)   \nonumber  \\
                             & = & \epsilon( k_x,k_y)   c^{a}_{m}( k_x,k_y)
\label{harper}
\end{eqnarray}
   where $ m=0,1,\cdots,q-1 $ and $ a,b $ are two sublattices of the honeycomb lattice.
   $ - \pi/q \leq k_x \leq \pi/q,  - \pi \leq k_y \leq \pi $ are inside the reduced Brillouin zone.
   The $ q $ minima of the $ 2 q $ bands $ \epsilon( k_x,k_y) $ was found to at $ (  k_x,k_y )=(0, 2\pi f l ), l=0,1,\cdots,q-1 $.

   In the following, we focus on $ q=3 $ case. One only need to find the
   eigenvalues and the corresponding eigenvectors of the $ 6 \times 6 $ matrix at the 3 minima
   $ (  k_x,k_y )=(0, 2\pi l/3 ) $ at $ l=0,1,2 $:
\begin{widetext}
\begin{equation}
 A_{l} = \left ( \begin{array}{cccccc}
     0   &  0 & 0  & 2e^{-i 2 \pi l/3 } & 1 &  0 \\
      0   &  0 & 0  &  0 & e^{-i 2 \pi l/3 }( 1+ e^{-i 2 \pi/3 } ) & 1  \\
        0   &  0 & 0  & 1 & 0 & e^{-i 2 \pi l/3 } ( 1+ e^{i 2 \pi /3 })  \\
       2 e^{ i 2 \pi l/3 } & 0 & 1 &   0   &  0 & 0  \\
       1  &  e^{ i 2 \pi l/3 } (1+ e^{i 2 \pi /3 } ) &  0 &  0   &  0 & 0 \\
      0 & 1 &  e^{ i 2 \pi l/3 } ( 1+ e^{-i 2 \pi /3 } )  & 0  &  0 & 0
   \end{array}   \right )
\end{equation}
\end{widetext}
    In fact, we only need the eigenvector $ c^{a}_{m} (l=0)= c^{a}_{m}, c^{b}_{m} (l=0)= c^{b}_{m} $ at $ l=0 $.
    The lowest eigenvalue is $ \epsilon= -( 1+ 2\cos 2 \pi/9 ) t $, the corresponding eigenvector is
    $  ( c^{a}_{m}, c^{b}_{m} )= [ (2\cos 4\pi/9 + 2\cos 2\pi/9 +1), 1/2 - i \sqrt{3}/2, 2 \cos 2\pi/9,
       ( 2 \cos 4\pi/9 + 2 \cos 2\pi/9 +1 ), 2\cos 2\pi/9, 1/2 -i \sqrt{3}/2  ] $.
       Note that $ c^{b}_{0}= c^{a}_{0}, c^{b}_{1}=c^{a}_{2}, c^{b}_{2}=c^{a}_{1} $.
    The eigen-vectors at $ l=1,2 $ can be achieved by
    the magnetic translation \cite{pq1,tri} along $ \vec{a}_{1} $: $ c^{a}_{m} (l) =c^{a}_{m} \omega^{-ml} , c^{b}_{m} (l)=  c^{b}_{m}\omega^{-ml} \omega ^{l}  $
    where $ \omega = e^{i 2 \pi f } $.
    The eigenfunctions at the three minima $ (  k_x,k_y )=( 0, 2\pi l/3 ) $ are
    $  \psi^{a}_{l}( \vec{x} )= \sum^{q-1}_{m=0} c^{a}_{m} (l) e^{i 2 \pi f ( m a_1+ l  a_2 ) },
    \psi^{b}_{l}( \vec{x} )= \sum^{q-1}_{m=0} c^{b}_{m} (l) e^{i 2 \pi f ( m a_1+ l a_2 )} $ where
    $ \vec{x}= a_1 \vec{a_1} + a_2 \vec{a_2} $ belongs to the sublattice $ a $ and $ \vec{x} + \vec{\delta },  \vec{\delta }= 1/3 \vec{a_1} + 2/3 \vec{a_2} $
    belongs to the sublattice $ b $.
    Then one can write the total vortex field as
    the expansion  $ ( \Phi^{a}( \vec{x} ), \Phi^{b}( \vec{x} ) )^{T}=  \sum^{2}_{l=0} ( \psi^{a}_{l}( \vec{x} ), \psi^{b}_{l}( \vec{x} ) )^{T} \xi_{l} $
    where the $ T $ means the transpose and  the $ \xi_{0}, \xi_{1}, \xi_{2} $ are the three vortex eigen-modes in Eqn.\ref{permut}, namely:
\begin{widetext}
\begin{eqnarray}
   \Phi_{a}( \vec{x} ) & = & \sum^{2}_{m=0}  c^{a}_{m} e^{ i 2 \pi m a_1/3 }[  \xi_0
   + \xi_1 \omega^{-m} e^{ i 2 \pi a_2/3 }
   +  \xi_2  \omega^{ m} e^{ - i 2 \pi a_2/3 } ]   \nonumber  \\
    \Phi_{b}( \vec{x} ) & = & \sum^{2}_{m=0}  c^{b}_{m} e^{ i 2 \pi m a_1/3 }[ \xi_0
   + \xi_1 \omega^{-m + 1 }  e^{ i 2 \pi a_2/3 }
   + \xi_2  \omega^{ m -1 } e^{ - i 2 \pi a_2/3 } ]
\label{fields}
\end{eqnarray}
\end{widetext}
    for the vortex fields at sublattice $ a $ and sublattice $ b $ respectively.

    Various kinds of mean field solutions of the Eqn.\ref{tri} in the
    permutative representation $ \phi_{0}, \phi_{1}, \phi_{2} $  will be discussed in Sec. II-A and II-B.
    Plugging these mean field solutions into Eqn.\ref{permut}, then Eqn.\ref{fields}, one can determine the vortex fields in the whole dual honeycomb lattice. Therefore, one can  extract the corresponding valence bond order in the direct triangular lattice.
    The physical quantities can only be the gauge invariant quantities: the densities at different sites in sublattices $ a $ and $ b $:
\begin{equation}
   | \Phi_{a}( \vec{x} ) |^{2}, ~~~~~| \Phi_{b}( \vec{x} ) |^{2}
\label{density}
\end{equation}
     and the bond quantities between sublattice $ a $ and sublattice $ b $ :
\begin{equation}
    \Phi^{\dagger}_{b}( \vec{x} ) e^{i 2 \pi f a_1} \Phi_{a}( \vec{x} ) = K - i I
\label{bond}
\end{equation}
    where $ \vec{x} $ belongs to the same unit cell shown in Fig.1b. All the other bonds having no phase factor from the gauge field.
    The real part $ K $ gives the kinetic energy between the two sites. The imaginary part $ I $  gives the current between the two sites.
    Note that the "-" sign in Eqn.\ref{bond} is important which make the definition of the current to be consistent with
    that of gauge factors in the Fig.1b. Because the boson is a vortex in the vortex field, so the vortex current
    $  \chi_{p}= \sum_{p} I $ around any plaquette $ p $  ( a hexagon )  leads to
    the boson density deviation from the filling factor $ f=1/3 $ at the center of the hexagon $ p $.


\subsection{ Ising limit $ v > 0 $.  }

 If $ v>0 $, the system is in the Ising limit, only one of the 3
 vortex fields condenses. So it is 3-fold degenerate.
 For example, substituting $ \phi_0=1, \phi_1=\phi_2=0 $ into Eqns.\ref{permut},\ref{fields},\ref{density},\ref{bond}
 one can evaluate the vortex densities, kinetic energy and the current in the dual honeycomb lattice.
 For simplicity, we only show the currents in the Fig.2 which are obviously conserved.  By counting the segments of currents
 along the bonds surrounding the $X, Y, Z $ lattice points, paying special attentions to
   their counter-clockwise or clock wise directions, we can calculate
   the densities at these points $ n_{x}=1/3 + 6 x I > 1/3, n_{y}=n_{z}= 1/3 - 3 x I < 1/3
   $ with the constraint $ n_x + n_y + n_z= 1 $ where the $ I $ is the current flowing around the $ X $ lattice site in Fig.2.
   The $ x \sim n_{x}-n_{y} >0 $ can be thought as a CDW order parameter and can be tuned by the distance away from the SF to the CDW transition in the Fig.3a.
   When one tunes the density $ n_{y}=n_{z}=0 $ which stand for vacancies, then $ n_{x}= 1 $ which stands for one boson.
   This corresponds to the classical limit $ t=0 $ in Eqn.\ref{boson}. With the quantum fluctuations $ t >0 $, then $ n_{x} <1, n_{y}=n_{z} > 0 $.

\begin{figure}
\includegraphics[width=6cm]{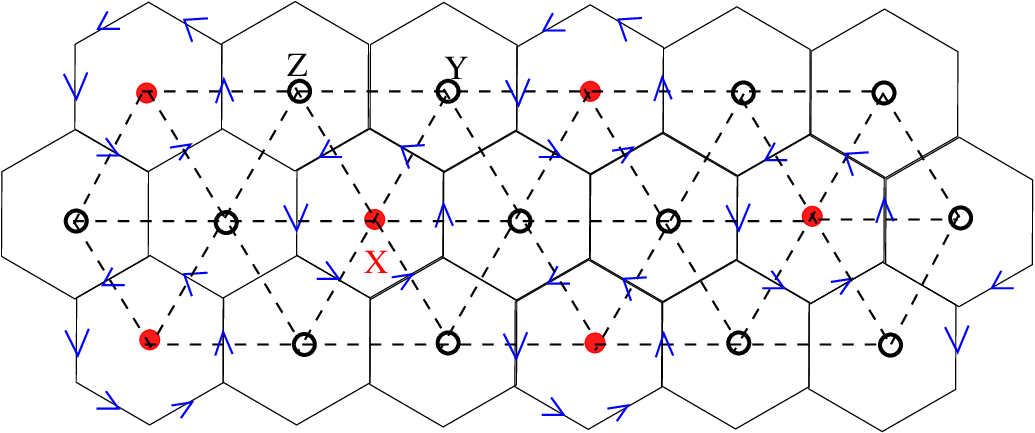}
\caption{ The X-CDW state in the Ising limit $ v > 0  $ at $ f=1/3 $.  It is 3 fold degenerate.
At $ f=2/3$, one can just reverse the current flow and perform a particle and hole
 transformation on $ f=1/3 $, namely, exchange particles ( red dots )  and vacancies ( black empty circles ). }
\label{triising}
\end{figure}

 Shown in Fig.2 is a CDW state corresponds to bosons occupying the $ X $ sublattice, its density can be written as:
\begin{equation}
    \rho_{CDW}( \vec{r} )= A_{n} \cos \vec{Q}_{n} \cdot \vec{r} + 1/3
\label{xcdw}
\end{equation}
   where $  \vec{Q}_n= 2\pi/3(1, 1 ) $.  Putting $ \vec{r} \rightarrow \vec{r}-\vec{a}_1,  \vec{r} \rightarrow \vec{r}-(\vec{a}_1 + \vec{a}_2 ) $
   will correspond to the other  two X-CDW located at the other two sublattices $ Z $  or $ Y $.
   Note that putting $ \vec{r} \rightarrow \vec{r}-\vec{a}_2 $ would be the same
   as $ \vec{r} \rightarrow \vec{r}-\vec{a}_1 $ as expected. So the X-CDW is 3 fold degenerate corresponding to
   condensing one of the 3 vortex fields.

   Eqn.\ref{tri} is an expansion around the  uniform saddle point $
   < \nabla \times \vec{A} > = f = 1/3 $ which holds in the SF and the
   VBS ( to be discussed in section II-B ). In the CDW state, the fact that there are currents flowing on the dual honeycome lattice
   indicates that the average densities are re-distributed. Because of sharp change of the saddle point from Eqn.\ref{tri} in the SF side
   to Eqn.\ref{is0} in the CDW side, we expect that the CDW to superfluid transition at exactly
   $ 1/3 $ filling in Fig.3a is first order.

   {\sl (a) CDW supersolid slightly away from $ 1/3 (2/3) $ filling }

   Well inside the CDW phase, there is a large CDW gap $ \Delta_{CDW} $.
   When studying the physics at slightly away from $ f=1/3 $ filling along the dashed line in Fig.3a,  one has to choose a
   different saddle point where  $ <\nabla \times \vec{A}^{x}> = 1- 2 \alpha =n_x,
   $ for the sublattice $ X $ and $ < \nabla \times \vec{A}^{yz}> = \alpha=n_y= n_z $ for the two sublattices $ Y $ and $ Z $ should be used.
   The $ \alpha \rightarrow \alpha_{c} < 1/3 $ limit corresponds to approaching
   the CDW to the SF transition in Fig.3a from the CDW side, while the $ \alpha \rightarrow 0^{+} $ limit corresponds to the classical limit  $ t/V_{1}
   \rightarrow 0 $.  It is easy to see that there is only one vortex minimum $ \phi_{yz} $ in such a staggered dual magnetic
   field with $ \alpha <  \alpha_{c}  < 1/3 $, so the effective action well inside the CDW state is:
\begin{eqnarray}
  {\cal L}_{CDW} & = & | (  \partial_{\mu} - i A^{yz}_{\mu} ) \phi_{yz} |^{2} + r |
  \phi_{yz}|^{2} +   u  | \phi_{yz} |^{4} + \cdots     \nonumber  \\
     & + & \frac{1}{ 4 } ( \epsilon_{\mu \nu \lambda} \partial_{\nu} A^{yz}_{\lambda}
    - 2 \pi \delta f \delta_{\mu \tau})^{2}
\label{is0}
\end{eqnarray}
   where the vortices in the phase winding of $ \phi_{yz} $ should be interpreted as the
   the boson number. Note that the gauge field $ \vec{A}^{x} $ is always massive with a large CDW gap $ \Delta_{CDW} $,
   so was already integrated out in Eqn.\ref{is0}. In the direct picture discussed in \cite{bragg},
   the gauge field $ A^{yz}_{\mu} $ stands for the uniform  density fluctuation near
   $ \vec{q}=0 $, while the $ A^{x}_{\mu} $ stands for the staggered density fluctuation near $ \vec{Q}_{N}=( 2\pi/3, 0 ) $ which has a large
   CDW gap in Fig.22.

   Eqn.\ref{is0} has the structure identical to the conventional $ q=1 $ component
   Ginzburg-Landau model for a " superconductor "  in a "magnetic"  field.
   It is easy to see  that  if $ < \phi_{yz} > \neq 0 $, the system is in the CDW state where
   both $ \vec{A}^{yz} $ and $ \vec{A}^{x} $ are massive ( Fig.22a).
   If $ < \phi_{yz} > =0 $, the system is in the CDW supersolid ( CDW-SS ) state where there is a gapless superfluid
   mode represented by the dual gauge field $ A^{yz}_{\mu} $ in Eqn.\ref{is0} ( Fig.22b ), it still has the gapped mode near $ \vec{q}=\vec{Q}_{N} $,
   it also has the same lattice symmetry breaking patterns as the CDW state at $ 1/3 $ filling.   Therefore we show that
   the CDW solid to the CDW-SS transition  driven by the chemical potential is in the
   same universality class as that from a Mott insulator to a superfluid transition,
   therefore have exact exponents $ z=2, \nu=1/2, \eta=0 $ with logarithmic corrections ( Fig.3a).
   It is known the SF is stable against changing the chemical
   potential ( or adding bosons ) in Fig.3a. There
   must be a transition from the CDW-SS to the SF inside the window driven by
   the quantum fluctuation $ r $ in the Fig.3a. This transition is likely to be first order.

\begin{figure}
\includegraphics[width=8cm]{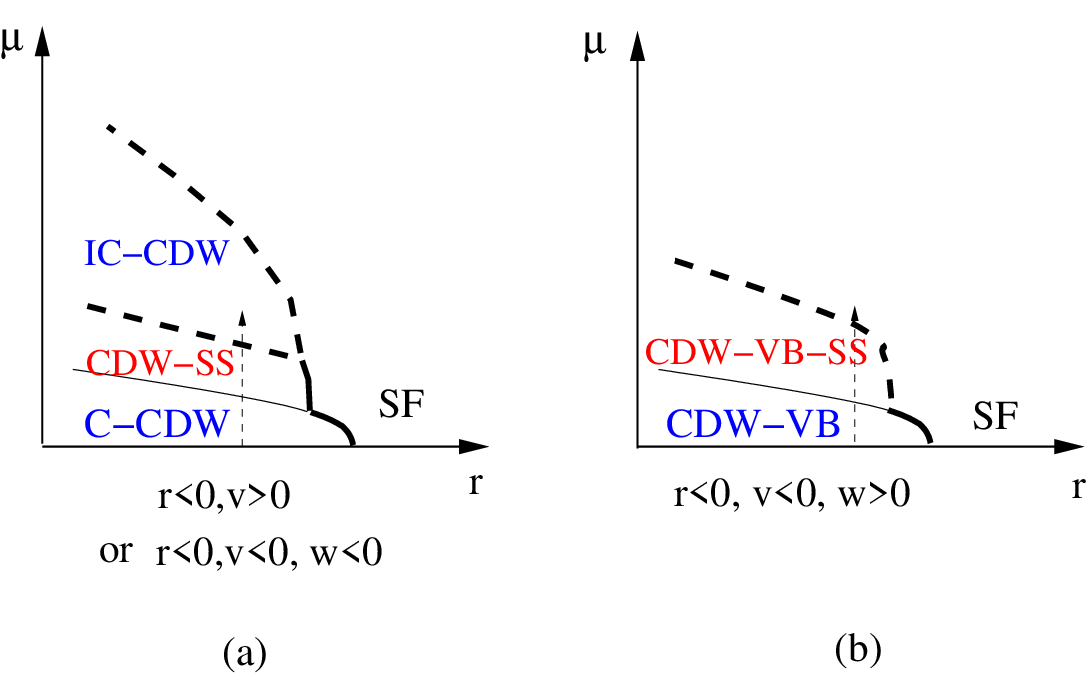}
\caption{Phase diagram of the EBHM in triangular lattice from the
DVM in Sec.II.   (a) There are also two different cases here (a1)
The Ising limit $ v > 0 $. There is a CDW  supersolid
   slightly above the $ 1/3 $ filling which  has the same lattice symmetry breaking as the X-CDW in the Fig.2.
   (a2) The Easy-plane limit $ v <  0, w <0 $. There is a stripe  supersolid
   slightly above the $ 1/3 $ filling which  has the same lattice symmetry breaking as the stripe in the Fig.4.
   (b)  The Easy-plane limit $  v <  0, w > 0 $. There is a CDW-VB-SS
   slightly above the $ 1/3 $ filling which  has the same lattice symmetry breaking as the CDW-VB in Fig.7.
   In both (a) and (b), when the filling factor is increased further from the $ 1/3 $
   filling, there is a first order transition from the CDW-SS in (a) ( CDW-VB-SS in (b) ) to In-commensurate CDW
   ( or In-commensurate CDW-VB in (b) ). The IC-CDW or IC-CDW-VB are stable only when there are sufficiently
   long range interactions. The thin ( thick ) line is the 1st ( 2nd ) order
   transition.  }
\label{triphases}
\end{figure}

   In the direct lattice picture, we can simply renormalize away the
sublattice  $ X $ where the bosons sit and focus on the effective
boson hopping on the two sublattices $ Y $ and $ Z $ which form a
honeycomb lattice whose dual lattice is a triangular lattice (
Fig.1a). Then the dual vortex action in the dual triangular lattice
is given by Eqn.\ref{is0}. From this direct picture, we can see why
a CDW SS can be realized much more easily in a triangular lattice
than in a square lattice: in a triangular lattice, the $ Y $ and $ Z
$ sublattices still form a connected honeycomb lattices where bosons
can move without going through the sublattice $ X $. While in a
square lattice, in a checkboard CDW ( see Fig.19a ), bosons can not
hop on sublattice $ A $ without going through sublattice $ B $. So
for a hard core boson, checkboard SS is unstable against phase
separation because bosons can not hop on to the sublattice B. But in
a soft core case, the checkboard SS could be stabilized because
bosons can hop on the sublattice B too. A striped SS in a square
lattice could be stabilized even in a hard core case, because boson
can hop easily along one direction, so it is must easier for the
bosons to overcome the barriers along the other direction to achieve
an effective hopping.


\subsection{ Easy-plane limit  $ v < 0 $. }

 If $ v<0 $, the system is in the easy-plane
 limit, all the three vortex fields have equal magnitude $ \phi_{l}= |\phi| e^{i \theta_{l}} $ and
 condense.  Well inside the VBS side, the mean field solution holds. The relative phases can be determined by the sign of $ w $.
 In the easy-plane limit, Eqn.\ref{tri} is similar to the action in Tri-layer quantum Hall systems \cite{tlqh}.
 Introducing the center of mass, left and right moving modes as
 done in Ref.\cite{tlqh}:
 $ \theta_{c} = \theta_{0}+\theta_{1}+\theta_{2}, \theta_{l}=\theta_{0}-
 \theta_{2}, \theta_{r}= \theta_{0}-2 \theta_{1}+\theta_{2} $, Eqn.\ref{tri} becomes:
\begin{eqnarray}
  {\cal L}_{EP}  &  =  & ( \frac{1}{3} \partial_{\mu} \theta_{c} - A_{\mu} )^{2}
      + \frac{1}{4} ( \epsilon_{\mu \nu \lambda} \partial_{\nu} A_{\lambda}
      - 2 \pi \delta f \delta_{\mu \tau})^{2} + \cdots    \nonumber  \\
       & + & \frac{1}{6} ( \partial_{\mu} \theta_{l} )^{2} + \frac{1}{18} ( \partial_{\mu} \theta_{r} )^{2}
                               \nonumber   \\
       & +  & 2 w  \cos \frac{ 3 \theta_{l} }{2} ( \cos \frac{ 3 \theta_{l} }{2} + \cos \frac{ 3 \theta_{r} }{2} )
\label{center}
\end{eqnarray}

 In the bipartite lattices in a square and honeycomb lattices \cite{pq1,univ}, in the easy-plane limit, the system is always in a VBS state.
 As shown in the following, the easy-plane limit in a triangular lattice is much involved, in fact, there is always a CDW components, so
 it never corresponds to a pure VBS state.

\subsubsection{ $ w < 0 $ case: Stripe solid phase }

    If $ w < 0 $, the mean field solution is $ \theta_1-\theta_0=  2 \pi m/3, \theta_2 - \theta_0= 2 \pi n/3  $ where $ m, n=0,1,2 $.
     So the state has a degeneracy $ 3 \times 3 $ corresponding to the 9 possible ways to condense the  3 vortex fields.
       Substituting $ \theta_0= \theta_1=\theta_2=0 $ into Eqns.\ref{permut},\ref{fields},\ref{density},\ref{bond},
 one can evaluate the vortex densities, kinetic energy and the current in the dual honeycomb lattice.
 For simplicity, we only show the currents in the Fig.4 which are conserved.  By counting the segments of currents
 along the bonds surrounding the three rows label, paying special attentions to
   their counter-clockwise or clock wise directions, we can calculate
   the densities at these points $ n_{1}=1/3 + 4 x I > 1/3, n_{2}=n_{3}= 1/3 - 2 x I < 1/3
   $ with the constraint $ n_1+ n_2 + n_3 =1 $  where the $ I $ is the current flowing across the lattice.
   The $ x \sim n_{1}-n_{2} >0 $ can be thought as a stripe  order parameter and can be tuned by the distance away from the SF to the stripe
   transition in the Fig.3a.
   When one tunes the density $ n_{2}=n_{3}=0 $ which stand for vacancies, then $ n_{1}= 1 $ which stands for one boson.
   This corresponds to the classical limit $ t=0 $ in Eqn.\ref{boson}. With the quantum fluctuations $ t >0 $, then $ n_{1} <1, n_{2}=n_{3} > 0 $.

   The density in stripe phase aligned along $ \vec{a}_1 $ in Fig.4 is:
\begin{equation}
    \rho_{stripe}( \vec{r} )= A_{s} \cos \vec{Q}_{s} \cdot \vec{r} + 1/3
\label{striped}
\end{equation}
   where $  \vec{Q}_s= 2\pi/3(0,1) $.  Putting $ \vec{r} \rightarrow \vec{r}-\vec{a}_2,  \vec{r} \rightarrow \vec{r}- 2 \vec{a}_2  $
   will shift the stripe along the $ \vec{a}_2 $ by one and two units. Putting $  \vec{Q}_s= 2\pi/3(1,0) $ and  $  \vec{Q}_s= 2\pi/3(1,-1 ) $
   correspond to the other two stripes  aligned along $ \vec{a}_2 $ and $ \vec{a}_d $.
   So the stripe phase is $ 3 \times 3 =9 $ fold degenerate corresponding to the  $ 3 \times 3 $ possibilities of $ m, n=0,1,2 $.

   Well inside the stripe phase, it is legitimate to set $ \theta_c= 3 \theta_0 + \frac{2 \pi}{3}( m+n),
   \theta_l= -\frac{2 \pi}{3} n, \theta_r= \frac{2 \pi}{3}( n-2m) $ in Eqn.\ref{center}, the last term reaches its minimum $ 4w $.
   It is this $ w $ term which drives the formation of the stripe CDW in the Fig.4.
   There is a large stripe CDW gap $ \Delta_{sCDW} \sim w  $.

{\sl (a) Stripe supersolid slightly away from $1/3 (2/3) $ filling. }

   When studying the physics at slightly away from $ f=1/3 $ filling along the dashed line in Fig.3a, then Eqn. \ref{center} reduces to Eqn.\ref{is0}.
   Then the discussions following the Eqn.\ref{is0} also follow with the underlying CDW state as the stripe CDW state shown in Fig.4.
   The supersolid state is a stripe supersolid. The excitation spectra across the stripe to the stripe supersolid transition is also
   given by the Fig.22   with $ \Delta_{stripe-CDW} \sim w  $.

   Alternatively, starting from the new saddle point for the dual gauge gauge fields corresponding to Fig.4:
   $ <\nabla \times \vec{A}^{1}> = 1- 2 \alpha =n_1 $ for the row 1 and $ < \nabla \times \vec{A}^{23}> = \alpha=n_2= n_3 $
   for the row 2 and 3, one can also construct
   the same effective action Eqn.\ref{is0} after making the replacement $ x \rightarrow 1, yz \rightarrow 23 $.

\begin{figure}
\includegraphics[width=6cm]{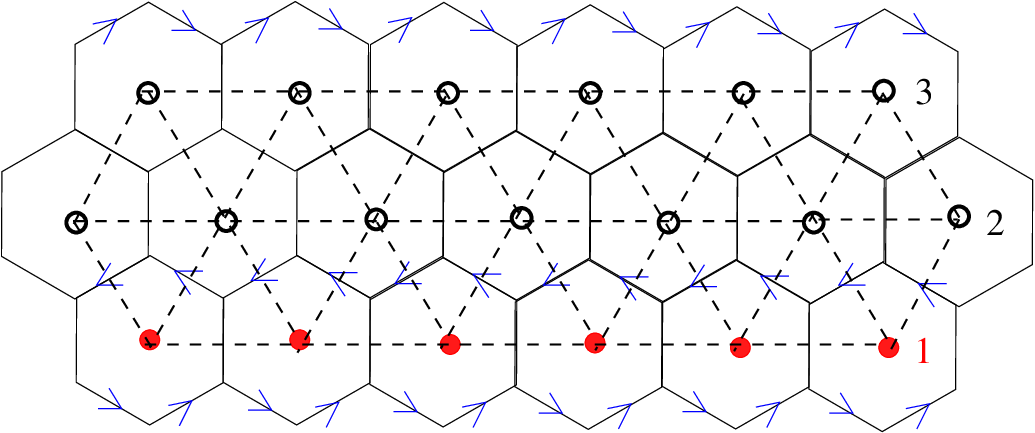}
\caption{The period 3 stripe phase in the easy plane limit $ v<0, w< 0 $ at $ f=1/3 $.  It is $ 3 \times 3=9 $ fold degenerate.
 At $ f=2/3$, one can just reverse the current flow and perform a particle and hole
 transformation in $ f=1/3 $. }
\label{stripe}
\end{figure}

   It is interesting to see that one is not able to get the stripe solid state from the DVM in bipartite lattices in \cite{univ} ( see also appendix B and C ),
   here one can get the stripe solid state from the DVM in a triangular lattice,
   surprisingly from a easy plane limit instead of a Ising limit.

\subsubsection{  $ w > 0 $ case: CDW-VBS phase }

\begin{figure}
\includegraphics[width=8cm]{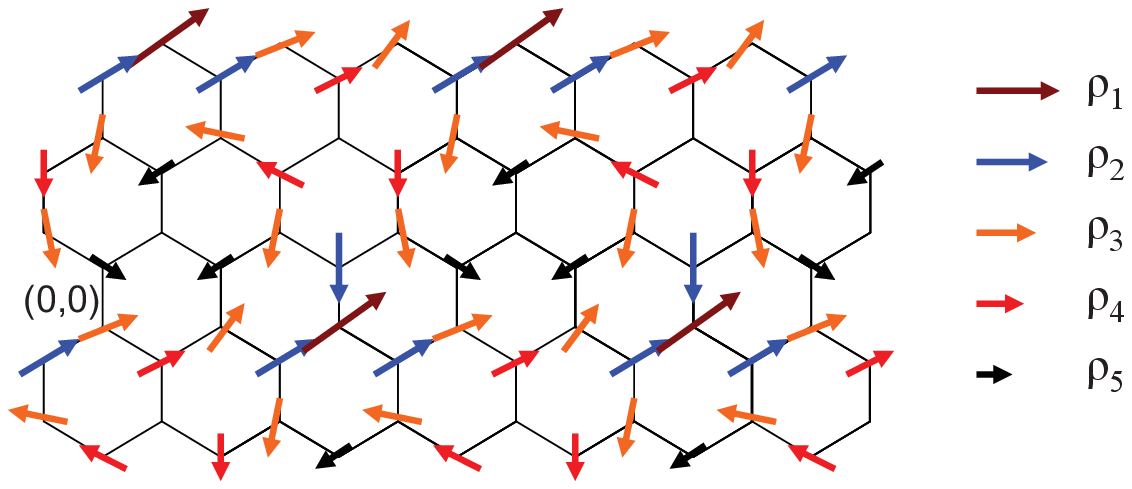}
\caption{The vortex fields of the  CDW-VB phase in a honeycomb
lattice dual to a triangular lattice at $ f=1/3 $ in the easy plane
limit $ v<0, w> 0 $. Shown is $ \theta_0=0, \theta_1= 2 \pi/9,
\theta_2= -2 \pi/9 $ case.  The $ (0,0) $ sets the origin. There are
5 different non-zero magnitudes shown  with 5 different lengths: $
\rho_1=   3 \sin(4\pi/9), \rho_2= 4 \sin(\pi /3) \cos(2\pi/9)
\cos(\pi/9), \rho_3=   2 \sin(\pi/3) \cos( \pi /18), \rho_4=   2
\sin(\pi/3) \cos(2 \pi /9), \rho_5= \sin(\pi/3) $. Note particularly
the lattice points where the vortex fields vanish ! No arrows are
drawn in these points. The values of the vortex field are listed in
the appendix D. } \label{figvortex}
\end{figure}

\begin{figure}
\includegraphics[width=8.3cm]{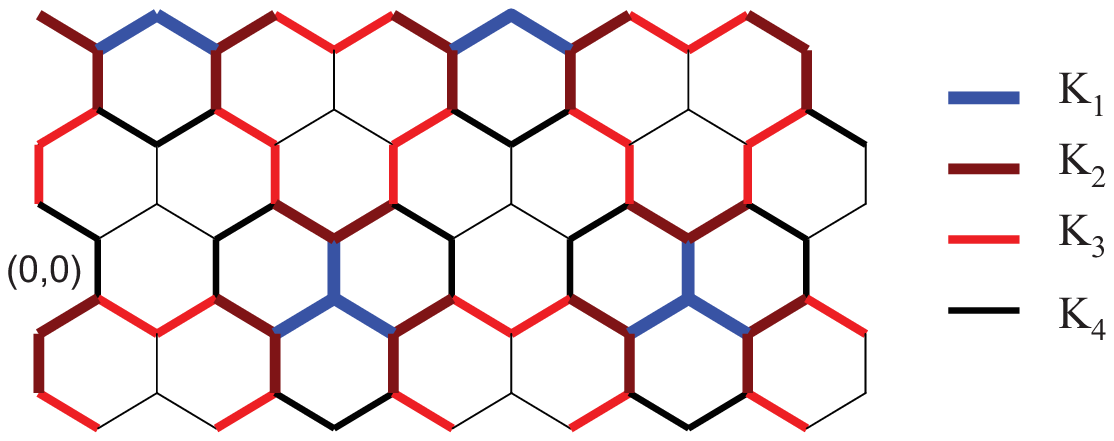}
\caption{The vortex Kinetic energies of the  CDW-VB phase in a
honeycomb lattice dual to a triangular lattice at $ f=1/3 $ in the
easy plane limit $ v<0, w> 0 $. Shown is $ \theta_0=0, \theta_1= 2
\pi/9, \theta_2= -2 \pi/9 $ case.  The $ (0,0) $ sets the origin.
There are 4 different non-zero kinetic energies shown  with 4
different colors: $ K_1=12 \sin(4 \pi/9) \sin(\pi/3) \cos(2 \pi /9)
\cos(\pi/9), K_2=8 \sin^2(\pi /3) \cos(2\pi/9) \cos(\pi/9)
\cos^2(\pi/18), K_3=4 \sin^2(\pi/3) \cos(2\pi/9) \cos^2(\pi/18),
K_4=2 \sin^2(\pi/3) \cos(5\pi/18) $. Note particularly the bonds
where the kinetic energies vanish ! These bonds emanate from the
lattice points where the vortex fields vanish in the
Fig.\ref{figvortex}. } \label{figreal}
\end{figure}

\begin{figure}
\includegraphics[width=7cm]{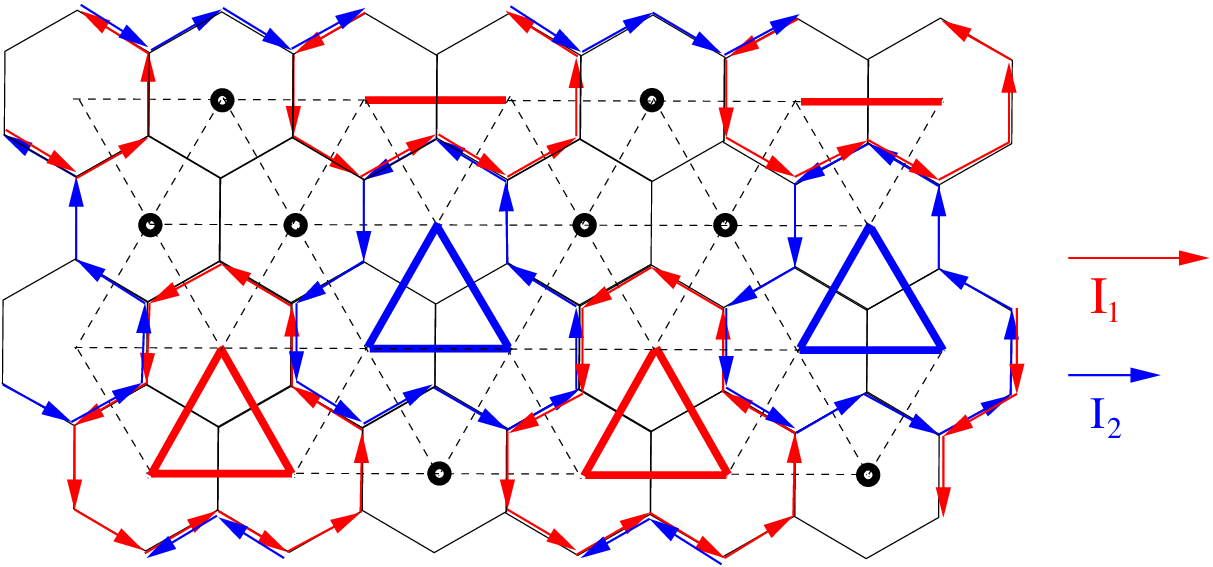}
\caption{The CDW-VB phase in a triangular lattice at $ f=1/3 $ in
the easy plane limit $ v<0, w> 0 $.  It is $ 3 \times 3 \times 2=18 $-fold degenerate.
Shown is $ \theta_0=0, \theta_1=
2 \pi/9, \theta_2= -2 \pi/9 $ case.  The $ (0,0) $ sets the origin.
There are two different vortex currents flowing in the dual
honeycomb lattice. The red current is $ I_{1}= \sin \frac{ 3 \pi}{9}
+ \sin \frac{ 2 \pi}{9} $, the blue current $ I_{2}= \sin \frac{
\pi}{9} + \sin \frac{ 2 \pi}{9} $. The currents are conserved at all
the lattice points. The two different currents in a dual honeycomb
lattice imply 3 different densities $ n_r,n_g,n_b $ in a triangular
lattice listed in the text. At $ f=2/3$, one can just reverse the
current flow and perform a particle and hole
 transformation on $ f=1/3 $. For $ \theta_0=0, \theta_1= - 2 \pi/9, \theta_2= 2 \pi/9 $, one only need to rotate the figure by $ 180^{\circ} $.}
\label{figcurrent}
\end{figure}

   If $ w>0 $, one of the 18 equivalent solutions is $ \theta_1-\theta_0= \theta_0 - \theta_2= 2 \pi/9 $, then $ \theta_1- \theta_2= 4 \pi/9 $.
   So the state has a degeneracy $ 18 $ corresponding to the 18 possible ways to condense the  3 vortex fields.
   Substituting this solution into Eqns.\ref{permut},\ref{fields}, we find the explicit values of the vortex fields at the
   two sublattices $ a $ and $ b $ of the dual honeycomb lattice graphically shown in  Fig.5 and listed in the appendix D.
   Substituting the vortex field values in Eqn.\ref{fieldvalues} into Eqns. \ref{density},\ref{bond}
   one can evaluate the vortex densities in Fig.\ref{figvortex}, kinetic energy in Fig.\ref{figreal} and the currents in
   Fig.\ref{figcurrent}. It is important to stress that there are two
   independent currents $ I_1 $ and $ I_2 $ flowing in the
   Fig.\ref{figcurrent}.  By counting the number of currents along the bonds surrounding the red, green and black lattice points, paying special attentions to
   their counter-clockwise or clockwise directions, we can calculate
   the densities at these points $ n_{r}=1/3 + 2 x [ I_{1} + ( I_{1}- I_{2} ) ] > 1/3, n_{g}=1/3 + 2 x [ I_{2} + ( I_{2}- I_{1} ) ] > 1/3,
   n_{b}=1/3-2 x [ I_{1}+ I_{2} ]  < 1/3 $ with the constraint $ n_r+ n_g + n_b =1 $  where $ I_{1}= \sin \frac{ 3 \pi}{9} +  \sin \frac{ 2 \pi}{9} > I_{2}= \sin \frac{ \pi}{9} +  \sin \frac{ 2 \pi}{9} $.
   The $ x $ can be tuned by the distance away from the SF to the CDW+VB transition in Fig.3b. When one tunes the density $ n_{b}=0 $, then
   $ n_{r}= \frac{ I_{1} }{ I_{1} + I_{2} } > n_{g} = \frac{ I_{2} }{ I_{1} + I_{2} } $ with $ n_{r}/n_{g} = I_{1}/I_{2}, n_{r}+ n_{g}=1 $.
   From Fig.7, we can count the degeneracy of this CDW+VBS phase. One can just focus on the upward red valence bond triangle in the Fig.7:
   In the $ 3 \times 3 $ unit cell, there are $ 3 \times 3 $ ways to put the upward red triangle,
   one can also reflect the upward red triangle with respect the $ \vec{a}_1 $ axis
   to downward one, so the total degeneracy is  $ 3 \times 3 \times 2 = 18 $ which corresponds to $ 18 $ equivalent solutions at $ w > 0 $.

   At $ f=2/3 $ case, one only need to change $ n_{r} \rightarrow  \tilde{n}_{r} = 1-n_{r}=2/3 - 2 x [ I_{1} + ( I_{1}- I_{2} ) ] < 2/3,
   n_{g} \rightarrow \tilde{n}_{g}= 1-n_{g}=2/3 - 2 x [ I_{2} + ( I_{2}- I_{1} ) ] < 2/3, n_{b} \rightarrow \tilde{n}_{b}= 1-n_{b}=2/3 + 2 x [ I_{1}+ I_{2} ] > 2/3 $
   withe constraint $ \tilde{n}_r+ \tilde{n}_g + \tilde{n}_b =2 $.
   When one tunes the density $ n_{b}=1 $, the $ n_r=\frac{ I_{2} }{ I_{1} + I_{2} } <  n_{g}= \frac{ I_{1} }{ I_{1} + I_{2} } $ with $ n_{r}+ n_{g}=1 $.
   When one tunes the density $ \tilde{n}_{r}=0 $, then
   $\tilde{n}_{g}= \frac{ 2 (I_{1} -I_2 ) }{  2I_{1} - I_{2} },  \tilde{ n}_{b} = \frac{ 2 I_{1} }{  2I_{1} - I_{2} } $ with $ \tilde{n}_{g} + \tilde{n}_{b}=2 $.

   As shown in the vortex field Fig.\ref{figvortex}, one can see that the vortex
   fields vanish  both at the centers of the red loop and the green loop. So the vortex density vanish at these centers.
   Furthermore, both the kinetic energy and the current emanating from
   the centers vanish.  This fact indicates that there are local SF ( or equivalently local VBS order ) around the
   these two dual lattice points as shown by the red and green triangles in Fig.\ref{figcurrent}.  It is interesting to compare this with the known fact
   that  interacting bosons in a kagome lattice at $ f=1/2 $ \cite{five,ka1} are always in a SF state  due to the localization of the vortices
   ( a flat vortex band )\cite{subir}.
   The difference is that here there is only a local SF around the the centers of the red loop and the green loop, while there is a global
   SF across the whole lattice in the latter case.
   However, the vortex field at the dual lattice points surrounded by the three black circles in the Fig.7 is non-vanishing, so only the current
   emanating from the dual lattice point vanishes, but the kinetic energy  does not vanish as shown in Fig.\ref{figreal}.
   This fact indicates that there are local CDW around the dual lattice point
   as shown by the black circles in the Fig.\ref{figcurrent}. So this state has both VBS and CDW which can be
   dubbed as CDW-VB state, it is a hybrid state unique to a frustrated lattice.

    It is interesting to compare the vortex motions in the dual lattice with
    those of the bosons in the direct lattice in the Fig.\ref{figcurrent}: the
    vortex motion can have both kinetic energies and the currents,
    however the boson motion has only kinetic energies, but vanishing  currents due to the time reversal symmetry of the original
    EBHM Eqn.\ref{boson}. This should not be too surprising, because the  vortex-charge duality is not self-dual anyway.


   Well inside the CDW-VB phase, it is legitimate to set $ \theta_c= 3 \theta_0,
   \theta_l= -\frac{2 \pi}{9}, \theta_r= -\frac{2 \pi}{3} $ in Eqn.\ref{center}, the last term reaches its minimum $ -\frac{w}{2}  $.
   It is this $ w $ term which drives the formation of the CDW-VB in the Fig.7.
   There is a large CDW-VB gap $ \Delta_{CDW-VB} \sim | w |  $.

{\sl (a) CDW-VB supersolid slightly away from $1/3 (2/3) $ filling. }

   When studying the physics at slightly away from $ f=1/3 $ filling along the dashed line in Fig.3b, then Eqn. \ref{center} reduces to
   Eqn.\ref{is0}, the discussions following the Eqn.\ref{is0} also apply with the underlying state as the CDW-VB state shown in Fig.\ref{figcurrent}.
   The supersolid state is a CDW-VB supersolid.  The excitation spectra across the CDW-VB to the CDW-CB SS transition is also given by the Fig.22
   with $ \Delta_{CDW-VB} \sim  | w  |  $.

   Alternatively, starting from the new saddle point for the dual gauge gauge fields corresponding to Fig.7:
   $ <\nabla \times \vec{A}^{r}>  =n_r  $ for the red site, $ < \nabla \times \vec{A}^{g}> =  n_g $ for the green site and
   $ < \nabla \times \vec{A}^{b}> =  n_b $ for the black site, one can construct
   a similar effective action as  Eqn.\ref{is0}.

  We conclude that the transitions from the X-CDW, stripe-CDW, CDW-VB solid to the corresponding supersolids  driven by the chemical
  potential in the Fig.3 is also in the same universality class as that from a Mott insulator to a  superfluid.

\subsection{ Contrast the orders of the CDW+VBS against the bubble
phase in the easy plane limit $ v<0, w > 0 $. }

  As mentioned in the introduction, using the density operator formalism (DOF) which will be commented in the appendix C,
  the authors in \cite{tri} got the bubble CDW phase in the same easy plane limit $ v<0, w > 0 $ shown in the Fig.25.
  In this subsection, we will determine the ordering of the bosons in the
  direct triangular lattice in the CDW+VBS and the bubble phase in Fig.25 respectively.
  It is easy to see that the bubble phase  also has the degeneracy $ 18 $.
  Despite the two phases have the same degeneracy, we will show here that
  they have completely different symmetry breaking patterns, therefore are two different phases.

  The most general form of the boson density in the direct lattice
  can be similarly expanded as the vortex density in Eqn.\ref{squaredensitybond}:
\begin{equation}
 \rho( \vec{x} )   =   \sum^{q-1}_{r,s=0} \rho_{rs}  e^{i 2 \pi f(rx+sy ) }
\label{directtri}
\end{equation}
    In the triangular lattice at filling factor $ f=1/3 $, we can
    write the above equation explicitly:
\begin{eqnarray}
 \rho( \vec{x} )  &  =  &  \rho_{00} +  \rho_{01} e^{i \frac{ 2 \pi}{3} y } +  \rho_{02} e^{- i \frac{ 2 \pi}{3} y }         \nonumber  \\
                  & + &  \rho_{10}  e^{i \frac{ 2 \pi}{3} x }+  \rho_{11} e^{i \frac{ 2 \pi}{3} (x +y) } +  \rho_{12} e^{i \frac{ 2 \pi}{3} (x-y ) }
                  \nonumber   \\
                  & + &  \rho_{20}  e^{- i \frac{ 2 \pi}{3} x }+  \rho_{21} e^{-i \frac{ 2 \pi}{3} (x - y) } +  \rho_{22} e^{-i \frac{ 2 \pi}{3} (x+y ) }
 \label{directtri3}
\end{eqnarray}
     Because $ \rho( \vec{x} ) $ is real, one can see that $
     \rho_{00}=\rho^{*}_{00},
     \rho_{01}=\rho^{*}_{02},  \rho_{10}=\rho^{*}_{20},
     \rho_{11}=\rho^{*}_{22}, \rho_{12}=\rho^{*}_{21} $. So there
     are $ 9 $ real numbers to be determined by the density
     distributions in the direct lattice.

{\sl (a) The CDW order in the CDW+VBS phase. }

  Substituting $ n_r= 1/3 + 2 x ( 2 I_1 -I_2 ), n_g=1/3+ 2x( 2 I_2 -I_1
  ), n_b=1/3-2x ( I_1+I_2) $ into Eqn.\ref{directtri3}, we find $
  |\rho_{01}|=|\rho_{02}|=|\rho_{10}|=|\rho_{20}|=|\rho_{12}|=|\rho_{21}|=
  2 x \sqrt{ \frac{I^{2}_1-I_{1} I_{2} + I^{2}_2 }{3}}, |\rho_{11}|=|\rho_{22}|= 0 $ and
  the corresponding phases $ \theta_{01} = \theta_{12} = -\theta_{02}= -\theta_{21}= \frac{
  \pi}{18}, \theta_{10}= -\theta_{20}=\frac{5 \pi}{18} + \pi $. Then the
  density of the CDW+VBS phase is given by:
\begin{eqnarray}
 \rho_{CV}( \vec{x} )  &  =  &  1/3 + 4 \delta \sqrt{ \frac{I^{2}_1-I_{1} I_{2} + I^{2}_2
 }{3}} [ - \cos ( \frac{ 2 \pi }{3} x + \frac{ 5 \pi}{18} )   \nonumber \\
 & + & \cos ( \frac{ 2 \pi }{3} y  + \frac{ \pi}{18}
 ) + \cos ( \frac{ 2 \pi }{3} (x-y)  + \frac{ \pi}{18} )]
\label{directtricv}
\end{eqnarray}
   which shows the CDW in the CDW-VBS phase has the 3 ordering wave
   vectors $ \vec{Q}_{\alpha}= 2\pi/3(1,0), 2\pi/3(0,1), 2\pi/3(1,-1),\alpha=1,2,3 $.
   We also used $ \delta $ to stand for the CDW-VBS order parameter
   in order not to confuse with the coordinate $ x $.

{\sl (b) The VBS order in the CDW+VBS phase. }

       There are 3 different boson densities $ n_r, n_g, n_b $.
       At any given lattice point $ \vec{x} $, there are 3 bonds
       along $ \vec{a}_1, \vec{a}_2, \vec{a}_d $. In general, from symmetry points of view, there
       are 6 different bond strengths $ B_{rr}, B_{rg},B_{rb}, B_{gg}, B_{gb}, B_{bb} $ along any given direction $ \alpha=1,2,d $.

       The bond strength along $ \alpha=1,2,d $ can be similarly expanded as in Eqn.\ref{directtri}
       ( In a square lattice, see  Eqn.\ref{squaredensitybond} ):
\begin{equation}
  B_{\alpha} ( \vec{x} )   =   \sum^{q-1}_{r,s=0} B^{\alpha}_{rs}  e^{i 2 \pi f(rx+sy ) }
\label{directtrib}
\end{equation}
    In a triangular lattice at filling factor $ f=1/3 $, we can
    write the above equation explicitly:
\begin{eqnarray}
 B_{\alpha}( \vec{x} )  &  =  &  B^{\alpha}_{00} +  B^{\alpha}_{01} e^{i \frac{ 2 \pi}{3} y } +  B^{\alpha}_{02} e^{- i \frac{ 2 \pi}{3} y }         \nonumber  \\
                  & + &  B^{\alpha}_{10}  e^{i \frac{ 2 \pi}{3} x }+  B^{\alpha}_{11} e^{i \frac{ 2 \pi}{3} (x +y) } +  B^{\alpha}_{12} e^{i \frac{ 2 \pi}{3} (x-y ) }
                  \nonumber   \\
                  & + &  B^{\alpha}_{20}  e^{- i \frac{ 2 \pi}{3} x }+  B^{\alpha}_{21} e^{-i \frac{ 2 \pi}{3} (x - y) } +  B^{\alpha}_{22} e^{-i \frac{ 2 \pi}{3} (x+y ) }
 \label{directtri3b}
\end{eqnarray}
     Because $ B_{\alpha} ( \vec{x} ) $ is real, one can see that $
     B^{\alpha}_{00}=B^{\alpha * }_{00},
     B^{\alpha}_{01}=B^{\alpha *}_{02},  B^{\alpha}_{10}=B^{\alpha*}_{20},
     B^{\alpha}_{11}=B^{\alpha *}_{22}, B^{\alpha}_{12}=B^{\alpha *}_{21} $. So there
     are $ 9 $ real numbers to be determined by the 6 bond
     distributions in the lattice for any bond orientation  $ \alpha $. So only 6 out of 9 real numbers are independent. We expect
     Eqn.\ref{directtri3b} takes similar form for different $
     \alpha $, the only difference comes the shifts of phases inside the $ \cos $ function.

     At the mean field level in Fig.7, one can set  $  B_{rg}=B_{rb}= B_{gb}= B_{bb} =0  $ and only keep $ B_{rr} $ and $  B_{gg} $.
     The non-zero bonds are given by the red and green triangles in
     the Fig.7, so one can assign the index $ \alpha $ to $ B $ according to  the Fig.7.
     From Fig.7, it is obvious to see that two vortex bonds along $ \vec{a}_1, \vec{a}_d $  are identical. So we only need to consider
     the directions along $ \vec{a}_1, \vec{a}_2 $. It is reasonable to
     assume that $ B_{rr}= c \delta  I_1 > B_{gg}= c \delta I_2  $ where $ c $ is an unknown constant, so the ratio of
     the two bonds is given by $ I_1/I_2 $, their magnitude is also  proportional to the CDW-VBS order parameter $ \delta $.
     We find the bond along $ \vec{a}_1 $:
\begin{eqnarray}
 B_{1}( \vec{x} ) &  =  &  B_{d}( \vec{x} )= c \delta ( I_1 + I_2 ) ( 1 + 2 \cos  \frac{ 2 \pi }{3} (x+y)
 )   \nonumber   \\
 & + & 2 c \delta \sqrt{ I^{2}_1-I_{1} I_{2} + I^{2}_2 }[  \cos ( \frac{ 2 \pi }{3} x + \frac{ 2 \pi}{9} )   \nonumber \\
 & + & \cos ( \frac{ 2 \pi }{3} y  - \frac{ 2 \pi}{9}
 ) + \cos ( \frac{ 2 \pi }{3} (x-y)  - \frac{ 2 \pi}{9} )]
\label{directtricvb1}
\end{eqnarray}

   When comparing with the CDW order in Eqn.\ref{directtricv}, we can see that in addition to the 3 ordering wave
   vectors $ \vec{Q}_{\alpha}= 2\pi/3(1,0), 2\pi/3(0,1), 2\pi/3(1,-1),\alpha=1,2,3
   $ with the magnitude $ \sim 2 c \delta \sqrt{ I^{2}_1-I_{1} I_{2} + I^{2}_2 } $, there is also a new VBS ordering wave vector $ \vec{Q}_{4}=2\pi/3(1,
   1) $ with a larger magnitude $ \sim  2 c \delta ( I_1 + I_2 ) $. The ratio
   of the two magnitudes is $ ( I_1 + I_2 )/\sqrt{ I^{2}_1-I_{1} I_{2} + I^{2}_2 } > 1 $.
   It is this new ordering wave vector which makes the detection of the
   VBS order inside the CDW-VB phase possible \cite{bragg}.

   Similarly, we find the bond along $ \vec{a}_2 $:
\begin{eqnarray}
 B_{2}( \vec{x} ) &  =  & c \delta ( I_1 + I_2 ) ( 1 + 2 \cos  [\frac{ 2 \pi }{3}
 (x+y)-\frac{ 2 \pi }{3} ])   \nonumber   \\
 & + & 2 c \delta \sqrt{ I^{2}_1-I_{1} I_{2} + I^{2}_2 } [  \cos ( \frac{ 2 \pi }{3} x - \frac{ 4 \pi}{9} )   \nonumber \\
 & + & \cos ( \frac{ 2 \pi }{3} y  - \frac{ 2 \pi}{9}
 ) + \cos ( \frac{ 2 \pi }{3} (x-y)  - \frac{ 8 \pi}{9} )]
\label{directtricvb2}
\end{eqnarray}

   In fact, when shifting $ \vec{x} \rightarrow  \vec{x} - \vec{a}_{1}
   $ in $ B_{1}( \vec{x} ) $, we get the $ B_{2}( \vec{x} ) $, namely $ B_{2}( \vec{x} )=  B_{1}( \vec{x}- \vec{a}_{1} ) $. This
   is expected by looking at the translational symmetry breaking pattern in Fig.7.

{\sl (c) The bubble phase. }

  The density distributions of the bubble phase was shown in Fig.25 in the appendix C.
  Substituting its density distribution $ n_r= 1/3 + 4x I, n_g= n_b=1/3-2x I $ into Eqn.\ref{directtri3}, we find $
  |\rho_{01}|=|\rho_{02}|=|\rho_{10}|=|\rho_{20}|=|\rho_{12}|=|\rho_{21}|=
  \frac{ 2 }{\sqrt{3}} x I, |\rho_{11}|=|\rho_{22}|= 0 $ and
  $ \theta_{01} = \theta_{12} = -\theta_{02}= -\theta_{21}= - \frac{
  \pi}{6}, \theta_{10}= -\theta_{20}=-\frac{ \pi}{2} $. Then the
  density of the bubble phase is given by:
\begin{eqnarray}
 \rho_{B}( \vec{x} )  &  =  &  1/3 + \frac{ 4 }{\sqrt{3}} \delta I [ \cos ( \frac{ 2 \pi }{3} x -\frac{  \pi}{2} )   \nonumber \\
 & + & \cos ( \frac{ 2 \pi }{3} y  - \frac{ \pi}{6}
 ) + \cos ( \frac{ 2 \pi }{3} (x-y)  - \frac{ \pi}{6} )]
\label{directtribd}
\end{eqnarray}

  Although  Eqn.\ref{directtricv} and
  Eqn.\ref{directtribd} have the same ordering wave-vectors $   \vec{Q}_{\alpha}= 2\pi/3(1,0), 2\pi/3(0,1),
  2\pi/3(1,-1),\alpha=1,2,3 $, the three phases inside the $ \cos $ functions are different.
  Most importantly, there is no VBS order in the bubble phase.

  In short, the main differences between the bubble phase and the CDW+VB phase are:
  In the former, there is only one vortex current flowing in the dual honeycomb lattice shown in Fig.25.
  It leads to 2 different boson densities  in the direct triangular lattice as given by Eqn.\ref{directtribd}.
  In the latter,  there are two different vortex currents  $ I_1, I_2 $ flowing in  the dual honeycomb lattice listed in the Fig.7.
  It leads to 3 different boson densities $ n_r, n_g, n_b $ in the direct triangular lattice as given by Eqn.\ref{directtricv}.
  This concludes that the two states have completely symmetry breaking patterns, therefore two completely different states.
  Some general problems associated with the density operator formalism will be critically examined in the appendix C.

\subsection{ Triangular valence bond (TVB) state and absence of TVB supersolid in a triangular lattice  }

 It is intuitive and constructive to draw a triangular VBS state (TVB) at $ f=1/3 $ in a triangular lattice in the
 Fig.8. This is a plaquette state similar to the plaquette state at $ f=1/2 $ in a square lattice shown in
 Fig.20b. In contrast to the dimer VBS in a square lattice, there is no dimer VBS at a triangular lattice ( see Fig 8b).
 It is easy to see that the uniform saddle point $ < \nabla \times \vec{A} > = f = 1/3 $ holds in both the SF and the
 triangular VBS (TVB). As shown in the previous two sections, from the dual vortex theory, one is not able to get the
 TVB state at $ f=1/3 $ shown in Fig.8
 in either the Ising limit or easy-plane limit. Fig.7 contains the VB component, but also a CDW component.
 Because the excitation of this triangular VBS state necessarily involves a density excitation shown in Fig.11b, drawing the insights gained from
 the direct first order transition from a pure VB state to a SF in a Kagome lattice Fig.12 to be discussed in Sec.III-2,
 we conclude that there is no corresponding TVB supersolid (TVB-SS) in a triangular lattice, in sharp contrast
 to the bipartite lattices discussed in \cite{univ}
 and in the appendix B and C. There can only be a direct first-order transition from the TVB to the SF.

\begin{figure}
\includegraphics[width=7cm]{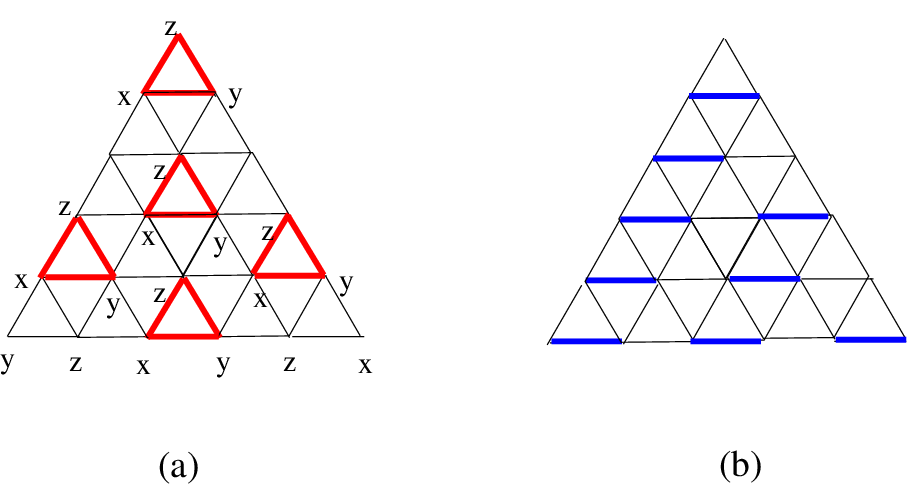}
\caption{ (a) Triangular valence bond (TVB)  solid at $ f=1/3 $. It is 3-fold degenerate.
(b)  A dimer VBS at $ f=1/2 $. There is no such dimer state at $ f=1/3 $. }
 \label{fig8}
\end{figure}

\subsection{ Implications on QMC simulations in triangular lattice }

{\sl (1) Nearest Neighbor ($ NN $) EBHM in a triangular lattice }

  The simplest EBHM in Eqn.\ref{boson} is the
  hard core bosons hopping in a triangular lattice with only nearest neighbor (nn) interaction  $ U=\infty, V_{1} > 0 $:
\begin{equation}
  H_{nn}   =  -t \sum_{ < ij > } ( b^{\dagger}_{i} b_{j} + h.c. )  +   V_{1} \sum_{ <ij> } n_{i} n_{j} - \mu \sum_{i} n_{i}
\label{bosonnn}
\end{equation}

  This model  was studied extensively by QMC in \cite{ss12,ss3,ss4} where the hard core constraint $ n_i= b^{\dagger}_{i} b_{i} \leq 1 $
  considerably reduces the size of Hilbert space  in QMC simulations. The QMC results in \cite{ss3,ss4} leads to Fig.9a.
  When comparing the Fig.3a achieved by the DVM in the Ising limit $ r<0, v>0 $ in this paper with the Fig.9a achieved by the QMC,
  one can see that the CDW state at $ 1/3 $ is just the X-solid state in Fig.2, the CDW-SS  slightly above $ 1/3 $ filling ( SS-i )
  is the interstitial induced supersolid ( SS-i ).
  The IC-CDW state in Fig.3a can not be realized in Fig.9a, because it can only be stabilized by
  much longer-range interactions than $ V_1 $ at such dilute interstitial densities.
  The dual vortex effective action to describe the transition from the SF to
  the X-solid is given by Eqn.\ref{tri} in the Ising limit $ r<0, v >0  $.
  The transition was found to be very weak first order by QMC  in \cite{ss3,ss4}.
  The effective action to describe the transition from the X-solid to the
  SS-i slightly away from $ 1/3 $ is given by Eqn.\ref{is0} and was found in Sect.II-A to be a second order
  transition  in the same universality class as that from a Mott insulator to a  superfluid.
  However, this transition was not studied by QMC  yet.
  The transition from the SF to the SS-i is likely to be a first order one.
  Note that sometimes it is difficult to distinguish a 2nd order transition from a weakly first order one
  from QMC.  More refined QMC simulations are needed to address  the nature of the SF to SS-i transition.

{\sl (2) Next Nearest Neighbor ($ NNN $) EBHM in a triangular
lattice: Stripe phase and Stripe-SS phase }

  The next simplest EBHM in Eqn.\ref{boson} is the
  hard core bosons hopping in a triangular lattice with $ NN $ and $ NNN $  interactions $ U=\infty, V_{1} >  0, V_{2} >0 $:
\begin{eqnarray}
  H_{nnn} & = & -t \sum_{ < ij > } ( b^{\dagger}_{i} b_{j} + h.c. )    +  V_{1} \sum_{ <ij> } n_{i} n_{j}  \nonumber  \\
     &  +  & V_{2} \sum_{ <<ik>> } n_{i} n_{jk} - \mu \sum_{i} n_{i}
\label{bosonnnn}
\end{eqnarray}
  which was studied by QMC in \cite{trinnn}.  A period-3 striped soild state is found at $ f=1/3 $ and
  a stripe-supersolid was found at slightly large than $ 1/3 $. The QMC results in \cite{trinnn} leads to Fig.9b.
  When comparing the Fig.3a in the easy plane limit $ r<0, v<0, w<0 $ achieved by the DVM in this paper with the Fig.9b achieved by the QMC,
  one can see that the CDW state at $ 1/3 $ is just the stripe state in
  Fig.4, the CDW-SS  slightly above $ 1/3 $ filling is the stripe supersolid.
  The IC-CDW state in Fig.3a can not be realized in Fig.9b, because it can only be stabilized by
  much longer-range interactions than $ V_2 $ at such dilute interstitial densities.
  The dual vortex effective action to describe the transition from the SF to the stripe  solid  is given by Eqn.\ref{tri}
  in the easy plane limit $ r<0, v<0, w<0 $,  it was found to be strongly first order by QMC  in \cite{trinnn}.
  The effective action to describe the transition from the stripe solid to the stripe supersolid  slightly away from $ 1/3 $
  is given by Eqn.\ref{is0} and was found in Sect.II-B-1 to be a second order transition in the same universality class as that from a Mott insulator to
  a superfluid. However, the universality class of the 2nd phase transitions in Fig.9b was not studied in \cite{trinnn}.

{\sl (3) Searching for the CDW-VB phase in the $ NNN $ EBHM model in
a triangular lattice }

  A meta-stable bubble solid phase in Fig.25 was also found in the QMC in \cite{trinnn}.
  This bubble solid phase has a higher energy than the stripe solid phase in Fig.4.
  The very interesting CDW-SS phase in Fig.7 was not searched in the QMC in \cite{trinnn} in any parameter regimes.
  In order to identify this phase, in addition to the QMC calculations of the superfluid density and the density structure factor
  in \cite{trinnn}, a bond structure factor  need also be studied to identify the VB ordering in the Fig.7.

       For the CDW-VBS phase in Fig.7 and Eqn.\ref{directtricv}, Eqn.\ref{directtricvb1} and Eqn.\ref{directtricvb2}, there are 3 CDW ordering wave
       vectors $ \vec{Q}_{\alpha}= 2\pi/3(1,0), 2\pi/3(0,1), 2\pi/3(1,-1),\alpha=1,2,3 $.
       In addition to the same 3 ordering wave vectors, the VBS order
       has its own new ordering wavevector $ \vec{Q}_{4}= 2\pi/3(1,0) $ with a bigger magnitude than that at $ \vec{Q}_{\alpha}, \alpha=1,2,3
       $. So the divergence in the bond structure factor   at $ \vec{Q}_{4}= 2\pi/3(1,0) $  can be used to
       determine the VBS order in the CDW-VBS phase in Fig.5.

\vspace{0.25cm}

\begin{figure}
\includegraphics[width=8cm]{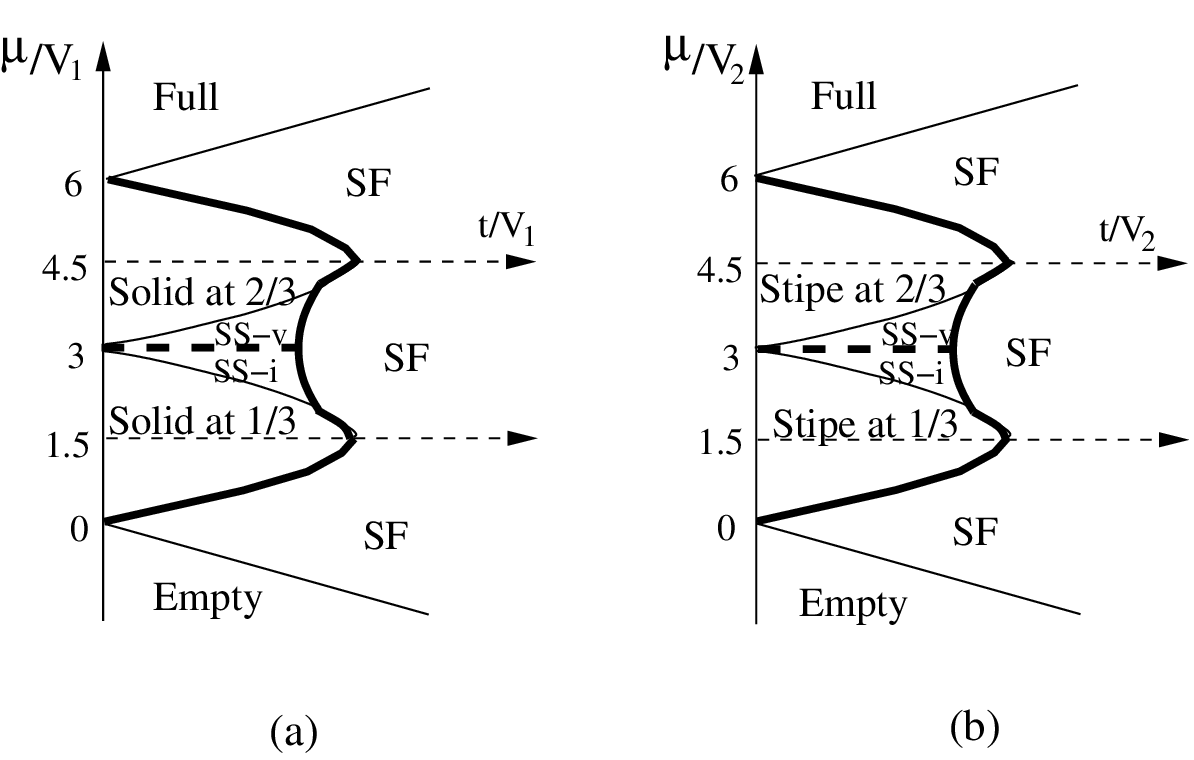}
\caption{ (a)  QMC results in \cite{ss3,ss4} on
    triangular lattice with $ U=\infty, V_{1} > 0 $. Slightly above ( below ) the $ 1/3 (2/3) $ filling,
     there is a SS-i ( SS-v) which has the same lattice symmetry breaking patterns as
    the solid at $1/3 $  ( $ 2/3 $ )  where
    the bosons occupying ( un-occupying ) one of
    the 3 sublattices $ X,Y,Z $ in Fig.1a. This solid phase is just the
    X-CDW shown in Fig.2.
    SS-i and SS-v are related by the P-H transformation.
    Exactly at half filling where $ \mu/V_{1}=3 $, SS-i and
    SS-v coexist at any possible ratio, there is a first order transition from SS-i to SS-v driven by the
    chemical potential $ \mu $  across the thick dashed line.
    (b) Suggested phase diagram  on a
     triangular lattice with $ U=\infty, V_{1}=0, V_{2} > 0 $ by the DVM in this paper. The Solid at $1/3 $
     is the period-3 stripe phase studied numerically first in
     \cite{trinnn}. This phase is shown in Fig.4.
     The SS-i slightly above $ 1/3 $ filling is the corresponding period-3 stripe
     supersolid phase studied by recent QMC in \cite{trinnn}.
     If the solid at $ 1/3 $ is the CDW-VB phase in Fig.7,
     then the corresponding possible CDW-VB SS phase maybe {\em unstable} against phase separation in the hard core case,
     but maybe stable in the soft core case.
     The thin ( thick ) line is the 1st ( 2nd ) order transition.
     As explained in the text, the first order transition is a
     strongly first order one.
     The universality class of the 2nd order phase transitions studied  by the DVM in this paper was not studied in \cite{ss3,ss4,trinnn}. }
 \label{fig9}
\end{figure}

\vspace{0.25cm}


\section{ Solids and supersolids in a Kagome lattice near $ 1/3 ( 2/3 )
$ }
  The Kagome lattice ( Fig.10a ) has 3 sublattices $ a, b $
  and $ c $, its dual lattice is the dice lattice ( Fig.10a).
     Due to too strong quantum fluctuations, as shown in \cite{gan} the spin wave expansion does not
     work anymore in Kagome lattice.
     Here we are trying to investigate the phases and
     quantum phase transitions when slightly away from the $ 1/3 $
     filling from the DVM.
     Because of the P-H symmetry at $ U=\infty $, the results are equally valid near $2/3 $
     filling.
     As shown in \cite{five} at $ q=2 $, because the lowest dual vortex
     band in the dice lattice is completely flat,
     so the dual vortices are completely localized, so it has to be a
     superfluid, in sharp contrast to $ q=2 $ on triangular lattice Fig.9.
     For $ q=3 $, from Table 1 of \cite{five}, we can see there are two minima at $
     (0, 0) $ and $ (-2 \pi/3, 2 \pi/3 ) $.
     Let's label the two eigenmodes at the two minima as
     $ \phi_{l}, l=0,1 $. The
     effective action invariant under all the MSG transformations upto sixth order terms was
     written down in \cite{ka1}.
     Again, inside the superfluid phase,  moving {\em slightly} away from the $ 1/3 $
     filling $ f=1/3 $ corresponds to adding a small {\em mean} dual magnetic field $ \delta f= f-1/3 $
     in the action derived in \cite{ka1}. Upto sixth order, the action is  $ {\cal
     L}_{SF}   ={\cal L}_{0} + {\cal L}_{1} + {\cal L}_{2} $:
\begin{eqnarray}
    {\cal L}_{0} & = & \sum_{l} | (  \partial_{\mu} - i A_{\mu} ) \phi_{l} |^{2} + r | \phi_{l} |^{2}
    +  \frac{1}{4} ( \epsilon_{\mu \nu \lambda} \partial_{\nu} A_{\lambda}
                    - 2 \pi \delta f \delta_{\mu \tau} )^{2}      \nonumber  \\
    {\cal L}_{1} & =  & u ( | \phi_{0} |^{2} + | \phi_{1} |^{2}  )^{2}
       - v  ( | \phi_{0} |^{2} - |\phi_{1} |^{2} )^{2}
                     \nonumber  \\
     {\cal L}_{2} & = &  w [ (\phi^{*}_{0} \phi_{1} )^{3} +h.c.]
\label{ka}
\end{eqnarray}
     where $ A_{\mu} $ is a non-compact  $ U(1) $ gauge field.

     In fact, the form of this action is {\em exactly the same} as that in
     honeycomb lattice at $ q=2 $ first derived in \cite{univ}. However,
     the physical meanings of the two eigenmodes  $ \phi_{l}, l=0,1 $ are completely
     different in the two lattices, this fact leads to completely different physics in the two lattices.

     Ref.\cite{ka1} studied the vortex environments in the dual dice lattices, but did not evaluate
     any gauge invariant physical quantities. The vortex environments are gauge dependent, so may not be used to
     characterize the symmetry breaking patterns. Ref.\cite{demler} studied the bosons hopping on a Dice lattice at an integer filling.
     These bosons have only on-site interaction, but are subject to some artificial gauge fields which can be generated
     in cold atom experiments. They identified the  vortex fields and the gauge invariant currents at $ f=1/3 $.
     These artificial gauge fields are static. While the gauge field in all the dual vortex actions  Eqn.\ref{ka},\ref{kacdwvb},\ref{kapm}
     on the dual Dice lattice are quantum fluctuating which are responsible for the very existences of the CDW-VB-SS.

\subsection{ Ising limit, $ v > 0 $. }

  If $ v>0 $, the system is in the Ising limit, only one of the 2  vortex fields condense.
  So it is 2-fold degenerate..
  Let us take $ \langle \phi_{0} \rangle \neq 0, \langle \phi_{1} \rangle =0 $.
  Using Eqns.\ref{density},\ref{bond}, one can evaluate
     both the vortex density, the kinetic energy and current in the dual dice lattice. This has been done in \cite{demler} in a very different context.
     There is no current flowing in the dual lattice which shows there is no CDW order. All the densities at the direct lattice are fixed at $ 1/3 $.
     It is important to stress that the vortex fields at the dual center of the red triangles in the Fig.10a vanish, so both currents and the
     kinetic energies emanating from these dual centers are vanishing, this fact indicates that there is a local VBS order around  red triangles
     as shown in the Fig.10. This VBS state has a degeneracy $ 2 $ with the other degenerate state corresponding to
     reflecting the red VBS triangles  across the bow-ties.

     The fact that one achieved the VBS in the Ising limit in a Kagome lattice
     is just opposite to those in the bipartite and triangular lattices discussed in \cite{univ} and the last section
     where one gets the CDW state in the Ising limit. This crucial
     difference is responsible for the absence of supersolid in the
     Ising limit in the kagome lattice shown in the following.
     Because the saddle point structure stays the same
     across the SF to the triangle-VBS ( TVB ) transition at $ f=1/3 $ shown in the Fig.12,
     Eqn.\ref{ka} still holds in the TVB side, so the transition is a {\em weak } first order one
     which breaks both the $ U(1) $ and the $ Z_2 $ exchange symmetry between $ \phi_0 $ and $ \phi_{1} $.
     In contrast, in the Ising limit of both the bipartite and triangular lattice, different saddle points for the dual gauge fields
     need to be chosen.

{\sl (a) Absence of a TVB supersolid slightly away from $1/3 (2/3) $ filling. }

     Slightly away from $1/3 $ filling, there must be a direct first order transition from
     the TVB to the superfluid as shown in Fig.12, there is no TVB supersolid intervening between the TVB and the
     SF, in sharp contrast to the cases in bipartite and triangular  lattices.


\begin{figure}
\includegraphics[width=7cm]{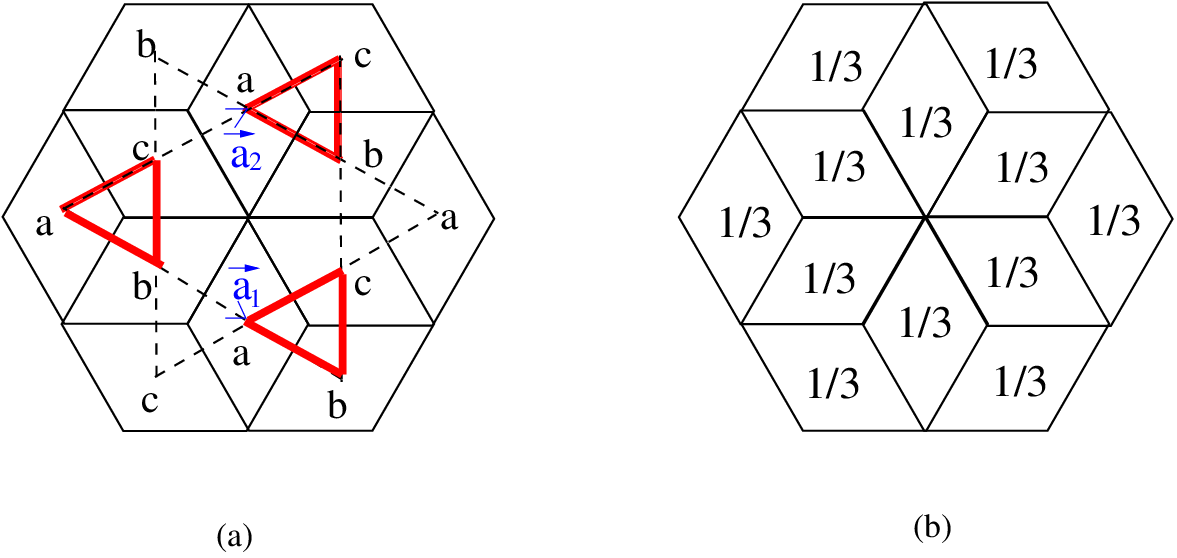}
\caption{  The VBS around the up triangle ( triangular valence bond
(TVB) ) in a Kagome lattice at $ f=1/3 ( f=2/3 ) $ in the Ising
limit $ v > 0 $.  It is 2-fold degenerate.
The thick triangle means one at $ f=1/3 $   ( or
two at $ f=2/3 $ ) boson(s) is (are) moving around the up triangle.
(b) The saddle point of the dual gauge field in a magnetic field
unit cell in the dice lattice. The saddle point stays to be uniform
just like that in the SF phase. } \label{kagomeising}
\end{figure}

    Note that due to the lack of dimer VBS on the Kagome  lattice at $ f=1/3 $, any excitation above the TVB state in the Fig.10 will have to
    break the TVB into a density excitation shown in Fig.11.  Because this density "defect"  could happen in any position of the lattice, so
    the moving of this through the whole lattice leads to the excitation spectrum in the TVB shown in the upper right inset in Fig.12.
    This gap is of density origin represented by the Higgs mechanism of the gauge field $ A_{\mu} $ fluctuations in Eqn.\ref{ka}.
    This is in sharp contrast to the dimer VB or plaquette VB state in  bipartite lattices such as square and honeycome lattices.
    In these bipartite lattices, there are VBS excitation gaps due to the
    operator $ \lambda \cos 3 \theta $ ( $ \lambda \cos 4 \theta $ ) in a honeycomb ( square ) lattice
    which only involves dimer flips instead of any
    density fluctuation shown in Fig.21. Here, such an operator is lacking in the Ising limit in a Kagome lattice, so there is no such a VBS gap.

\begin{figure}
\includegraphics[width=7cm]{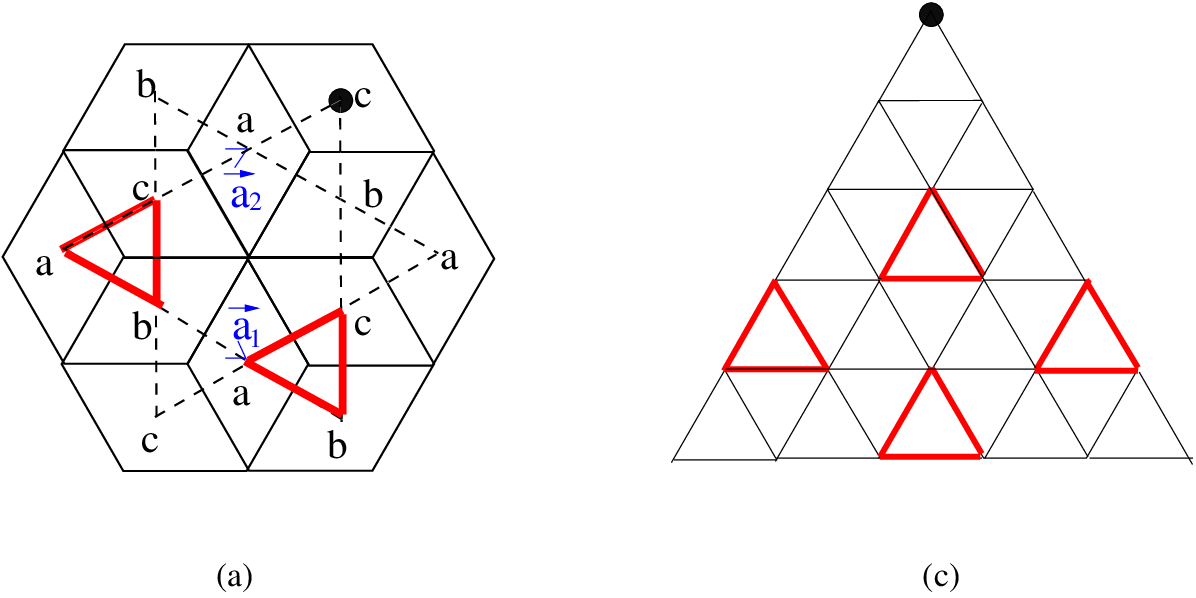}
\caption{  A local density excitation above the TVB state in Fig.10.
Its translational moving through the whole lattice leads to the
excitation spectrum in the upper right inset in Fig.12. Compare it
with the local bond flip excitations in a VBS in Fig.21. }
\label{kagomeisingex}
\end{figure}

\begin{figure}
\includegraphics[width=7cm]{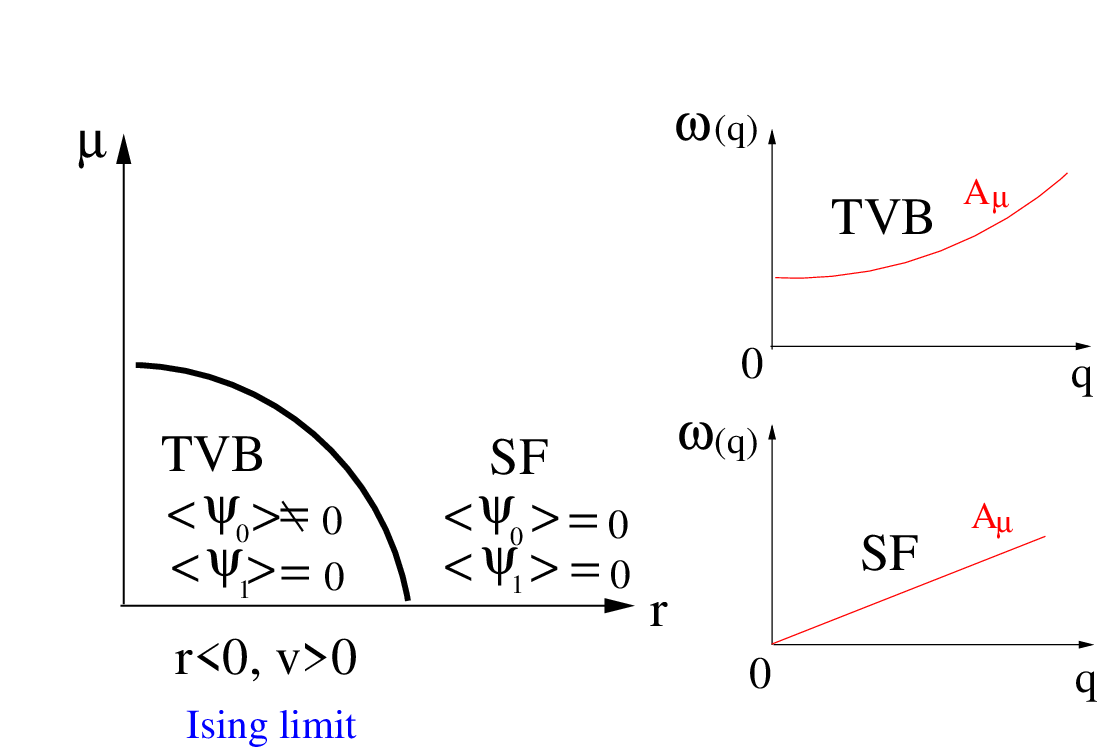}
\caption{  The phase diagram of the chemical potential $ \mu $
versus $ r $ in a Kagome lattice at and near $ f=1/3 $ in the Ising
limit $ v > 0 $. There is a direct first order transition from the
TVB to the superfluid. There is no supersolid intervening between
the TVB and the SF in any case. Lower right inset is the gapless
excitation due to the gauge field $ A_{\mu} $ in the SF. Upper right
inset is the gapped excitation due to the Higgs mechanism of the
gauge field $ A_{\mu} $ inside the TVB shown in Fig.10.}
 \label{fig11}
\end{figure}

\subsection{ Easy-plane limit $ v < 0 $. }

     If $ v<0 $, the system is in the easy-plane
     limit, Eqn.\ref{tri} is similar to the action in
     Bi-layer quantum Hall systems \cite{blqh}, the two vortex fields have equal magnitude $ \phi_{l}= |\phi| e^{i
     \theta_{l}} $ and  condense. The relative phases can be determined by the sign of $ w $\cite{univ}.
     When moving from $ 1/3 $ to $1/2 $ where the bosons has to be in a SF state,  Eqn.\ref{ka} can be rewritten as:
\begin{eqnarray}
  {\cal L}_{EP}  &  =  & ( \frac{1}{2} \partial_{\mu} \theta_{+} - A_{\mu} )^{2}
      + \frac{1}{4} ( \epsilon_{\mu \nu \lambda} \partial_{\nu} A_{\lambda}
      - 2 \pi \delta f \delta_{\mu \tau})^{2} + \cdots    \nonumber  \\
       & + & \frac{1}{2} ( \partial_{\mu} \theta_{-} )^{2} +  2 w  \cos 3 \theta_{-}
\label{kapm}
\end{eqnarray}
     where $ \theta_{\pm} = \theta_{0} \pm \theta_{1} $.

     In the following, we discuss both cases separately.

\subsubsection{ $ w < 0 $ case: 6-fold CDW order }

\begin{figure}
\includegraphics[width=3cm]{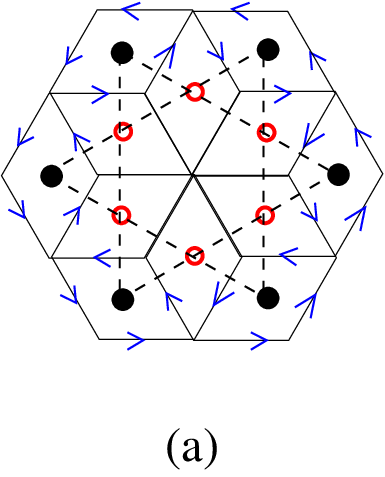}
\hspace{0.5cm}
\includegraphics[width=3cm]{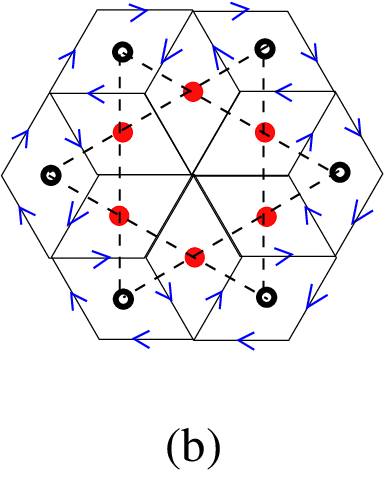}
\caption{The 6-fold CDW state  in a Kagome lattice in the easy-plane
limit $ v < 0, w < 0 $
 (a) $ f=1/3 $ (b) $ f=2/3 $. It is 3-fold degenerate.
The red or black solid circle is a boson, the red or black empty circle is an vacancy.
(a) and (b) are related by time reversal symmetry and particle-hole
symmetry. Note that the vortex filed in the center of the Dice
lattice is non-vanishing, the vortex currents flowing in the 6 bonds
emanating from the dual center vanish due to the cancelation of
currents flowing around the 6 neighboring parallelograms. }
\label{kagoeasy1}
\end{figure}

     When $ w < 0 $, $ \theta_{-} =\theta_0-\theta_1= 2 n \pi/3, n=0, 1,2 $.
     So the state has a degeneracy $ 3 $ corresponding to the 3 possible ways to condense the  2 vortex fields.
     Using Eqns.\ref{density},\ref{bond}, one can evaluate
     both the vortex density, the kinetic energy and current in the dual dice lattice. This has been done in \cite{demler} in a very different context.
     The current was shown in the Fig.\ref{kagoeasy1}.  It is important to point out that the vortex at the dual center of the 6 black spots is non-vanishing,
     so only the current is zero due to the cancelations of the currents flowing around the neighboring black points, but the kinetic energy is positive.
     This fact indicates the there is a local CDW order around the dual center as shown in the Fig. \ref{kagoeasy1}.
     By counting the segments of currents
     along the bonds surrounding the red and black lattice points, paying special attentions to
   their counter-clockwise or clock wise directions, we can calculate
   the densities at these points $ n_{b}=1/3 + 4 x I > 1/3, n_{r}= 1/3 - 2 x I < 1/3
   $ with the constraint $ n_b+ 2 n_r=1 $
   where the $ I $ is the current flowing around the red lattice site in Fig.\ref{kagoeasy1}.
   The $ x \sim n_{b}-n_{r} >0 $ can be thought as a CDW order parameter and can be tuned by the distance away from the SF to the CDW transition in the Fig.3a.
   When one tunes the density $ n_{r}=0 $ which stand for vacancies, then $ n_{b}= 1 $ which stands for one boson.
   This corresponds to the classical limit $ t=0 $ in Eqn.\ref{boson}. With the quantum fluctuations $ t >0 $, then $ n_{b} <1, n_{r} > 0 $.

   Because the centers of all the hexagons in a Kagome lattice makes a triangular lattice,
   one can see that the centers of the Fig.\ref{kagoeasy1} occupies one of the 3 sublattices of this underlying triangular lattice,
   so the degeneracy of the 6-fold CDW order is 3  corresponding to the 3 possible ways to condense the 2 vortex fields.

   Very similar calculations can be made for the filling factor $ f=2/3 $ which is
   related to $ f=1/3 $ by a particle-hole transformation
   $ n_{b} \rightarrow \tilde{n}_b=1-n_b, n_{r} \rightarrow \tilde{n}_r=1-n_r $ with the constraint
   $ \tilde{n}_b+ 2 \tilde{n}_r= 2  $ shown in Fig.13b.

{\sl (a) 6-fold CDW supersolid slightly away from $1/3 (2/3) $ filling. }

   Well inside the 6-fold CDW phase, it is legitimate to set $ \theta_{-} =\theta_0-\theta_1= 2 n \pi/3 $ in Eqn.\ref{kapm},
   the last term reaches its minimum $ 2 w $.
   It is this $ w $ term which drives the formation of the 6-fold CDW in the Fig.13.
   There is a 6-fold CDW gap $ \Delta_{CDW} \sim w  $.
   When studying quantum phases at slightly away from $ f=1/3 $ filling along the dashed line in Fig.3a, then Eqn. \ref{kapm} with $ w < 0 $  reduces to
   Eqn.\ref{is0}.
   The discussions following the Eqn.\ref{is0} also hold with the underlying CDW state as the 6 fold CDW state shown in Fig.13.
   The supersolid state is a 6 fold CDW supersolid. The excitation spectra across the 6-fold CDW to the 6-fold
   CDW supersolid transition is also given by the Fig.22
   with $ \Delta_{CDW} \sim w  $.
   Alternatively, starting from the new saddle point for the dual gauge gauge fields
   which is identical to that of the CDW-VB phase to be discussed in the following and shown in Fig.15b,
   one can also derive Eqn.\ref{kacdwvb} and use its following discussions.

\subsubsection{ $ w > 0 $ case:  CDW+VBS order: }

\begin{figure}
\includegraphics[width=3cm]{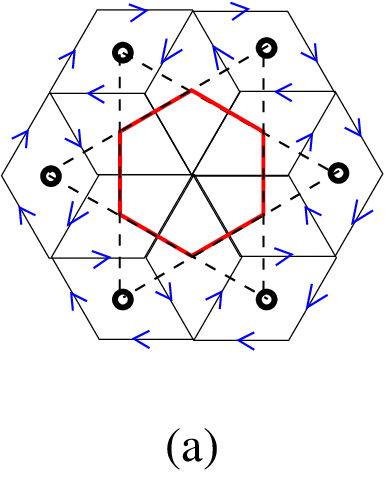}
\hspace{0.5cm}
\includegraphics[width=3cm]{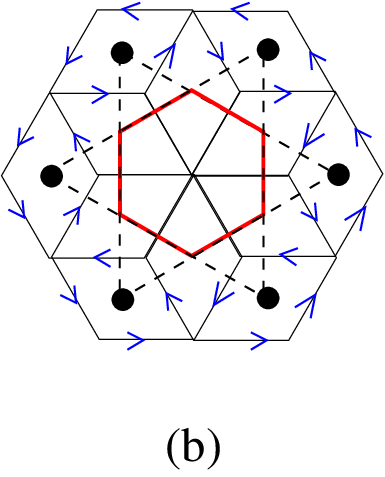}
\caption{The charge density wave + VBS state in a Kagome lattice in
the easy-plane limit $ v < 0, w > 0 $ at (a) $ f=1/3 $ (b) $ f=2/3
$. It is 3-fold degenerate.
The black solid ( empty ) circle is a boson ( an vacancy ). The
vortex at the center of the dual dice lattice vanishes, so there are
3 bosons moving around the red hexagon surrounding the center
leading to a local VB order (a) and (b) are related by time reversal
symmetry and particle-hole symmetry.  } \label{kagoeasy2}
\end{figure}

     When $ w > 0 $, $ \theta_{-} =\theta_0-\theta_1= ( 2 n + 1 ) \pi/3, n=0, 1,2 $.
     So the state has degeneracy $ 3 $ corresponding to the 3 possible ways to condense the  2 vortex fields.
     Using Eqns.\ref{density},\ref{bond}, one can evaluate
     both the vortex density, the kinetic energy and current in the dual dice lattice. This has been done in \cite{demler} in  a different context.
     The current was shown in the Fig.\ref{kagoeasy2}.  It is important to point out that the vortex at the dual center of the red hexagon vanishes
     which indicates the there is a local VB order around the dual center as shown by the red hexagon in the Fig.\ref{kagoeasy2}.
     Again, it is interesting to compare this with the known fact
     that  interacting bosons in a kagome lattice at $ f=1/2 $ \cite{five,ka1} are always in a SF state
     due to the localization of the vortices ( a flat vortex band ) \cite{subir}.
     The difference is that here there is only a local SF around the the center of the red hexagon, while there is a global
     SF across the whole lattice in the latter case.  By counting the segments of currents
     along the bonds surrounding the red and black lattice points, paying special attentions to
   their counter-clockwise or clock wise directions, we can calculate
   the densities at these points $ n_{b}=1/3 - 4 x I < 1/3, n_{r}= 1/3 + 2 x I > 1/3 $ with
   the constraint $ n_b+ 2 n_r=1 $ at the filling factor $ f=1/3 $
   where the $ I $ is the current flowing around the black lattice site in Fig.\ref{kagoeasy2}.
   The $ x \sim n_{r}-n_{b} >0 $ can be thought as a CDW-VB order parameter and can be tuned by the distance away from the SF to the CDW-VB transition in the Fig.3b.
   When one tunes the density $ n_{b}=0 $ which stand for vacancies, then $ n_{r}= 1/2 $ which stands for 3 bosons are hopping around the the red hexagon to form a
   local VB order.

   As said in the $ w < 0 $ case, the centers of all the hexagons in a Kagome lattice makes a triangular lattice,
   the centers of the Fig.\ref{kagoeasy2} occupies one of the 3 sublattices of this underlying triangular lattice,
   so the degeneracy of the CDW+VBS is  also 3  corresponding to the 3 possible ways to condense the 2 vortex fields.

   Very similar calculations can be made for the filling factor $ f=2/3 $ which is
   related to $ f=1/3 $ by a particle-hole transformation
   $ n_{b} \rightarrow \tilde{n}_b=1-n_b, n_{r} \rightarrow \tilde{n}_r=1-n_r $ with the constraint
   $ \tilde{n}_b+ 2 \tilde{n}_r= 2  $ shown in Fig.\ref{kagoeasy2}b.
   In fact, the 6-fold CDW state Fig.13 has the {\sl same} symmetry as the CDW-VB state in Fig.14.
   The only difference is the local CDW order versus a local VBS order.
   Taking the filling factor $ f=2/3 $ for a example, one can still distinguish
   the 6-fold CDW in Fig.13b from the CDW-VBS in Fig.14b: in the
   former, the boson density around the hexagon is always larger
   than that in the 6 corners $ n_r > n_b $, while in the latter,
   the boson density around the hexagon is always smaller
   than that in the 6 corners $ n_r < n_b $. This fact can indeed be
   used to distinguish the two states Fig.13 and Fig.14 in the QMC
   to be discussed in the Sect.III-D. In fact, the chirality $ \chi_p $ around
   the large hexagon in the 6-fold CDW in Fig.13b is just opposite
   to that the  CDW-VBS in Fig.15b, so one may use the chirality $
   \chi_p $ of the vortex current as the order parameter to distinguish the two states as $
   w < 0 $ changes to $ w > 0 $, so the sign of the $ w $ can be
   effectively identified as the vortex  chirality of the two states.

   It is interesting to compare the easy plane limit of a triangular lattice discussed in the last section with
   that in a Kagome lattice discussed in this section: in the $ w <
   0 $ case, both have a CDW state: stripe CDW and the 6-fold CDW,
   while in $ w > 0 $ case, both have a CDW-VBS state which has a VBS
   component, so the CDV-VBS state may be a common and robust state
   in any frustrated lattices.

   Well inside the CDW-VB phase, it is legitimate to set $ \theta_{-} =\theta_0-\theta_1= ( 2 n + 1 )\pi/3 $ in Eqn.\ref{kapm},
   the last term reaches its minimum $ - 2 w $.
   It is this $ w $ term which drives the formation of the CDW-VB phase in the Fig.\ref{kagoeasy2}.
   There is a  CDW-VB gap $ \Delta_{CDW-VB} \sim w  $.

{\sl (a) CDW+VB supersolid slightly away from $1/3 ( 2/3 )$ filling. }

   When studying the physics at slightly away from $ f=1/3 $ filling along the dashed line in Fig.3a,
   then  Eqn.\ref{kapm} reduces to \ref{is0}. The discussions
   following Eqn.\ref{is0} apply. In fact, one can bring out more interesting physics
   by studying the effects of the in-commensurate
   fillings directly in the CDW-VB side. In this side, one can start from the new saddle point for the dual gauge gauge
   fields corresponding to Fig.15b which is
 $ < \nabla \times \vec{A}_{CDW} > =n_b= 1- \alpha, < \nabla \times \vec{A}_{VBS} > =n_r=( 1 + \alpha )/2 $
 as shown in Fig,3b. The lowest energy dual vortex band in such a
 saddle point is found to be:
\begin{eqnarray}
  E(k_{1},k_{2} ) & = & - \sqrt{ 6 + 2 A(k_{1},k_{2} ) },~~~ -\pi < k_{1},k_{2} < \pi   \nonumber  \\
  A(k_{1},k_{2} ) & =  & \cos k_{1}- \cos( k_{1}-\alpha \pi ) + \cos k_{2}- \cos( k_{2}+ \alpha \pi )   \nonumber  \\
      & +  & \cos(k_{1}- k_{2})- \cos(k_{1}- k_{2}- 2 \alpha \pi )
\end{eqnarray}
     It is easy to see that there is a symmetry $ E(k_{1},k_{2} )=  E(k_{2} + \pi(1+\alpha) ,k_{1}+ \pi (1-\alpha )
     ) $. For $ 0 < \alpha < 0.404 $, there are two minima at $
     k_{1}= \alpha \pi/2-arcsin( \frac{ \sin(\alpha \pi/2)}{2 \cos(
     \alpha \pi ) }), k_{2}=-k_{1} $ and $
     k_{1}= \alpha \pi/2 + arcsin( \frac{ \sin(\alpha \pi/2)}{2 \cos(
     \alpha \pi ) }) + \pi, k_{2}=-k_{1} $. The two minima are related
     by the symmetry.

     When $ \alpha=1/3 < 0.404 $, it reduces to $ q=3 $ case where there are
     two minima at $ (0,0), (-\frac{2 \pi}{3},  \frac{2 \pi}{3} ) $ which is the two minima
     at the superfluid side. The $ \alpha \rightarrow 1/3^{-} $ case corresponds to
     approaching to the transition from the CDW+VB to the
     SF in Fig.3b from the left hand side.
     For $ \alpha =0 $ which corresponds to $ t/V_{1} \rightarrow 0 $ limit in Fig.17a,
     $  A(k_{1},k_{2} ) = 2 \cos(k_{1}- k_{2}) $, then  the minima is along the line $
     k_{1}=k_{2} $, the band is flat along this direction. At $ q=2 $, the
     band is completely flat as shown in \cite{five}. In fact, when $ \alpha=0 $, all the 6
     parallelograms around the center in the Fig.15b have $ q=2 $, the other 6 have $ q=1 $, so it is
     not a surprising that the band is flat along one direction.
     Of course, for $ \alpha =1 $, it reduces to the zero magnetic
     case $ q=1 $ where there is only one minimum at $ (0,0) $.

     Now  for  $ 0 < \alpha < 1/3^{-} $, because the two minima are smoothly connected to those in the SF side, let's still label the
     two minima by $ \phi_{l}, l=0,1 $. Then the MSG
     inside the SF state $ T_{1}, T_{2}, R_{\pi/3}, I_{1} $ is
     reduced to $ T_{\vec{a}_{1} + \vec{a}_{2}}, R_{\pi/3}, I_{\vec{a}_{1} + \vec{a}_{2}} $
     inside the CDW+VBS state in Fig.15. Note that only the translational symmetries are reduced, but
     the rotational symmetry remains the same  as inside the SF state.
     The effective action slightly away from $ 1/3 $ invariant under
     the MSG transformations inside the CDW+VBS state upto sixth order terms
     is  $ {\cal L}_{CDW+VBS} = \tilde{{\cal L}}_{0} + \tilde{{\cal L}}_{1} + \tilde{{\cal L}}_{2} $:
\begin{eqnarray}
    \tilde{{\cal L}}_{0} & = & \sum_{l} | (  \partial_{\mu} - i A^{VBS}_{\mu} ) \phi_{l} |^{2} +  \tilde{ r} | \phi_{l} |^{2}
                                   \nonumber  \\
   & + &  \frac{1}{4} ( \epsilon_{\mu \nu \lambda} \partial_{\nu} A^{VBS}_{\lambda}
                    - 2 \pi \delta f \delta_{\mu \tau} )^{2}      \nonumber  \\
    \tilde{{\cal L}}_{1} & =  &  \tilde{u} ( | \phi_{0} |^{2} + | \phi_{1} |^{2}  )^{2}
       - \tilde{v}  ( | \phi_{0} |^{2} - |\phi_{1} |^{2} )^{2}
                     \nonumber  \\
     \tilde{{\cal L}}_{2} & = & \tilde{w} [ (\phi^{*}_{0} \phi_{1} )^{3} +h.c.]
\label{kacdwvb}
\end{eqnarray}
     where $ A^{VBS}_{\mu} $ is a non-compact  $ U(1) $ gauge field.

     Note that $ A^{CDW}_{\mu} $ is always massive, so was already integrated out in Eqn.\ref{kacdwvb}.
     Because the system is already in the CDW+VBS, so we expect $
     \tilde{r} < 0 $ and $ \tilde{v} < 0 $.  Then we can set  $ \phi_{l}= |\phi| e^{i
     \theta_{l}} $. The relative phases can be determined by the sign of $
     \tilde{w} $ which has the same sign as $ w >0 $.
     Because Eqn.\ref{ka} in the SF side and Eqn.\ref{kacdwvb}
     in the CDW+VB side take the same form despite the reduced  translational symmetry in the CDW+VB
     state, also the $ w $ term in Eqns.\ref{ka} and \ref{kacdwvb} is only weakly irrelevant, so we
     expect the SF to the CDW+VB transition  driven by $ r $ is very weakly first
     order. This is in sharp contrast to the strong 1st order CDW to
     the SF transition in a triangular lattice in Fig.3a and Fig.9a.
     This picture achieved in the DVM can lead to some additional insights on why the transition was found to
     be a very weak first order one in a Kagome lattice \cite{ka2}.

    When moving from $ 1/3 $ to $1/2 $ where the bosons have to be in a SF state, Eqn.\ref{kacdwvb}
    reduces to Eqn.\ref{kapm} which, in turn, will reduce to an effective action similar to Eqn.\ref{is0}.
    The phase diagram  may be similar to Fig.3b in the triangular lattice.
    The SS slightly away from 1/3 filling has the same CDW and VB order as the
    solid at $ 1/3 $, so let's call this kind of novel SS as CDW-VB-SS.
    Then the transition from the CDW+VB to the
     CDW-VB-SS transition driven by the chemical potential can be
     similarly discussed as the VBS to the VB-SS in bipartite
     lattices \cite{univ}, so it is is the same universality class
     as that from the Mott insulator to the SF.
    The CDW-VB-SS to the SF transition is 1st order.
    Although the CDW-VB-SS has not been found stable in the hard core case \cite{ka2,ka3}, it should be stable in
    the soft core case. Of course, there is no P-H symmetry anymore
    in the soft core case. This conclusion may inspire
    more accurate QMC to search for this novel CDV-VB-SS state in the soft-core case.
    The excitation spectra across the CDW-VB to the CDW-VB supersolid transition is also given by the Fig.22
    with $ \Delta_{CDW} \sim w  $ or equivalently by Fig.25 with $ \Delta_{VBS} \sim w  $.

\begin{figure}
\includegraphics[width=3cm]{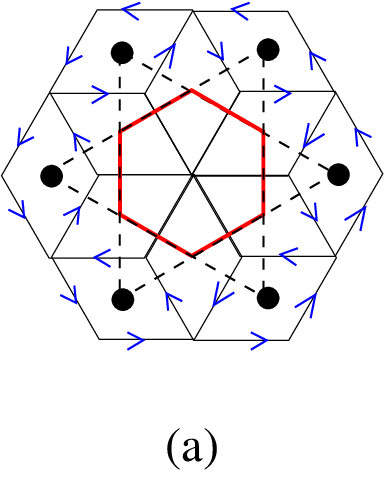}
\hspace{0.5cm}
\includegraphics[width=3cm]{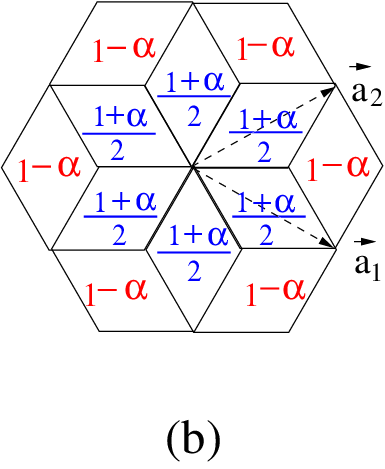}
\caption{( a ) The CDW+VBS in a Kagome lattice at $ f=2/3 $
identified in Fig.14. The black dot means one ( or $ 1-\alpha $ )
boson occupying the site. The red hexagon means 3 ( or $ 3(1+\alpha)
$ ) bosons hopping around the hexagon. The Kagome  has 3 sublattices
labeled as $ a, b, c $. ( b ) The saddle point structure of the dual
gauge field in a magnetic field unit cell in the dice lattice. The $
\vec{a}_{1} $ and $ \vec{a}_2 $ are the two basis vectors in the
dual dice lattice }
 \label{kagoeasy2saddle}
\end{figure}

     However, Eqn.\ref{kacdwvb} may break down as $ \alpha \rightarrow
     0 $ which corresponds to $ t/V_{1} \rightarrow 0 $ limit,
     because the two minima become very shallow. At $ \alpha=0
     $, the band is flat along $ k_{1}=k_{2} $, so we expect the regime of the CDW + VB-SS shrinks to
     zero as $ t/V_{1} \rightarrow 0 $ as shown in Fig.17a.

\subsection{ Stripe-CDW order: }

      As shown in the previous two sections, one is not able to get the stripe state at $ f=1/3 $ in Fig.16 from the dual vortex theory
     in either the Ising limit or easy-plane limit. But it is very easy to get this state from the $ NNN $ EBHM Eqn.\ref{bosonnnn} in a Kagome lattice.
     If we add very strong $ NNN $ interaction $ V_{2} $, the bosons will go to the
     $ NNNN $  which belong to $ a $ again ( Fig.16), then the bosons will
     simply take one of the 3 sublattices such as $ a $  to form a CDW alone ( Fig.16). Obviously, just like the 6-fold CDW order and the CDW+VBS order,
     the stripe phase is also 3 fold degenerate. Slightly away from $ 1/3 $, there is a stripe SS as shown in  Fig.17b.

     In the direct lattice Fig.16a, even if
     taking away the sublattice $ a $, the bosons can still move
     easily along the chain $ bcbc... $, so they need only overcome the
     barrier along the chain $ baba... $ to achieve an effective
     hopping along $ baba... $.  So the superfluid stiffness along the $ bc $ chain is  larger than that
     along the $ ba $ chain. In the stripe CDW state in Fig.16a, a
   different saddle point where  $ <\nabla \times \vec{A}^{a}> = 1-2 \alpha
   $ for the sublattice $ a $ and
   $ < \nabla \times \vec{A}^{bc}> =\alpha $ for the two sublattices $ b $ and $ c $ should be used.
   The $ \alpha \rightarrow \alpha_{c} < 1/3 $ limit corresponds to approaching
   the stripe CDW to the SF transition in Fig.17b from the stripe CDW side, while
   the $ \alpha \rightarrow 0^{+} $ limit corresponds to $ t/V_{1}
   \rightarrow 0 $ limit in Fig.17b.
   It is easy to see that there
   is only one vortex minimum $ \phi_{bc} $ in such a staggered dual magnetic
   field with $ \alpha <  \alpha_{c}  < 1/3 $, so the effective action inside the stripe CDW state should be the same as
   Eqn. \ref{is0} with $ yz \rightarrow bc $.
   Because of sharp change of the saddle point from Eqn.\ref{ka} in the SF side
   to Eqn.\ref{stripe} in the stripe CDW side, so the transition from the SF to the
   stripe CDW  along the horizontal axis  in Fig.17b is likely to be a strong first
   order.  Then the stripe-CDW to the stripe-SS transition
   driven by the chemical potential along the vertical axis is s second order transition
\begin{figure}
\includegraphics[width=7cm]{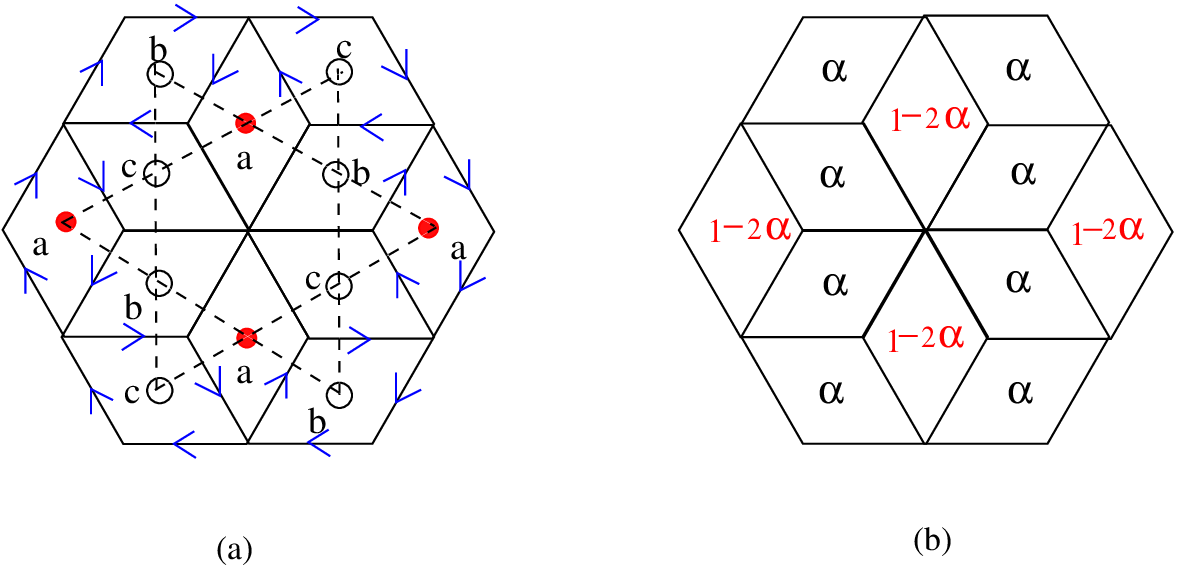}
\caption{( a ) The stripe CDW in a Kagome lattice at $ f=1/3 $. The
red dot means one ( or $ \alpha > 1/3 $ ) boson occupying a site in
the sub-lattice $ a $. It is 3-fold degenerate. ( b ) The saddle point structure of the dual
gauge field in a magnetic field unit cell in the dice lattice.
Compare the saddle point with Fig.15b } \label{fig15}
\end{figure}


\subsection{ Implications on QMC simulations in a Kagome lattice }

 {\sl (1) Nearest Neighbor ($ NN $) EBHM in a Kagome lattice, the CDW-VBS versus the 6-fold CDW state.  }

 Eqn.\ref{boson} ( namely, Eqn.\ref{bosonnn} ) in a Kagome lattice with $ U=\infty, V_{1} > 0 $
 was studied by two recent QMC simulations \cite{ka2,ka3}. It was found that the solid state at
 exactly $ f= 2/3 $ filling has both CDW and VB order. The VB order
 corresponds to boson hopping around a hexagon ( shown in Fig.14b ). This is a unique
 feature of the Kagome lattices. This feature can be intuitively
 understood without any calculation: with very large $ V_{1} $ interaction at $
 2/3 $ filling, if a boson taking
 sublattice $ c $, then all its 4 $ NN $  which belong to sublattices $ a $ and $ b $ are excluded, so bosons will go to
 the 4 $ NNN $  which still belong to $ a $ and $ b $ instead of $ c $ ( Fig.16a). This is in sharp contrast to triangular
 lattice at $ 2/3 $ filling in Fig.1a. So the bosons may move around the
 hexagon to form some VB order in addition to the CDW order.
 This state is just the CDW-VB state in Fig.14b. As explained in
 Sec.III-B-2, this CDW-VB state in Fig.14 can distinguished from the 6-fold
 CDW state Fig.13 by checking $ n_r < n_b $ or equivalently the chirality $ \chi_p $ around the large hexagon in Fig.14b. Unfortunately, this
 check was not explicitly spelled out in \cite{ka2,ka3}. The bond-bond
 correlation functions calculated in Ref.\cite{ka2} can not distinguish the two states.
 Although the  QMC in Ref.\cite{ka3} found the the SF to the CDW+VB transition is a 2nd order transition through a noval
 deconfined quantum critical point \cite{senthil}, the QMC in Ref.\cite{ka2} found a weakly 1st order
 transition from a double-peaked histogram of the boson kinetic energy.
 The dual vortex effective action to describe the transition from the SF to the CDW+VB is given by Eqn.\ref{ka}
 in the easy plane limit $ r<0, v<0, w > 0 $. As explained below
 Eqn.\ref{kacdwvb}, the picture achieved in the DVM can lead to some insights on why the transition
 could be be a very first order one as observed in \cite{ka2}.

 The effective action to describe the transition from the stripe solid to the stripe supersolid  slightly away from $ 1/3 $
 is given by Eqn.\ref{is0}.   Drawing the insights gained from the square lattice  \cite{univ},
 we conclude that the CDW-VB-SS is unstable against the phase
 separation in the hard core limit studied in \cite{ka2,ka3},  but may be stable in the soft core case.
 A QMC in the soft case has not been performed so far.

 {\sl (2) Next Nearest Neighbor ($ NN $) EBHM in a Kagome lattice: Stripe CDW and Stripe supersolid  }

 It may be interesting to do QMC with $ U=\infty, V_{1} > 0, V_{2} >0 $ to test Fig.17b.
 The Hamiltonian is the same as Eqn.\ref{bosonnnn}, but in a Kagome lattice.
 As elucidated in Sec. III-C, when both $ V_{1} $ and $ V_{2} $
 are large, a stripe-CDW state where bosons simply take one of the 3
 sublattices can be stabilized. Because the $ NN $  and the $ NNN $ neighbors
 are similar as can be seen from Fig.16a, the vertical axis in Fig.17b
 need to be replaced by $ \mu/( V_{1}+V_{2} ) $, the horizontal axis
 stay as $ t/V_{1} $ at fixed large $ V_{2} $ ( Fig.17b ).
 While the conclusion achieved in
 \cite{five} that the system has to be a superfluid state at $ 1/2 $ remains
 robust. So we conclude that when moving from $ 1/3 $ to $ 1/2 $,
 the system evolves from the stripe CDW to stripe-SS, then  to the SF. The first is
 the second order, the second is a first order transition.
 Again, longer-range interactions favor the description in terms of the dual
 vortices and the stability of SS, so the Stripe-SS will surely
 be stable even in the hard core case ( Fig.17b ).

 {\sl (3) Searching for the TVB state: possible ring exchange interactions }

      So far, the TVB state has not been identified in any QMC simulations.
      The ring exchange interaction $ - K_{k} \sum_{ijkl} ( b^{\dagger}_{i}
      b_{j} b^{\dagger}_{k} b_{l} + h.c. ) $  where $ i, j, k, l $ label
      the 4 corners of a bow tie in the Kagome lattice maybe needed to stabilize
      this TVB phase \cite{sandvik}.


\begin{figure}
\includegraphics[width=8cm]{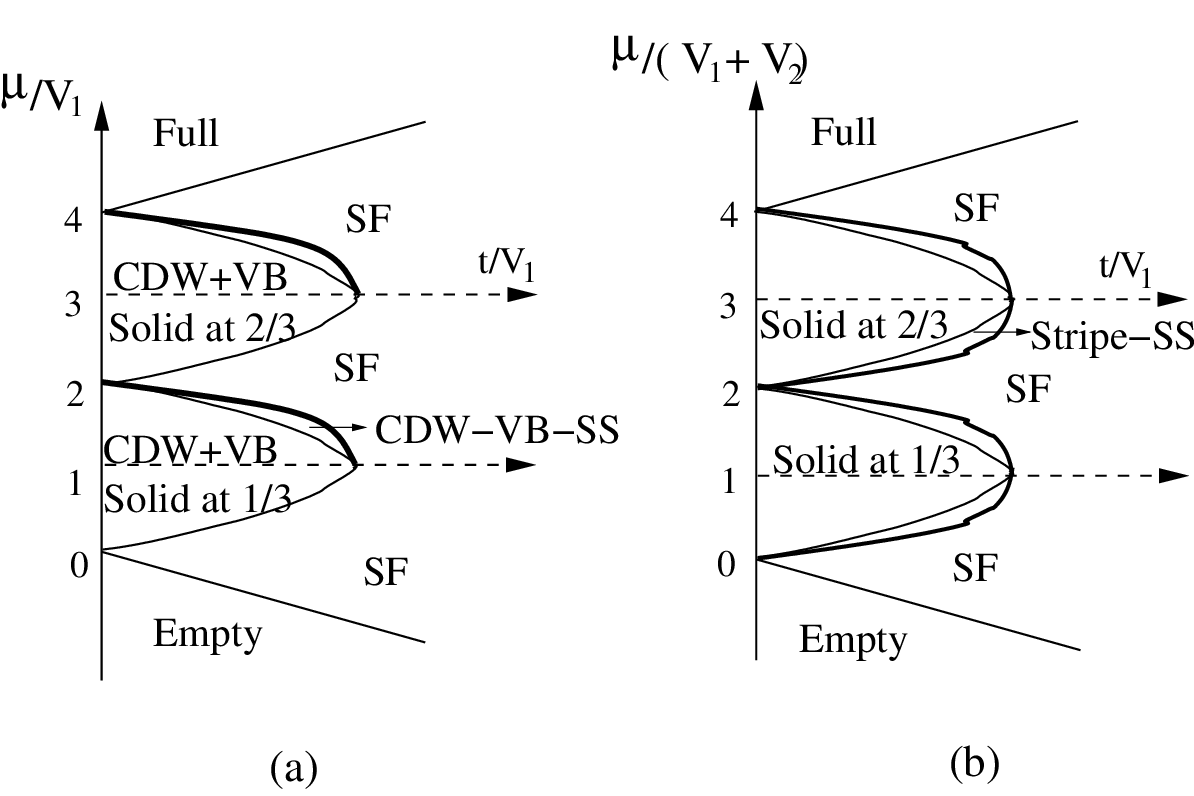}
\caption{( a ) Suggested zero temperature phase diagram in Kagome
lattice $ U =\infty, V_{1}>0 $. In the soft core case $ U < \infty
$, there maybe a narrow window of CDW-VB-SS sandwiched between the
CDW-VB solid and the SF. Of course, there is no P-H symmetry anymore
in the soft core case. In the hard core case studied in
\cite{ka2,ka3}, the CDW-VB-SS is unstable, there is a direct weakly
first order transition from the CDW-VB solid to the SF. Thin (thick)
line is 2nd (first ) order transition. It was shown in \cite{five}
that at half filling $ f=1/2 $ where $ \mu/V_{1}=2 $, it can only be
a SF state, in sharp contrast to $ q=2 $ on triangular lattice in
Fig.9. (b) Very similar phase diagram also works for $ U=\infty,
V_{1}>0, V_{2}>0 $. In this case, $ V_{1} $ in (a) need to be
replaced by $ V_{1}+ V_{2} $, the solid at $1/3 (2/3) $ is simply a
CDW where the bosons occupying ( un-occupying ) one of the 3
sublattices $ a,b,c $ in Fig.16a. The SS is simply a stripe SS where
the superfluid stiffness $ \rho_{s1} \gg \rho_{s2} $. The Stripe-SS
will surely be stable even in the hard core case. As shown in the
text, the first order transition in (a) is a weakly first order one,
but strongly first order one in (b) } \label{fig16}
\end{figure}

\section{ Conclusions }

        The DVM developed in \cite{pq1,five,univ,tri} and pushed further in this paper is a magnetic space group ( MSG )  symmetry-based approach
        which, in principle, can be used to classify all the possible phases
        and phase transitions. But if a particular phase identified by the DVM will become a stable
        ground state or not depends on the specific values of all the
        possible parameters in the EBHM in Eqn.\ref{boson}. This
        kind of question can only be addressed by a microscopic approach such as Quantum Monte-Carlo
        (QMC) simulations on a specific Hamiltonian. The DVM can guide the QMC to search for particular phases
        and phase transitions in a specific microscopic model. Finite size
        scalings in QMC in a microscopic model can be used to confirm the phases and the universality
        classes of phase transitions discovered by the DVM.
        The two methods are complementary to each other. Both
        methods have their own advantages and shortcomings.
        The combination of both methods are
        needed to completely understand quantum phases  and phase transitions in the EBHM Eqn.\ref{boson}.

     In the bipartite lattices, so far, there are several established examples of nice comparisons
     between the phenomenological DVM and the microscopic QMC simulations on a specific EBHM:
    (1)  The transition from the SF to the X-CDW ( Fig.19a ) in the Ising limit $ v >0  $ in a square lattice
         was studied by the DVM \cite{pq1} and the QMC on the EBHM Eqn.\ref{bosonnn}
         with the $ NN $  $ V_1 $ interaction in a square lattice at and near $ f=1/2 $ filling with both hard \cite{squarehard} and
         soft  core \cite{squaresoft} bosons.
    (2)  The transition from the SF to the X-CDW ( Fig.19b ) in the Ising limit $ v >0  $ in a honeycomb lattice was studied by the DVM \cite{univ}
         and the QMC on the EBHM Eqn.\ref{bosonnn} with  $ NN $  $ V_1 $ interaction in a honeycomb lattice at $ f=1/2 $
         with both hard  and soft core bosons\cite{honey}.
     Unfortunately, there are still many cases where the DVM and QMC simulations can not be directly compared with each other.
     For example,  the DVM is not able to get the stripe-solid phase
     in a square lattice studied extensively by QMC  \cite{squarehard} on the EBHM of hard core bosons with the $ NNN $ $ V_2 $
     interaction Eqn.\ref{bosonnnn} at $ f=1/2 $.
     Interestingly, the stripe solid phase in Fig.4 in a triangular lattice has been achieved by the DVM in the easy plane limit $ v < 0, w < 0 $.
     However, the stripe solid phase Fig.16a in a Kagome lattice can not be achieved from the the DVM yet.
     It remains an open problem how to achieve the strip solid phases
     in a square and a Kagome lattice by the DVM. The above two connections are established only in the Ising limit.
     On the other hand, so far, it has not been possible to study any of the VBS phases identified in the easy-plane limit of the DVM listed in the Fig.20 and Fig.23
     by QMC simulations on any microscopic models.
     This maybe  due to the fact that a ring exchange interaction may be needed to stabilize all these VBS phases. But it is still not known
     what are the specific forms of these ring exchange interactions.

     In the frustrated lattices, so far, there are also several  established examples of nice comparisons
     between the phenomenological DVM and  the microscopic QMC simulations on a specific model. For example,
     The DVM \cite{ka1,five} predicted that the Kagome lattice at $ 1/2 $ must be a superfluid due to the complete flat band of
     the dual vortices in the dual Dice lattice \cite{subir}. No insulating phases can be possible at $ 1/2 $. This result
     has been confirmed by QMC \cite{ka2}. In the following, we list 3 more connections:
    (1)  The transition from the SF to the X-CDW ( Fig.2 ) in a triangular lattice
         at $ f=1/3 $  was studied by the DVM in \cite{tri}
         in the Ising limit $ v >0  $  of the effective action  Eqn.\ref{tri0} and also by the  QMC \cite{ss12,ss3,ss4} on the EBHM of hard core
         bosons with the $ NN $  $ V_1 $ interaction Eqn.\ref{bosonnn}.
    (2)  The transition from the SF to the Stripe-CDW  ( Fig.4)  in a triangular lattice
         at $ f=1/3 $  was studied by the DVM in \cite{tri} in the easy plane limit $ v < 0, w< 0 $
         of the effective action Eqn.\ref{tri0}  and also by the
         QMC \cite{trinnn} on the EBHM of hard core bosons with the $ NNN $ $ V_2 $  interaction Eqn.\ref{bosonnnn} at $ f=1/3 $ in a triangular lattice.
    (3)  The transition from the SF to the CDW-VB phase in a Kagome lattice at $ f=2/3 $  was thoroughly studied by the DVM in this paper
         in the easy plane limit $ v < 0, w > 0 $ of the effective action Eqn.\ref{ka}
         and the QMC in \cite{ka1,ka2} on the EBHM of hard core bosons with the $ NN $ $ V_1 $ interaction Eqn.\ref{bosonnn}  at $ f=2/3 $ in a Kagome lattice.

         In this paper, by using the simple and effective method developed to identify the insulating
         phases in a direct lattice in terms of vortex degree of freedoms in the dual lattice,
         we confirmed the first and the second connections, also established the third explicitly for the first time.
         The first connection is on the Ising limit.
         It was known that the transition is first order in the Ising limit.
         The last two are the only two known examples so far realized in the easy plane limit where  it was believed there is some chance to realize
         the DCQCP \cite{senthil} proposed from the DVM \cite{pq1}.
         The last one is the only known example involving some VB order where there is a better chance to realize the DCQCP proposed from the DVM \cite{pq1}.
         In terms of the realization of possible VB orders, the advantages of the frustrated lattices over the bipartite lattices are
         that local VBS order can be realized
         in the former even without any ring exchange interaction. Unfortunately, both cases as possible examples to realize the DCQCP came up as negative.
         The first is a first order transition. From the DVM in Sect.II-B-1, it is related to a sharp change
         of the saddle point structure of the dual gauge field from the SF to the stripe CDW phase.
         The second is a very weak first order transition which needs some high power QMC technique such as double-peaked histogram
         to distinguish the very weak first order from a possible second order transition. From the DVM in Sect.III-B-2,
         we showed that the form of effective action in the CDW-VB side is identical to that in the SF side despite the reduced translation
         symmetry in the CDW-VB side.
         This gives an intuitive explanation why the transition is a very weak first order as observed in the QMC in \cite{ka2}.

         We also inspired further comparisons between the two approaches. For example, the CDW+VB order in a triangular lattice at $ 1/3 $ filling in Fig.7
         was not even searched in the QMC
         in \cite{trinnn} in any parameter regimes. We suggest to measure the bond-bond correlations to identify this very interesting state by
         refining the QMC in \cite{trinnn}.
         We also suggest the QMC simulations of the  EBHM Eqn.\ref{bosonnnn} with the $ NNN $  $ V_2 $ interaction in a Kagome lattice at $ 1/3 $
         to search for a stripe solid phase  and a stable stripe supersolid phase
         slightly away from $ 1/3 $. It is still not known how to realize the TVB phase in the triangular lattice Fig.8  and in a Kagome lattice Fig.10
         by a microscopic model.
         However, it was shown that the excitation spectra above both TVB phases is necessarily of a density wave origin, so there is no
         TVB supersolid phase in both triangular and Kagome lattice even away from $ 1/3 $ filling.
         We expect more examples along these comparisons are needed to have
         a complete picture of quantum phases and phase transitions in various lattices at various filling factors
         from the two complementary approaches.

    In summary, we used the DVM developed in \cite{pq1,univ}
    to study various kinds of insulating phases and phase transitions in the two most common
    frustrated lattices such as triangular and kagome lattices at and slightly away $ 1/3 $ and $ 2/3 $ fillings.
    At the commensurate fillings $ 1/3 (2/3) $, we developed a systematic way to identify the symmetry breaking insulating states
    uniquely and completely by using gauge invariant vortex density, kinetic energy and current on the dual lattices.
    We reproduced several known phases found by the previous methods, also identified several new phases.
    The two of the most important new phases are the CDW+VBS in the
    triangular lattice in Fig.7 and the CDW + VBS in the Kagome
    lattice in Fig. 14, both happen in the same easy-plane limit with $ v<0, w>0 $. These new phases have both CDW and VBS
    orders and are unique to frustrated lattices. It is the  frustration which leads to the very existence of these phases.
    This consistency between the triangular lattice and the Kagome lattice hints that the bubble phase
    identified in Ref.\cite{tri} can not be the correct phase in the easy-plane limit with $ v<0, w>0 $.

    It is interesting to compare the vortex motion in the dual lattice with
    that of the boson in the direct lattice in Fig.7 and Fig.14: the
    vortex motion can have both kinetic energies and the currents,
    however the boson motion  has only kinetic energies, but no boson
    currents due to the time reversal symmetry of the original boson
    EBHM. This should not be too surprised, because the
    vortex-charge duality is not self-dual anyway.
    Local correlations of vortices on the dual lattice can be
    mapped to the local correlation of bosons in the direct
    lattice precisely. It is this delicate local relation which is lacking in the
    previous Density operator formalism (DOF) used in \cite{pq1,tri}.
  As detailed in the appendix C, there is an intrinsic problem with the DOF: the DOF is NOT periodic under  $ f \rightarrow f+1 $.
  Of course, the original EBHM in Eqn.\ref{boson}
  is not periodic under $ f \rightarrow f+1 $, because all the interaction terms do depend on the filling factors $ f $. However,
  the universal symmetry breaking patterns of the insulating states should be periodic under $ f \rightarrow f+1 $,
  although the microscopic details such as the energy of the states may not be periodic under $ f \rightarrow f+1 $.
  We believe that so far,  the scheme developed in this manuscript is the only correct and complete method to
  identify the symmetry breaking patterns of the bosons in the original lattice  in terms of the vortex degree of freedoms in the dual lattice.

    We contrast different symmetry breaking patterns  of the insulating phases in both Ising limit and two different and Easy
    plane limits in the two frustrated lattices.
    There is always a first order transition between the Ising limit
    and one of the two easy-plane limit, also between the two
    different easy-plane limit.
    By  analyzing carefully the  saddle point structures of the dual gauge fields in the translational symmetry breaking sides and
    pushing the effective action slightly away from the commensurate fillings,
    We classified all the possible types of supersolids and analyze their stability conditions.
    By using these effective actions, we studied different kinds of solids, different kinds of
    supersolids and the universality class of quantum phase transitions from the solids to the supersolids.
    We also compared our results achieved by the DVM with some available QMC
    results and make implications on possible future QMC simulations.
    Although supersolids on lattices are different than that in continuous systems such as
    $^{4} He $ \cite{qgl} or exciton supersolid in electron-hole bilayer system
    in \cite{ess}, they may share  some common properties and shed considerable lights on each other.
    The results achieved in this paper should have direct impacts on
    possible near future experiments of  ultra-cold atoms with long range interactions loaded on optical lattices.
    Assuming these novel phases can be realized in these ultra cold atom experiments, the
    optical scattering and atom spectroscopy detections of these
    phases  are discussed in details in \cite{bragg}.

{\bf Acknowledgements }

    We thank Fuchun Zhang and Zidan Wang for the hospitality
    during our visit at Hong Kong university which initiated the collaboration.
    J.Ye  thank G.G. Batrouni and  S. V. Isakov for  very helpful  discussions,
    Longhua Jiang for technical assistances in
    finding the saddle point structures in the CDW+VBS phase in the Kagome lattice,
    Han Pu for his hospitalities during
    his visit at Rice university. Y. Chen thanks Yaowu Guo for technical
    assistances.  JY was supported by NSF-DMR-1161497, NSFC-11074173,11174210,
Beijing Municipal Commission of Education under grant No.PHR201107121, at
KITP was supported in part by the NSF under grant No. PHY-0551164.  Y. Chen's research was supported by the National Natural
Science Foundation  of China (Grant No. 10874032), the State Key Programs of China
    (Grant no. 2009CB929204) and Shanghai Municipal Government.

\appendix

\section{ Applications to the CDW and VBS states in a square  lattice }

 In this appendix, we apply the same method in the Section II to study the symmetry breaking patterns of the insulating states in a square lattice.
 We only focus at the $ q=2 $ case.  We reproduced the results achieved in \cite{pq1} using the operator formalism. Obviously, our method leads to
 much more intuitive and physical identifications of these states.
 Fig.20a,b and Fig.19a have also been previously
 identified in \cite{disagree} in the context of $ Z_2 $ gauge theory of high temperature superconductors. However, we disagree with
 their arguments leading to Fig.20b of the plaquette state. We believe that this disagreement involves  different physical interpretation of
 a local VBS state. Our interpretations are very important for the identifications of all the important phases in frustrated lattices discussed in the main text.
 Furthermore, we also explicitly give elementary excitation spectra in the dimer and plaquette VBS state.

\begin{figure}
\includegraphics[width=6cm]{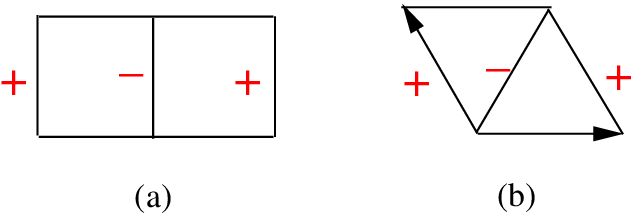}
\caption{The bond phase factors in a square and a honeycomb lattice
at half-filling. (a) a dual square lattice  (b) a dual triangular
lattice. } \label{fig17}
\end{figure}

 As a warm-up, we can first apply our method to look at the
 simplest superfluid to Mott transition at integer fillings. Inside the SF $ \langle \psi \rangle =0 $, in the Mott state, $ \langle \psi \rangle \neq 0 $.
 Then the vortex density $ |\psi(\vec{x})|^{2} $ is a constant in the
 Mott state. Also in any bond $ \psi^{\dagger}(\vec{x})
 \psi(\vec{x}) $ is always real, so no currents are flowing in the dual
 lattice, so no CDW or VBS in the direct lattice. The Mott state is
 a translational symmetric state.

 The Harper's equation at $ q=2 $ in a square lattice  \cite{five} leads to $ c_{m}(l=0) = c_{m} =( 1, \sqrt{2}-1 ) $
 ( The case at a general $ q $ will be discussed in the appendix C ). $ \omega= e^{ i 2 \pi f } = -1 $.
 Denoting the dual square lattice site by $ \vec{x} = (a_1, a_2 ) $.
 The vortex operator is:
\begin{eqnarray}
   \psi( \vec{x} ) & = &  \sum^{1}_{m=0}  c_{m} e^{ i \pi m a_1 }[  \xi_0
   + \xi_1 \omega^{-m} e^{ i \pi a_2 } ]     \nonumber  \\
   & = & \sum^{1}_{m=0}  c_{m} (-1)^{ m a_1 }[  \xi_0 + \xi_1 (-1)^{m} (-1)^{ a_2 } ]
\end{eqnarray}

   In the permutative representation:
\begin{eqnarray}
  \xi_0 & = &  ( \phi_0 + \phi_1 )/\sqrt{2}   \nonumber  \\
  \xi_1 & = &  -i( \phi_0 - \phi_1 )/\sqrt{2}
\end{eqnarray}

   In the Ising case $ v > 0 $:  $  \phi_0=1,  \phi_1=0 $ or vice versa.
   So the state has the degeneracy $ 2 $ corresponding to which of the 2 vortices condenses.
   The vortex fields, kinetic energies and currents are shown in Fig.19a

\begin{figure}
\includegraphics[width=4.5cm]{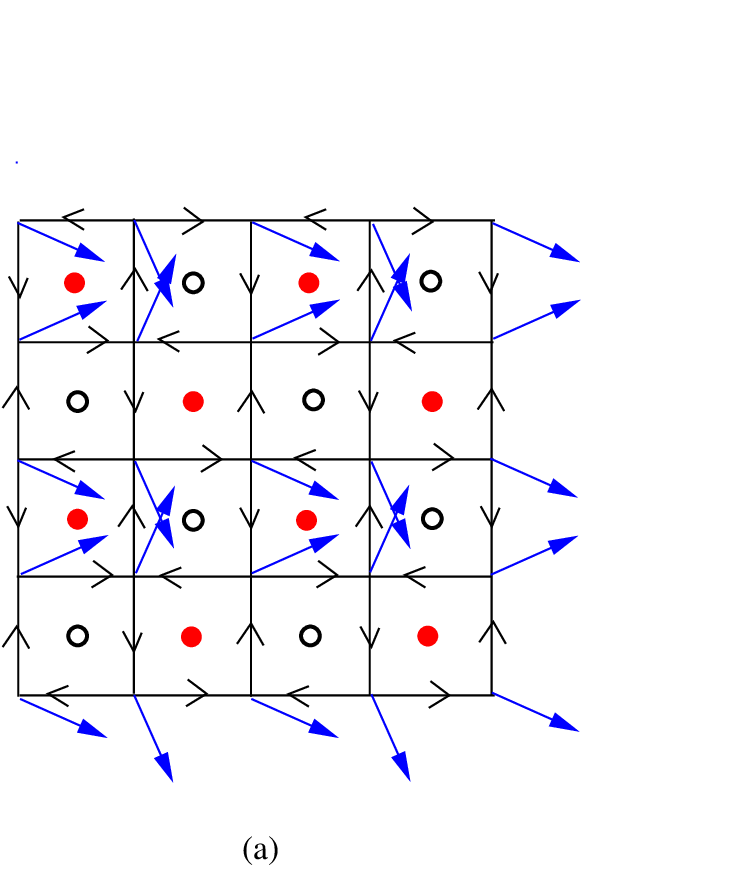}
\hspace{0.5cm}
\includegraphics[width=2.5cm]{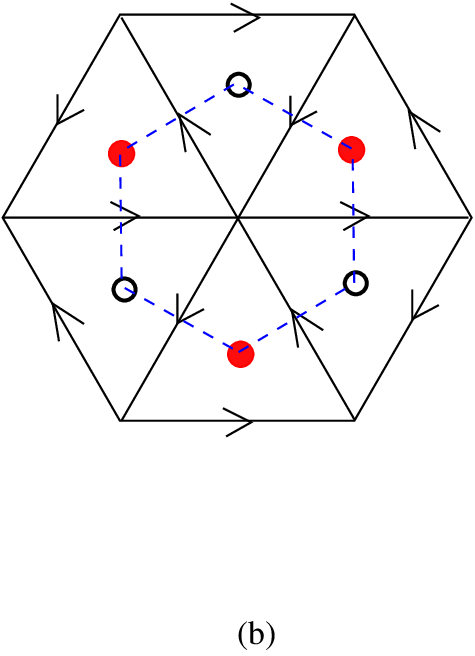}
\caption{ The X-CDW in the Ising limit in (a)  a square lattice.  It is 2 fold degenerate.
  The vortex fields
$ \psi( 0,0) =   1-( \sqrt{2}-1)i, \psi( 0,1) =  1+ ( \sqrt{2}-1)i,
  \psi( 1,0) =   ( \sqrt{2}-1)- i, \psi( 1,1) =  ( \sqrt{2}-1) + i $.
  The vortex current flowing counter clockwise is $ I=  \sqrt{2}-1 $. All the bonds have the same vortex kinetic energy $ K= \sqrt{2}-1 $.
  There are no frustrated bonds.
  Following the current counting methods in the main text, one can see the density at the red
  site is $ n_r= 1/2+4 x I $, that at the black spot is $ n_b= 1/2 -4 x I$. The $ x \sim n_r-n_b $ tunes the distance away from
  the SF to the CDW transition. When $ n_b=0 $, then $ n_r=1 $ which corresponds to the classical limit $ t/V_1=0 $.
  (b) in a honeycomb lattice. It is also 2 fold degenerate. The vortex current flowing counter clockwise is $ I=  \sqrt{3}/2 $.
  All the bonds have the same vortex kinetic energy $ K= 1/2 $.
  There are no frustrated bonds. The density at the red
  site is $ n_r= 1/2+3 x I $, that at the black spot is $ n_b= 1/2 -3 x I$. The $ x \sim n_r-n_b $ tunes the distance away from
  the SF to the CDW transition.}
\label{fig19}
\end{figure}

  In the easy plane limit \cite{pq1} $ v < 0 $, there are also two cases depending on the signs of the quartic term  $ \lambda \cos 4 \theta $:

  (A) If $ \lambda > 0 $, then $ \phi_0= \phi_1 e^{i n \pi/2 }, n=0,1,2,3 $.
      So the state has a degeneracy $ 4 $ corresponding to the 4 possible ways to condense the  2 vortex fields.
      The vortex fields, kinetic energies of the $ n=0 $ case with  $ \phi_0= \phi_1=1 $ is shown in Fig.20a.
      It is a dimer state.

  (B) If $ \lambda < 0 $, then $ \phi_0= \phi_1 e^{i ( n+1/2 ) \pi/2 }, n=0,1,2,3 $.
      So the state has a degeneracy $ 4 $ corresponding to the 4 possible ways to condense the  2 vortex fields.
      The vortex fields, kinetic energies of the $ n=0 $ case with  $ \phi_1=1 $ is shown in Fig.20b
      which is plaquette state.

\begin{figure}
\includegraphics[width=3cm]{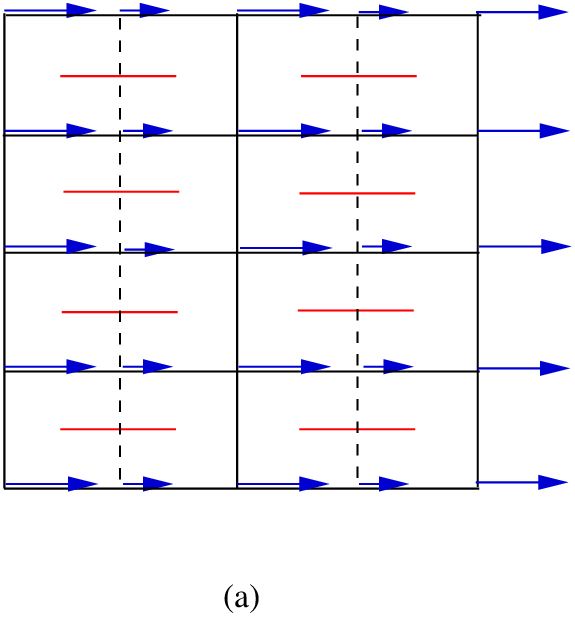}
\hspace{0.5cm}
\includegraphics[width=3cm]{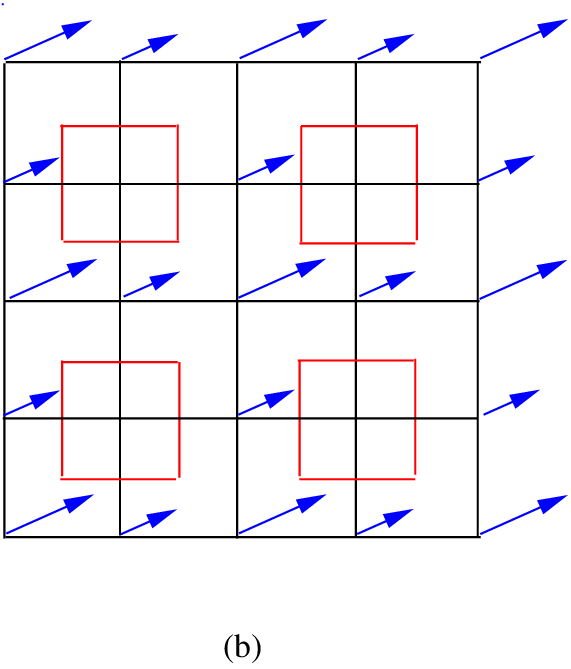}
\caption{ The ground states in the easy-plane limit $ v < 0 $ in a square lattice
(a) Dimer state at $ \lambda > 0 $. It is 4 fold degenerate. The long arrow is $ \sqrt{2} $,
the short arrow is $ \sqrt{2} ( \sqrt{2}-1) $. The dashed line is the frustrated vortex bond with strength $ K_{fv}=-2 (\sqrt{2}-1)^{2} $.
The un-frustrated vertical vertex bond is $ K_{v}=2 $. The horizontal bond is $ K_{h}=  2 (\sqrt{2}-1) $.
The red bond which is perpendicular to the frustrated vortex bond is the boson valence bond in the direct lattice.
The bosons are hopping back and forth along the red bond.
(b) Plaquette state at $ \lambda < 0 $. The vortex fields
$ \psi( 0,0) = 2  [1+ ( \sqrt{2}-1)i], \psi( 0,1) = \sqrt{2} [1+ ( \sqrt{2}-1)i],
  \psi( 1,0) = \sqrt{2} [1+ ( \sqrt{2}-1)i]= \psi( 0,1), \psi( 1,1) = 0 $. The arrows indicate both the magnitude and the phase of the vortex fields.
  Because the vortex field vanishes at $ (1,1) $, so there is a local boson superfluid ( or local boson VB order )
  around this dual lattice point as indicated by the red plaquette. The bosons are hopping around the red plaquette.
  The kinetic energies emanating from the zero vortex field points are zero.  All the other bonds have strength $ K= 8 ( \sqrt{2}-1) $.
  It is also 4 fold degenerate.
  There are no frustrated bonds.  There are no vortex currents in both (a) and (b) indicating no CDW order,
  so the average boson densities are fixed at $ f=1/2 $. }
\label{fig18}
\end{figure}

  The corresponding local low energy excitations in the dimer and plaquette states are shown in the Fig.21.
  Because the density remains uniform, so there is no associated density wave excitation. The translational moving of these local excitations
  through the whole lattice  leads to the excitations spectra shown in Fig.24.

\begin{figure}
\includegraphics[width=6cm]{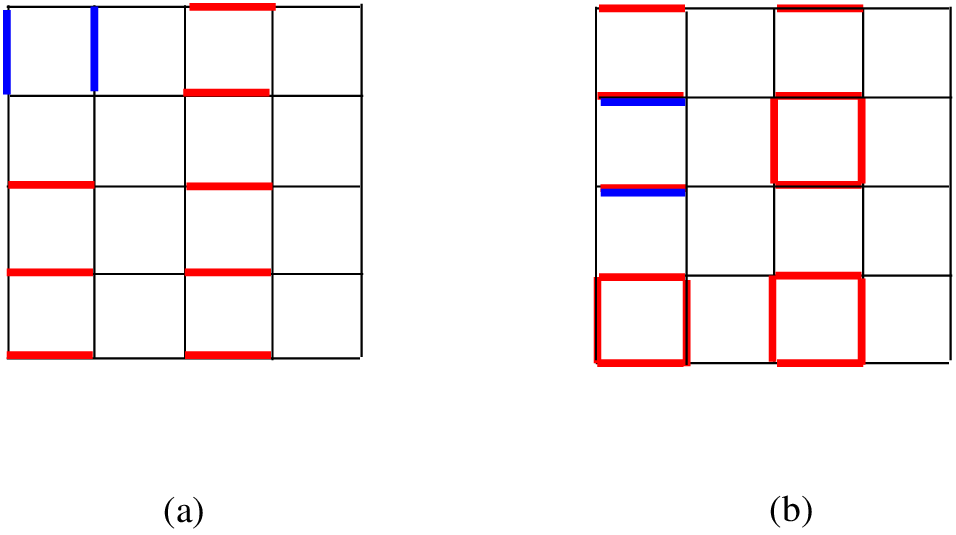}
\caption{ Local bond flip excitations in the dimer (a) and (b)
plaquette states. The flipped bonds are denoted as green bonds.
Their translational motions through the whole lattice lead to the
the VBS excitation spectrum in Fig.24a. Compare with a local density
excitation in the TVB in a triangular and Kagome lattice in Fig.11.
} \label{fig192}
\end{figure}

     It is interesting to point out that one is not able to get a stripe phase which also has the degeneracy $ 4 $.

\section{ Applications to the CDW and VBS states in a honeycomb  lattice }

    In this appendix, we apply the same method in the Section II to study the symmetry breaking insulating states in a honeycomb lattice
    whose dual lattice is a triangular lattice.  We only focus at the $ q=2 $ case. We reproduced the dimer state in the Fig.23a
    achieved in \cite{univ}. Unfortunately, the plaquette state in Fig.23b was not identified in \cite{univ}.
    We also work out the excitation spectra in all these insulating states and corresponding supersolid states.

    The Harper's equation at $ q=2 $ in a honeycomb lattice leads to the vortex field in the Eqn.13 in \cite{univ}:
\begin{equation}
   \psi( \vec{x} )  =  e^{ i \vec{k} \cdot \vec{x} } \phi_{+} ( \vec{x} ) + e^{ - i \vec{k} \cdot \vec{x} } \phi_{-} ( \vec{x} )
\end{equation}
   where $ \vec{k} = ( \pi/3, \pi/3 ) $.

   In the Ising case  $ v > 0 $:  $  \phi_{+}=1,  \phi_{-}=0 $ or vice versa.
   So it is 2 fold degenerate.
   The vortex fields, kinetic energies and currents are shown in Fig.19b.
   the corresponding excitation spectra are given in the Fig.22.

\begin{figure}
\includegraphics[width=6cm]{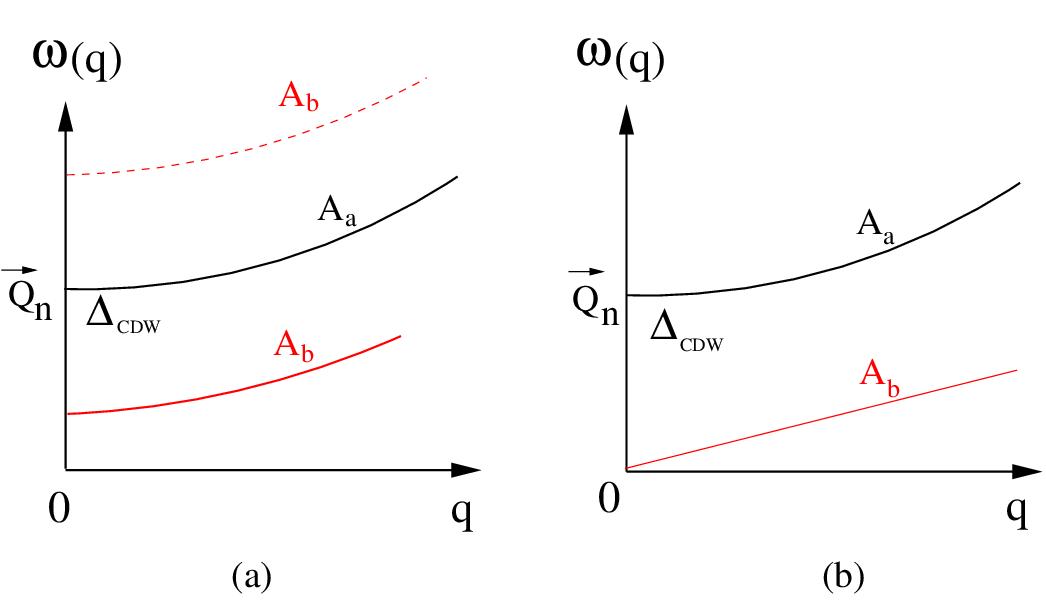}
\caption{ The excitation spectra  in (a) CDW. The $ \vec{Q}_n $ is
the CDW ordering wavevector. The lower $ A_{b} $ mode appears due to
the non-zero chemical potential $ \mu $ in Fig.3a.
          In the absence of this chemical potential, namely, $ \mu =0 $, then the upper $ A_{b} $ mode shown as a dashed red line is well above the CDW mode $ A_a $.
          (b) CDW+ SS state. The $ A_b $ is the gapless mode inside the CDW+ SS state.  }
\label{cdwex}
\end{figure}

   In the easy plane limit \cite{univ} $ v < 0 $, there are also two cases depending on the signs of the cubic term  $ w \cos 3 \theta $:

   (A) If $ w > 0 $, then $ \phi_{+}= \phi_{-} e^{i ( 2n + 1 ) \pi/3 }, n=0,1,2$.
       So the state has a degeneracy $ 3 $ corresponding to the 3 possible ways to condense the  2 vortex fields.
        The vortex fields, kinetic energies of
        the $ n=0 $ case with  $ \phi_{-}=1 $  is shown in Fig.23a.  It is a dimer state.

   (B) If $ w < 0 $, then $ \phi_{+}= \phi_{-} e^{i 2 n \pi/3 }, n=0,1,2 $.
       So the state has a
       degeneracy $ 3 $ corresponding to the 3 possible ways to condense the  2 vortex fields.
       The vortex fields, kinetic energies of the $ n=0 $ case with $ \phi_{-}=1 $  is shown in Fig.23b. It is plaquette state.

\begin{figure}
\includegraphics[width=3cm]{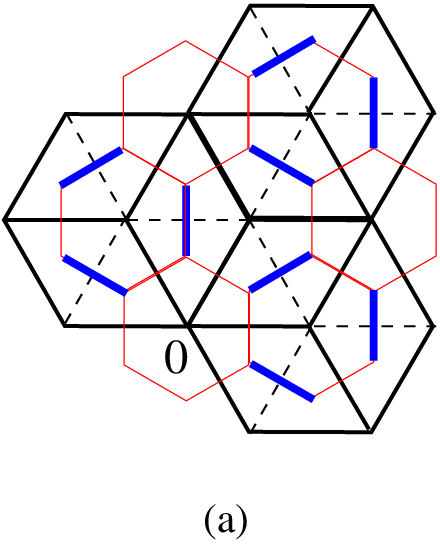}
\hspace{0.5cm}
\includegraphics[width=3cm]{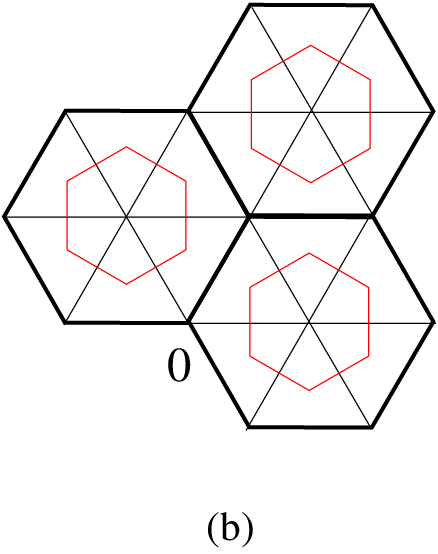}
\caption{Easy-plane limit in a honeycomb lattice
 (a) Dimer state.  The dashed line is the frustrated vortex bond with strength $ K_{f}=-1 $.
The un-frustrated vertex bond is $ K =2 $. The blue bond which is
perpendicular to the frustrated vortex bond is the boson valence
bond in the direct honeycomb lattice. The bosons are hopping along
the blue bond back and forth. Because one can put the blue bond along any of the three bond orientations, so the degeneracy is 3. (b) Plaquette state.
  Because the vortex field vanishes at $ (1,0) $, the kinetic energies emanating from the zero vortex field points are also zero,
  so there is a local boson superfluid  around this dual lattice point as indicated by the red hexagon.
  The bosons are hopping around the red hexagon. There are no frustrated bonds.  All the other  bonds have strength $ K= 3$.
  Because the centers of the red hexagon take one of the three sublattices in the dual triangular lattices, so the degeneracy is 3.
  There are no vortex currents in both (a) and (b) indicating no CDW order, so the average boson densities are fixed at $ f=1/2 $. }
\label{fig20}
\end{figure}

     Again, the corresponding local low energy excitations in the dimer and plaquette states in the honeycomb lattice
     can be similarly constructed as those in the square lattice Fig.21.
     Because the density remains uniform, so there is no associated density wave excitation.
     The translational moving of these local excitations
     through the whole lattice  leads to the excitations spectra shown in Fig.22.
     From Eqn.22 in \cite{univ}, we can get the excitation spectra across the VBS to the VB-SS transition in Fig.24.
     It is instructive to compare Fig.24 with its counterpart across the CDW to CDW-SS transition Fig.22.
\begin{figure}
\includegraphics[width=6cm]{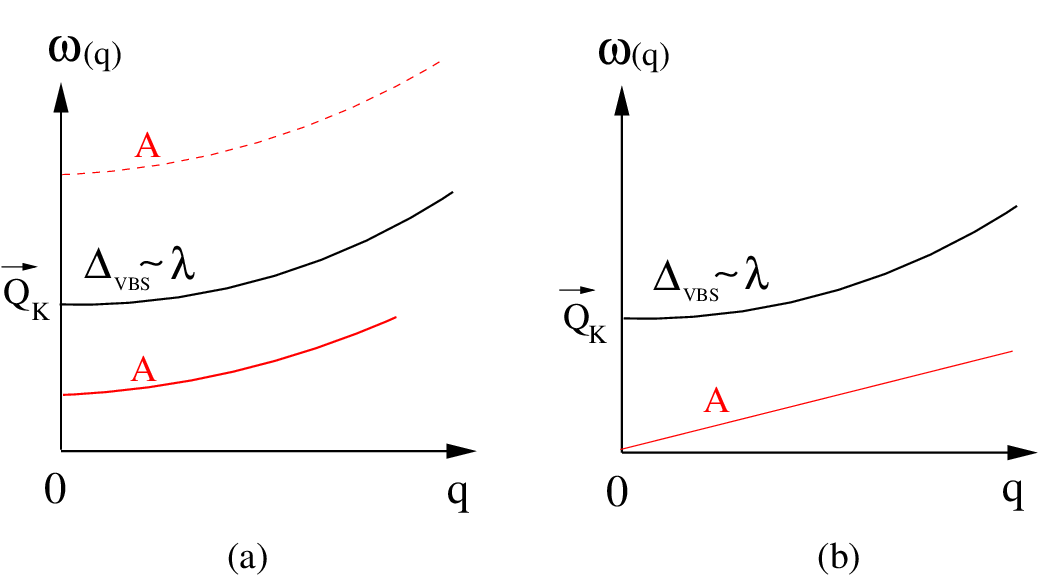}
\caption{ The excitation spectra  in (a) VBS. The $ \vec{Q}_{K} $ is
the VBS ordering wavevector. The VBS gap is given by the  operator $
 \lambda \cos 3 \theta $ in Eqn.22 in \cite{univ}.
          The lower branch $ A_b $ mode appears due to the non-zero chemical potential
          $ \mu $ in Fig.3b in \cite{univ}.
             In the absence of this chemical potential, namely, $ \mu =0 $, then the upper $ A_b $ mode shown as a dashed red line is well above the VBS mode.
  (b) VB-SS state. The $ A_b $ is the gapless mode inside the VB-SS state. }
\label{fig22}
\end{figure}

\section{ Comments on Density operator formalism (DOF) in a square latices }

In this appendix, we compare our gauge invariant density operator, kinetic energy and current methods with the density operator formalism developed
in \cite{pq1,tri}. We pointed out some potential problems with the density operator formalism in \cite{pq1,tri} and also stress
the different conclusions reached by the two methods.

In the dual vortex methods developed in \cite{pq1}, how to characterize the
CDW order of bosons on direct lattice in terms of vortex operators on dual lattice is a little
bit tricky.  In \cite{pq1}, the density operator parameter was constructed to be the most general gauge invariant bilinear combinations of
the vortex fields:
\begin{eqnarray}
 \tilde{\rho}( \vec{x} ) & =  & \sum^{q-1}_{r,s=-q} \tilde{\rho}_{r,s} e^{i 2 \pi f(rx+sy ) }  \nonumber \\
 \tilde{\rho}_{rs}  & = &  S( | \vec{Q}_{rs} |  ) \omega^{-rs/2} \sum^{q-1}_{l=0} \phi^{*}_{l}\phi_{l+s}  \omega^{-lr}
\label{pq1density}
\end{eqnarray}
    where $ \vec{Q}_{rs}=  2 \pi f( r,s ) $ is the ordering wavevector and $ S( | \vec{Q}_{rs} | ) $ is a general form factor which can not be determined
    from symmetry considerations and depend on the microscopic details. For simplicity without affecting the symmetry of the underlying state,
    it can be taken as a Lorentzian $ S(Q)= \frac{1}{ Q^{2}+1 } $. Then Eqn.\ref{pq1density} was evaluated at the direct lattice point, the link points
    and the dual lattice points to represent the boson density, kinetic energy and the ring exchange amplitudes respectively.
    As argued in \cite{univ}, the link points of a square lattice is still a square lattice with a lattice constant $ \sqrt{2} $,
    the dual lattice  points of a square lattice is still a square lattice withe same lattice constant. Putting the direct, link and dual lattices together
    forms a square lattice with a lattice constant $ 1/2 $.  This may be the reason why the density operator in Eqn.\ref{pq1density}
    is the sum from $ -q $ to $ q-1 $ and contains a factor $  \omega^{-rs/2} $ which is non-periodic under $ f \rightarrow f + 1 $.
    As pointed out in \cite{univ}, it is very hard to extend the density operator defined in \cite{pq1} in square lattice
to other lattices. In this appendix, we give a simple way to construct the density and bond operators  for any general $ q $ in
terms of the dual vortex operator and then we will discuss the relations and difference between the two methods.

\begin{figure}
\includegraphics[width=6cm]{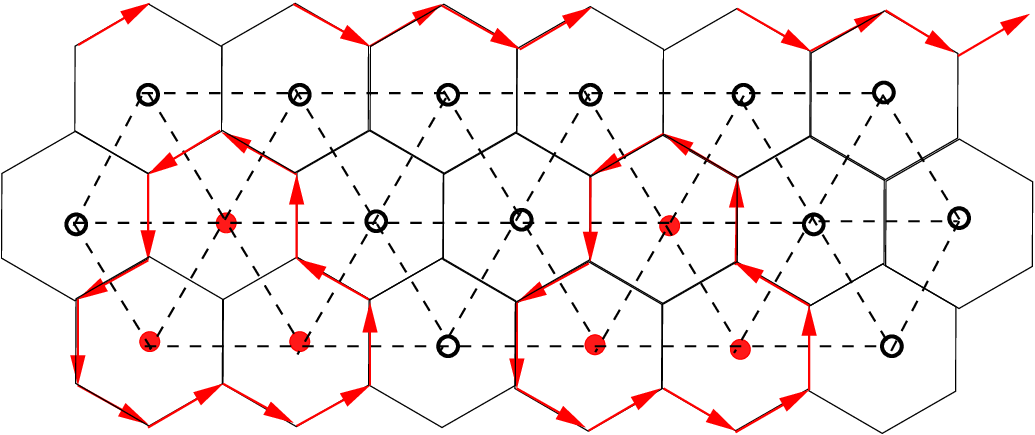}
\caption{The bubble CDW phase in a triangular lattice at $ f=1/3 $
 proposed in \cite{tri} in the easy plane limit $ v<0, w> 0 $.  It is 18 fold degenerate.
   Drawn are currents in this bubble phase.
   By counting the segments of currents
   along the bonds surrounding hexagons, one can calculate
   the densities at the red and black  points $ n_{r}=1/3 + 4 x I > 1/3, n_{b}= 1/3 - 2 x I < 1/3 $.
   The $ x \sim n_{r}-n_{b} >0 $ can be thought as a bubble CDW order parameter.
   When one tunes the density $ n_{b} =0 $ which stand for vacancies, then $ n_{r}= 1 $ which stands for one boson.
   Note the vortex field at the center of the red loop is non-vanishing, only the current emanating from the center
   vanish due to the cancelations of currents flowing around the 3 neighboring hexagons, but the kinetic energies are positive.
   This indicates a local CDW order.
   Compare this bubble phase to the CDW+VB phase in the Fig.7.
   There is only one current flowing here leading to 2 different boson densities.
   While there are two different currents flowing in the CDW+VBS phase in Fig.7 leading to 3 different boson densities and also local VBS order. }
\label{bubble}
\end{figure}

The q eigenstates  $ \chi_{l}; l = 0, 1,\cdots, q-1  $ which forms
a q dimensional representation of the magnetic space group ( MSG ) [38] can be written
as $ \chi_{l}(\vec{x}) =   \frac{1}{\sqrt{N}} \sum^{q-1}_{m=0} c_{m} \omega^{-m l} e^{ i 2 \pi f ( m x + l y) } $
 where  $ \vec{x} = (x, y) $. Then expand the
vortex operator $ \psi(\vec{x}) = \sum^{q-1}_{l=0} \phi_{l}(\vec{x}) \chi_{l}(\vec{x}) $  where $ \phi_{l}(\vec{x}) $
 are the q order parameters. Just like Eqns.\ref{density},\ref{bond} in the triangular lattices, one can construct the gauge invariant
 generalized density operator (GDO) $ \rho( \vec{x} )   =  \psi^{*}(\vec{x}) \psi(\vec{x}) $, the bond operators along the horizontal and vertical directions
 $ K_{h} ( \vec{x} )   =  \psi^{*}(\vec{x}+ \vec{a}_{1} ) \psi(\vec{x}),
 K_{v} ( \vec{x} )  =  \psi^{*}(\vec{x}+ \vec{a}_{2} ) e^{i 2 \pi f a_1} \psi(\vec{x}) $:
\begin{eqnarray}
 (\rho, K_{h}, K_{v} )( \vec{x} ) &  = &  \sum^{q-1}_{r,s=0} (\rho_{rs}, K^{h}_{rs}, K^{v}_{rs} )  e^{i 2 \pi f(rx+sy ) }  \nonumber  \\
 \rho_{rs}  & = &  A_{rs}[ \omega^{-rs} \sum^{q-1}_{l=0} \phi^{*}_{l}\phi_{l+s}  \omega^{-lr} ]   \nonumber   \\
 ~~~A_{rs}  & =  & \frac{1}{N} \sum^{q-1}_{m=0} c^{*}_{m} c_{m+r} \omega^{-m s}  \nonumber  \\
 K^{h}_{rs}  & = &  A^{h}_{rs}[ \omega^{-rs} \sum^{q-1}_{l=0} \phi^{*}_{l}\phi_{l+s}  \omega^{-lr} ]   \nonumber   \\
 ~~~A^{h}_{rs}  & =  & \frac{1}{N} \sum^{q-1}_{m=0} c^{*}_{m} c_{m+r} \omega^{-m (s+1) }
     \nonumber  \\
 K^{v}_{rs}  & = &  A^{v}_{rs}[ \omega^{-(r-1) s} \sum^{q-1}_{l=0} \phi^{*}_{l}\phi_{l+s}  \omega^{-lr} ]   \nonumber   \\
 ~~~A^{v}_{rs}  & =  & \frac{1}{N} \sum^{q-1}_{m=0} c^{*}_{m} c_{m+r-1} \omega^{-m s }
\label{squaredensitybond}
\end{eqnarray}
  By construction, all these operators transform like a density, a bond or a current operator and is also periodic under
$ f \rightarrow f + 1 $. We need evaluate these quantities only at dual lattice points to characterize
the CDW and VBS orders in the direct lattice. The simplest case at $ q=2 $ was discussed in the appendix A.


     Note that gauge invariant physical quantities
     constructed in Eqns.\ref{squaredensitybond} where the sum is from $ 0 $ to $ q-1 $ and contain only $ f \rightarrow f+1 $ periodic
     factors such as $  \omega^{-rs} $. However, in Eqn.\ref{pq1density}, the sum is from $ -q $ to $ q-1 $ and
     contains a factor $  \omega^{-rs/2} $ which is non-periodic under  $ f \rightarrow f+1 $.
     So the $ \tilde{\rho}_{rs} $ in Eqn.\ref{pq1density} is very different than the $ \rho_{rs} $ in Eqn.\ref{squaredensitybond}.
     Even more differently, Eqn.\ref{pq1density} contains no bond information.
     Of course, the original EBHM in Eqn.\ref{boson} is not periodic under $ f \rightarrow f+1 $, because all the interaction terms do depend on the filling factors $ f $, however, we believe the
     the symmetry of the insulating states should be periodic under $ f \rightarrow f+1 $. This is what the Eqn.\ref{pq1density} originally designed for:
     characterize the symmetry breaking patterns of the insulating states only without the ability to spell out the microscopic details.
     As argued in \cite{univ}, although Eqn.\ref{pq1density} seems work well in a square lattice and gives the same states
     shown in the appendix A, but it may not work in a honeycomb lattice. Similar density operators in a triangular were constructed in \cite{tri},
     they gave the same CDW state in the Ising limit $ v > 0 $ and the stripe CDW state in the easy plane limit $ v <0, w < 0 $, but it leads to the bubble CDW state
     in the other easy plane limit $ v<0, w > 0 $  in Fig.25 which, as shown in Sect.II-C,
     is completely different than the CDW-VB state shown in Fig.7. This disagreement calls for
     some re-examinations of the density operators constructed in \cite{pq1,tri}.
     No such density operators were constructed on honeycomb and Kagome lattices yet.

\section{ The vortex fields at a triangular lattice in the easy plane limit $ v<0, w > 0 $. }

   The vortex fields are shown graphically in the Fig.5 and
   have period 3 along both $ a_1 $ and $ a_2 $ directions.
   Their specific values at $ \vec{x}= ( a_1=0,1,2, a_2=0,1,2 ) $ are listed in the following two $ 3 \times 3 $ matrices:
\begin{widetext}
\begin{eqnarray}
 \Phi_{a} ( \vec{x} )   =  \left ( \begin{array}{ccc}
     2 \sin( \pi/3) \cos( \pi /18)e^{i \pi /9}       &  2 \sin( \pi /3) \cos( \pi /18)e^{-i 4 \pi /9}      &   0 \\
     2 \sin( \pi/3) \cos( \pi /18) e^{i 2\pi /9} &  0  &   2 \sin( \pi/3) \cos( \pi /18) e^{-i 5\pi /9} \\
     3 \sin(4\pi/9) e^{i \pi /6}  &  2 \sin( \pi/3) \cos( \pi /18) e^{-i 5\pi /9}  &  -2 \sin( \pi/3) \cos( \pi /18) e^{i \pi /9}
     \\   \end{array}   \right )  ~~~~~~~~~~~~~~~~~~~~\nonumber  \\
 \Phi_{b} ( \vec{x} )  =   \left ( \begin{array}{ccc}
      \sin(\pi/3)e^{-i \pi /6}      &  -2i \sin(\pi/3) \cos(2\pi /9)     &   2 \sin(\pi/3) \cos(2\pi /9) e^{i \pi /6}  \\
     -\sin(\pi/3)   e^{i  \pi /6}  &   - \sin(\pi/3)   e^{i  \pi /6}   & 2 \cos(2\pi /9) ( \sin( \pi /9)+ \sin( 2\pi/9) ) e^{i \pi /6}   \\
      -2i \cos(2\pi /9) ( \sin(4 \pi /9)+ \sin(2\pi /9) )   &  - 2 \sin(\pi/3) \cos(2\pi /9) e^{i \pi /6}  &   2 \cos(2\pi /9) ( \sin(4 \pi /9)+ \sin(2\pi /9) ) e^{i \pi /6} \\  \end{array}   \right )
\label{fieldvalues}
\end{eqnarray}
\end{widetext}


\begin{thebibliography}{99}
\bibitem{dof} Such a relation is not needed in the DOF which was
constructed from general symmetry principles in \cite{pq1,tri}. See appendix C.
\end{thebibliography}
\end{document}